\pdfoutput=1
\documentclass[PAPER, cernpreprint, UKenglish, texmf, orcidlogo]{atlasdoc}
 
\usepackage[multirow]{atlaspackage}
\usepackage{atlasbiblatex}
 
\usepackage{atlasphysics}

\graphicspath{{logos/}{figures/}}
 
\usepackage{ANA-STDM-2018-32-PAPER-defs}
 
\usepackage{bm}
\usepackage{tablefootnote}
 
\LEcontact{Bob Kowalewski bob.kowalewski@cern.ch}

\AtlasTitle{Measurement and interpretation of same-sign $W$ boson pair production in association with two jets in $pp$ collisions at $\sqrt{s} = 13$~TeV with the ATLAS detector}

\AtlasAbstract{
This paper presents the measurement of fiducial and differential cross sections for both the inclusive and electroweak production of a same-sign $W$-boson pair in association with two jets ($W^\pm W^\pm jj$) using $139$~fb$^{-1}$ of proton--proton collision data recorded at a centre-of-mass energy of $\sqrt{s}=13$~TeV by the ATLAS detector at the Large Hadron Collider. The analysis is performed by selecting two same-charge leptons, electron or muon, and at least two jets with large invariant mass and a large rapidity difference.
The measured fiducial cross sections for electroweak and inclusive $W^\pm W^\pm jj$ production are $2.92 \pm 0.22\, \text{(stat.)} \pm 0.19\, \text{(syst.)}$~fb and $3.38 \pm 0.22\, \text{(stat.)} \pm 0.19\, \text{(syst.)}$~fb, respectively, in agreement with Standard Model predictions.
The measurements are used to constrain anomalous quartic gauge couplings by extracting 95\%\ confidence level intervals on dimension-8 operators.
A search for doubly charged Higgs bosons \(H^{\pm\pm}\) that are produced in vector-boson fusion processes and decay into a same-sign $W$ boson pair is performed. The largest deviation from the Standard Model occurs for an \(H^{\pm\pm}\) mass near $450~\mathrm{GeV}$, with a global significance of 2.5 standard deviations.
}

\AtlasRefCode{STDM-2018-32}
\PreprintIdNumber{CERN-EP-2023-221}

\HepDataRecord{141650}
 
\AtlasJournal{JHEP}
\AtlasJournalRef{\JHEP 04 (2024) 026}
\AtlasDOI{10.1007/JHEP04(2024)026}

\hypersetup{pdftitle={ATLAS document},pdfauthor={The ATLAS Collaboration}}

\newcolumntype{Z}{>{\setbox0=\hbox\bgroup}c<{\egroup}@{\hspace*{-\tabcolsep}}}
 
\begin{document} 
 
\maketitle
 
\tableofcontents

\section{Introduction}

 
The study of massive vector boson scattering (VBS), $VV\to VV$ ($V=W,Z$), at the Large Hadron Collider (LHC) probes the mechanism of electroweak (EW) symmetry breaking (EWSB) in the Standard Model (SM), and provides unique sensitivity for new physics phenomena that affect the gauge sector~\cite{Chanowitz:1988ft,Bagger:1993zf,Bagger:1995mk}.
In the SM, couplings to the Higgs boson prevent the divergence of longitudinally polarised VBS amplitudes at high energies and unitarity violation at the TeV scale~\cite{Lee:1977yc,Lee:1977eg,Alboteanu:2008my,Rindani:2009gm}.
 
In proton$-$proton ($pp$) collisions, VBS at leading order (LO) involves two initial quarks, each of which radiates a vector boson. The two bosons subsequently interact and then decay. The two outgoing quarks fragment usually close to the beam direction.
The final state consists of two vector bosons and two jets ($j$), $VVjj$, and can also be produced from non-VBS processes.
 
\begin{figure}[htb]
\centering
\subfloat[]{\includegraphics[width=.26\textwidth]{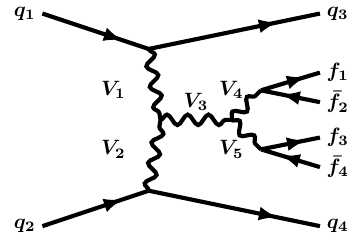}\label{fig:Vbscore_a}}
\subfloat[]{\includegraphics[width=.26\textwidth]{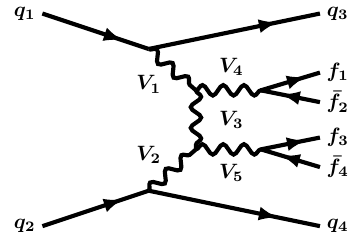}}
\subfloat[]{\includegraphics[width=.26\textwidth]{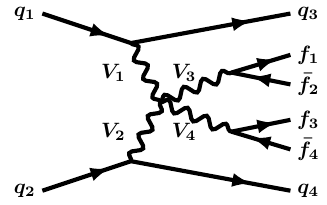}}\\
\subfloat[]{\includegraphics[width=.26\textwidth]{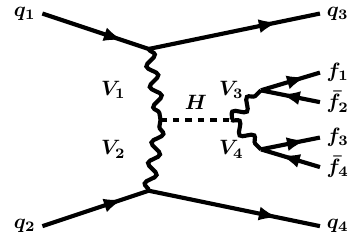}\label{fig:Vbscore_d}}
\subfloat[]{\includegraphics[width=.26\textwidth]{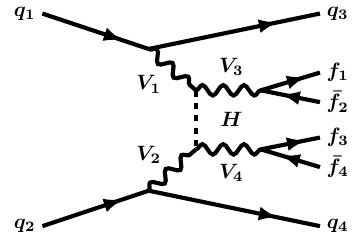}}
 
\caption{Representative Feynman diagrams for EW $VVjj$ production with a scattering topology
that includes either a triple-gauge-boson vertex with an internal electroweak gauge boson in the (a) $s$-channel or the (b) $t$-channel, (c) a quartic gauge boson vertex, or the exchange of a Higgs boson in the (d) $s$-channel or the (e) $t$-channel. The lines are labelled by quarks ($q$), vector bosons ($V=W/Z/\gamma$), the Higgs boson ($H$) and fermions ($f$). The $s$-channel diagrams, (a) and (d), are forbidden in the SM for \sswwjj final states. The diagram (d) is possible in extensions of the SM with a doubly charged Higgs boson. In these and following Feynman diagrams, not all boson combinations are allowed by the Standard Model.}
\label{fig:diagrams_ewk_vbs}
\end{figure}

\begin{figure}[htb]
\centering
\includegraphics[width=.28\textwidth]{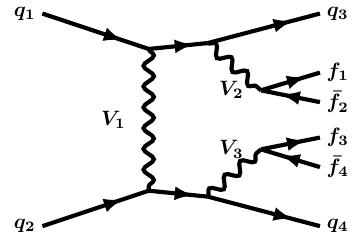}\hspace*{1cm}
\includegraphics[width=.28\textwidth]{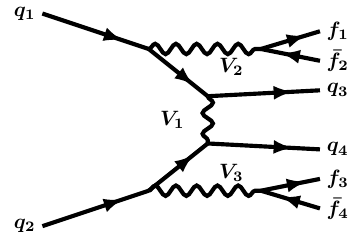}\\
\caption{Representative Feynman diagrams for EW $VVjj$ production without vector-boson scattering. The lines are labelled by quarks ($q$), vector bosons ($V=W/Z/\gamma$), and fermions ($f$).}
\label{fig:diagrams_ewk_nonvbs}
\end{figure}

\begin{figure}[htb]
\centering
\includegraphics[width=.24\textwidth]{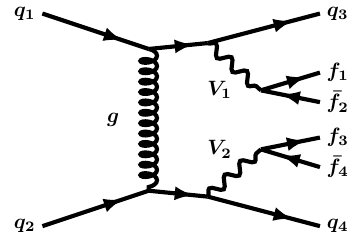}\hfill
\includegraphics[width=.24\textwidth]{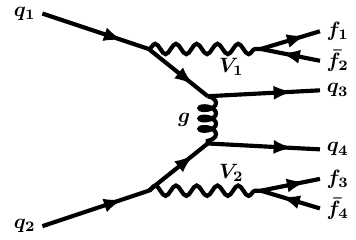}\hfill
\includegraphics[width=.24\textwidth]{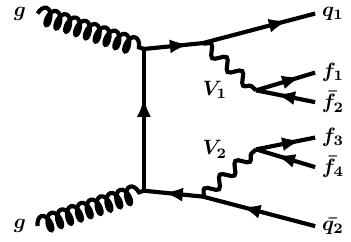}\hfill
\includegraphics[width=.24\textwidth]{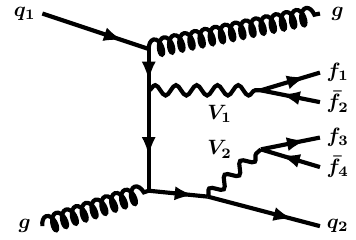}\\
\caption{Representative Feynman diagrams for QCD $VVjj$ production with strong interaction vertices. The lines are labelled by quarks ($q$), vector bosons ($V=W/Z/\gamma$), fermions ($f$), and gluons (g). The two diagrams on the right with gluons in the initial state are forbidden for \sswwjj production.}
\label{fig:diagrams_qcd}
\end{figure}

The production of $VVjj$ at LO has contributions both from modes that involve only the EW-interaction vertices (EW $VVjj$) and from modes that involve the strong interaction (quantum chromodynamics, QCD) vertices (QCD $VVjj$).
Representative LO Feynman diagrams for EW $VVjj$ are shown in Figures~\ref{fig:diagrams_ewk_vbs} and~\ref{fig:diagrams_ewk_nonvbs}.  The leading QCD $VVjj$ diagrams are shown in Figure~\ref{fig:diagrams_qcd}, where the two diagrams with gluons in the initial state are forbidden when there are two $W$ bosons with the same electric charge produced, \sswwjj. The EW production is further categorised into two components. The first component is EW VBS production (see Figure~\ref{fig:diagrams_ewk_vbs}), which involves the actual scattering of the two vector bosons. The scattering occurs via triple or quartic gauge vertices, the $s$- or $t$-channel exchange of a Higgs boson or a $W/Z/\gamma$ boson. The diagrams with bosons in the $s$-channel, shown in Figures~\ref{fig:Vbscore_a} and~\ref{fig:Vbscore_d}, are forbidden in the SM for \sswwjj final states. The diagram in Figure~\ref{fig:Vbscore_d} is possible in extensions of the SM with a doubly charged Higgs boson. The second component is EW non-VBS production (see Figure~\ref{fig:diagrams_ewk_nonvbs}) with EW vertices only, where the two bosons do not interact. The EW non-VBS component cannot be separated from the EW VBS component in a gauge-invariant way and is therefore considered as a contribution to the signal. Triboson production where one boson decays hadronically also results in the $VVjj$ final state. These processes only contain EW interactions and are separable in a gauge-invariant manner~\cite{Ballestrero:2018anz}. The resonant decay of a boson into two quarks can be suppressed by applying a requirement on the invariant dijet mass $\mjj$ arising from the two quarks. As a consequence, triboson processes are suppressed in the EW VBS phase-space region.
 
The \sswwjj final state has the largest ratio of electroweak to strong production cross sections among final states sensitive to VBS diboson processes~\cite{Bagger:1995mk}; this is because
quark--gluon and gluon--gluon initiated diagrams are absent
at LO accuracy in perturbative QCD and contributions from quark and (anti-)quark annihilation diagrams are suppressed. This production ratio is of order five in the fiducial phase-space region of the analysis. The $s$-channel VBS diagrams with trilinear
interactions are absent in the \sswwjj final state. In addition, electroweak diagrams not involving self interactions are suppressed~\cite{Szleper:2014xxa}, which enhances the sensitivity of this final state to gauge-boson self couplings.
 
\sswwjj scattering where both $W$ bosons decay leptonically (into $e\nu$ or $\mu\nu$, collectively denoted by $\ell\nu$) is a sensitive process for studying VBS, since the electric charges of $W$ bosons can be determined directly from leptons.
The final state consists of two leptons with the same electric charge, two neutrinos, and two jets with a large rapidity separation. The requirement of the presence of two leptons with the same electric charge significantly reduces SM backgrounds coming from top--anti-top quark pair production (\(t\bar{t}\)) and  $Z$ + jets events.
 
The first studies of the EW \sswwjj production were performed by the ATLAS and CMS experiments using Run~1 LHC $pp$ dataset with the centre-of-mass energy of $\sqrt{s}=8$~TeV~\cite{STDM-2013-06,STDM-2014-05,CMS-SMP-13-015}. They were followed by an observation of this process by both experiments using partial LHC Run~2 data at $\sqrt{s}=13$~TeV~\cite{CMS-SMP-17-004,ATLAS:2019cbr}. More recently, the CMS Collaboration published an updated measurement using full Run~2 data, corresponding to an integrated luminosity of $137$~fb$^{-1}$~\cite{CMS-SMP-19-012}.
 
This paper presents fiducial and differential EW and inclusive \sswwjj production cross section measurements at $\sqrt{s}=13$~TeV using $139$~fb$^{-1}$ of data recorded by the ATLAS detector at the LHC\@.
There are several changes with respect to the previous result~\cite{ATLAS:2019cbr}.  The current measurement uses improved baseline signal modelling, implements a tighter veto on additional leptons, includes a charge selector tool based on boosted decision trees for rejecting electron candidates where the charge is likely misidentified, employs a data-driven extraction of the $\mjj$ shape of the $WZ$ background, and incorporates updates to the estimation of non-prompt lepton background.
 
The distribution of reconstructed dilepton invariant mass $m_{\ell\ell}$ for \sswwjj event candidates is used in an Effective Field Theory (EFT) interpretation to constrain dimension-8 (D-8) operators~\cite{Eboli:2006wa}, to extract one-dimensional limits on these operators and to simultaneously constrain two operators (two-dimensional limits).
Using the distribution of the transverse mass of the dilepton system and the missing transverse momentum ($\mt$), a search for doubly charged Higgs bosons produced in vector-boson fusion (VBF) processes and decaying into a pair of same-sign $W$ bosons is performed in the context of the Georgi--Machacek (GM) model~\cite{Georgi:1985nv}, and 95\% Confidence Level (CL) limits are extracted on the model parameter describing contributions of the isotriplet scalar fields to the masses of the $W$ and $Z$ bosons, and on the corresponding cross section times branching ratio.

The data corresponding to many plots presented in this paper and a Rivet~\cite{Bierlich:2019rhm} routine are available on HEPData~\cite{Maguire:2017ypu}.


\section{Experimental setup}

 
\newcommand{\AtlasCoordFootnote}{
ATLAS uses a right-handed coordinate system with its origin at the nominal interaction point (IP)
in the centre of the detector and the \(z\)-axis along the beam pipe.
The \(x\)-axis points from the IP to the centre of the LHC ring,
and the \(y\)-axis points upwards.
Cylindrical coordinates \((r,\phi)\) are used in the transverse plane,
\(\phi\) being the azimuthal angle around the \(z\)-axis.
The pseudorapidity is defined in terms of the polar angle \(\theta\) as \(\eta = -\ln \tan(\theta/2)\).
Angular distance is measured in units of \(\Delta R \equiv \sqrt{(\Delta\eta)^{2} + (\Delta\phi)^{2}}\).}
 
The ATLAS detector~\cite{PERF-2007-01} at the LHC covers nearly the entire solid angle around the collision point.\footnote{\AtlasCoordFootnote}
It consists of an inner tracking detector surrounded by a thin superconducting solenoid, electromagnetic and hadron calorimeters,
and a muon spectrometer that incorporates three large superconducting air-core toroidal magnets.
 
The inner-detector system (ID) is immersed in a \qty{2}{\tesla} axial magnetic field
and provides charged-particle tracking in the region \(|\eta| < 2.5\).
The high-granularity silicon pixel detector covers the vertex region and typically provides four measurements (hits) per track,
the first hit normally being in the insertable B-layer (IBL) installed before Run~2~\cite{ATLAS-TDR-19,PIX-2018-001}.
It is followed by the silicon microstrip tracker (SCT), which usually provides eight measurements per track.
These silicon detectors are complemented by the transition radiation tracker (TRT),
which enables radially extended track reconstruction up to \(|\eta| = 2.0\).
The TRT also provides electron identification information
based on the fraction of hits (typically 30 in total) above a higher energy-deposit threshold corresponding to transition radiation.
 
The calorimeter system covers the pseudorapidity range \(|\eta| < 4.9\).
Within the region \(|\eta|< 3.2\), electromagnetic calorimetry is provided by barrel and
endcap high-granularity lead/liquid-argon (LAr) calorimeters,
with an additional thin LAr presampler covering \(|\eta| < 1.8\)
to correct for energy loss in material upstream of the calorimeters.
Hadron calorimetry is provided by the steel/scintillator-tile calorimeter,
segmented into three barrel structures within \(|\eta| < 1.7\), and two copper/LAr hadron endcap calorimeters.
The solid angle coverage is completed with forward copper/LAr and tungsten/LAr calorimeter modules
optimised for electromagnetic and hadronic energy measurements, respectively.
 
The muon spectrometer (MS) comprises separate trigger and
high-precision tracking chambers to measure the deflection of muons in the magnetic field generated by the superconducting air-core toroidal magnets.
The field integral of the toroids ranges between \num{2.0} and \qty{6.0}{\tesla\metre}
across most of the detector.
Three layers of precision chambers, each consisting of layers of monitored drift tubes, cover the region \(|\eta| < 2.7\),
complemented by cathode-strip chambers in the forward region, where the background is highest.
The muon trigger system covers the range \(|\eta| < 2.4\) with resistive-plate chambers in the barrel, and thin-gap chambers in the endcap regions.
 
Interesting events are selected by the first-level trigger system implemented in custom hardware,
followed by selections made by algorithms implemented in software in the high-level trigger~\cite{TRIG-2016-01}.
The first-level trigger accepts events from the \qty{40}{\MHz} bunch crossings at a rate below \qty{100}{\kHz},
which the high-level trigger further reduces in order to record events to disk at about \qty{1}{\kHz}.
 
An extensive software suite~\cite{ATL-SOFT-PUB-2021-001} is used in data simulation, in the reconstruction
and analysis of real and simulated data, in detector operations, and in the trigger and data acquisition
systems of the experiment.


\section{Data and simulation}
\label{sec:data_simulation}

 
The results presented in this paper are based on data from $pp$ collisions at a centre-of-mass energy of $\sqrt{s}=13~\TeV$. The data were collected between 2015 and 2018 (LHC Run-2) with the ATLAS detector and correspond to an integrated luminosity of $139$~fb$^{-1}$, with an uncertainty of~1.7\%~\cite{ATLAS-CONF-2019-021}, obtained using the LUCID-2 detector~\cite{Avoni:2018iuv} for the primary luminosity measurements. The number of $pp$ interactions per bunch crossing ranged between around 8 and 70, with the mean value of 33.7.

A set of single-electron\footnote{Throughout this paper, the term ``electron'' indicates both electrons and positrons.}~\cite{TRIG-2018-05} and single-muon triggers~\cite{TRIG-2018-01} were used, with transverse momentum ($\pT$) thresholds in the range $20$--$26$~GeV depending on the lepton flavour and data-taking period. All detector subsystems were required to be operational during data taking and to satisfy data quality requirements~\cite{DAPR-2018-01}.

Signal kinematic distributions were modelled using Monte Carlo (MC) simulation, while background processes were modelled using a mixture of MC and data-driven techniques.
All samples were produced using the ATLAS simulation infrastructure~\cite{SOFT-2010-01} and \textsc{Geant4}~\cite{Agostinelli:2002hh}. The effect of additional $pp$ interactions per bunch crossing (pile-up) is accounted for by overlaying the hard-scattering process with minimum-bias events generated with \PYTHIA[8.186]~\cite{Sjostrand:2007gs} using the \NNPDF[2.3lo] set of parton distribution functions (PDF)~\cite{Ball:2012cx} and the A3 set of tuned parameters~\cite{ATL-PHYS-PUB-2016-017}. Different pile-up conditions between data and simulation are taken into account by reweighting the mean number of interactions per bunch crossing in simulation to the number observed in data.  The samples generated with \MGNLO{} or \POWHEGBOX{} described below used the \EVTGEN~1.2.0 or 1.6.0 program~\cite{Lange:2001uf} for the properties of $b$- and $c$-hadron decays.
 
A detailed description of the MC samples is given below. A summary of the generators used for modelling signal and background SM processes in the signal region is given in Table~\ref{tab:mc_samples}.
 
\begin{table}[h!]
\caption{Summary of the MC samples used to simulate signal (upper part of the table) and background (lower part of the table) processes in the signal region. The notation $V$ is used to represent either $W$ or $Z/\gamma^*$.}
\label{tab:mc_samples}
\centering
\resizebox{\linewidth}{!}{
\begin{tabular}{cccccc}
\toprule
Process, short description & ME Generator + parton shower & Order & Tune & PDF set in ME\\
\midrule
EW, Int, QCD \sswwjj, nominal signal & \MGNLO{2.6.7} + \HERWIG{7.2} & LO & \HERWIG & \NNPDF[3.0nlo] \\
EW, Int, QCD \sswwjj, alternative shower & \MGNLO{2.6.7} + \PYTHIA{8.244} & LO & A14 & \NNPDF[3.0nlo] \\
\multirow{2}*{EW \sswwjj, NLO pQCD approx.} &  \SHERPA{2.2.11} \& \SHERPA{2.2.2}($WWW$) \& & +0,1j@LO & \SHERPA & \multirow{2}*{\NNPDF[3.0nnlo]} \\
& \POWHEGBOX{2}+\PYTHIA{8.235} ($WH$) & NLO & A14 & \\
EW \sswwjj, NLO pQCD approx. & \POWHEGBOX{v2} + \PYTHIA{8.230} & NLO (VBS approx.) & AZNLO & \NNPDF[3.0nlo] \\
QCD \sswwjj, NLO pQCD approx. & \SHERPA{2.2.2}  & +0,1j@LO & \SHERPA & \NNPDF[3.0nnlo]  \\
\midrule
QCD $VVjj$ & \SHERPA{2.2.2} & +0,1j@NLO; +2,3j@LO & \SHERPA & \NNPDF[3.0nnlo] \\
EW $W^{\pm}Z/\gamma^* jj$ & \MGNLO{2.6.2}+\PYTHIA{8.235} & LO & A14 & \NNPDF[3.0nlo] \\
EW $Z/\gamma^* Z/\gamma^*jj$ & \SHERPA{2.2.2} & LO & \SHERPA & \NNPDF[3.0nnlo] \\
QCD $V\gamma jj$ & \SHERPA{2.2.11} & +0,1j@NLO; +2,3j@LO & A14 & \NNPDF[3.0nnlo]\\
EW $V\gamma jj$ & \MGNLO{2.6.5}+\PYTHIA{8.240} & LO & A14 & \NNPDF[3.0nlo] \\
$VVV$ & \SHERPA{2.2.1} (leptonic) \& \SHERPA{2.2.2} (one $V\to jj$) & +0,1j@LO & \SHERPA & \NNPDF[3.0nnlo] \\
$t\bar{t}V$ & \MGNLO{2.3.3.p0} + \PYTHIA{8.210} & NLO & A14 & \NNPDF[3.0nlo] \\
$tZq$ & \MGNLO{2.3.3.p1} + \PYTHIA{8.212} & LO & A14 & \NNPDF[2.3lo] \\
\midrule
\midrule
\sswwjj EFT & \MGNLO[2.6.5] + \PYTHIA{8.235} & LO & A14 & \NNPDF[3.0nlo] \\
\midrule
$\mathrm{H}^{\pm\pm}_5$ & \MGNLO[2.9.5] + \PYTHIA{8.245} & LO & A14 & \NNPDF[3.0nlo] \\
\bottomrule
\end{tabular}
}
\end{table}
 
\subsection{\sswwjj\ samples}
\label{sec:sswwjj_samples}
 
The \sswwjj\ processes, $p p \to \ell^\pm \nu \ell^\pm \nu j j$,  are simulated at LO accuracy in perturbative QCD (pQCD) with \MGNLO{2.6.7}~\cite{Alwall:2014hca} separately for the QCD ($\mathcal{O}(\alpha^4\as^2)$), interference (Int, $\mathcal{O}(\alpha^5\as)$) and EW ($\mathcal{O}(\alpha^6)$) production modes, where $\alpha$ and $\as$ are the electroweak and strong coupling constants, respectively. For the modelling of the parton shower, hadronisation and underlying event, \HERWIG{7.2.1}~\cite{Bahr:2008pv,Bellm:2015jjp} (nominal sample) or \PYTHIA{8.244}~\cite{Sjostrand:2014zea} (alternative sample) are used. In the following, these samples are referred to as ``\MGHWshort'' and ``\MGPYshort''. The PDF set used was \NNPDF[3.0nlo]~\cite{Ball:2014uwa} implemented at next-to-leading-order (NLO) in pQCD and with a strong coupling constant $\as(M_Z) = 0.118$. For the nominal \MGHWshort sample, the \textit{dipole parton shower} mode~\cite{Platzer:2009jq} was used instead of the default angular-ordered mode.
The alternative \MGPYshort sample is produced using the \textit{dipole recoil} scheme~\cite{Cabouat:2017rzi}, since the default $\pT$-ordered shower used in \PYTHIA{8} for initial state radiation (ISR) emission,
where the recoil of an ISR emission is taken by the whole final state, produces VBS samples with too much radiation in the central region of rapidity~\cite{Ballestrero:2018anz}.
The dipole recoil scheme mitigates this behaviour by having only one final-state parton take the recoil of an emission.
A dedicated study of the effect of these settings can be found in Ref.~\cite{ATL-PHYS-PUB-2019-004}. The renormalisation and factorisation   scales are chosen to be  $\mu_{\mathrm{R}} = \mu_{\mathrm{F}} = \sqrt{p_{\mathrm{T}}^\mathrm{j_1} p_{\mathrm{T}}^\mathrm{j_2}}$, where $p_{\mathrm{T}}^\mathrm{j_1}$ and $p_{\mathrm{T}}^\mathrm{j_2}$ are the transverse momenta of the jets with the highest (leading) and second highest (subleading) transverse momentum, respectively.
 
A combination of MC samples is used to assess partial NLO pQCD corrections to the EW \sswwjj production. This combination is based on a \SHERPA{2.2.11} EW \sswwjj sample, which excludes triboson contributions, and is simulated with up to one additional parton at LO. The non-overlapping triboson $WWW$ and $WH$ ($H \to W^{+}W^{-}$) final states (in the decay mode  $\ell^\pm \nu \ell^\pm \nu j j$) are modelled with  \SHERPA{2.2.2} at NLO with up to two additional partons at LO, and \POWHEGBOX{v2} + \PYTHIA{8.235} samples~\cite{Alioli:2010xd}, respectively. This combination is referred to as ``\SHERPA~EW'' in the following. Partial NLO pQCD corrections to the QCD \sswwjj process are checked using a separate \SHERPA{2.2.2} sample, referred to as ``\SHERPA QCD'', which also models \sswwjj final states with up to one extra parton at LO. The \(\mu_{\mathrm{R}}\) and \(\mu_{\mathrm{F}}\) of both \SHERPA models are set equal to the invariant mass of \(W\) boson pair. These and the following background \SHERPA samples use dedicated parton shower tuning developed by the \SHERPA authors~\cite{Bothmann:2019yzt}.
 
An additional alternative EW \sswwjj\ sample, simulated with the \POWHEGBOX{v2} event generator interfaced to the \PYTHIA{8.230} parton shower model without deploying the dipole recoil scheme, was used for comparisons. This \POWHEG+\PYTHIA EW \sswwjj sample was generated at NLO in pQCD in the VBS approximation~\cite{Jager:2011ms}, using the \NNPDF[3.0nlo] PDF set in the matrix element. The \(\mu_{\mathrm{R}}\) and \(\mu_{\mathrm{F}}\) are fixed to
the \(W\) boson mass. The AZNLO set of tuned parameters~\cite{STDM-2012-23} is used, with the CTEQ6L1~\cite{Pumplin:2002vw} PDF set, for the modelling of non-perturbative effects. The \PHOTOSpp[3.61] program~\cite{Golonka:2005pn,Davidson:2010ew} is used to simulate quantum electrodynamic emissions, including lepton pair emissions.

\subsection{Background samples}
 
Diboson QCD $VVjj$ processes $Z/\gamma^*(\ell^\pm\ell^\mp)Z/\gamma^*(\ell^\pm\ell^\mp)$ and $W(\ell^\pm\nu)Z/\gamma^*(\ell^\pm\ell^\mp)$ are simulated with the \SHERPA{2.2.2} event generator~\cite{Gleisberg:2008ta} using matrix elements that contain all diagrams with four EW vertices. These processes are calculated for up to one additional parton at NLO and up to three additional partons at LO using \textsc{Comix}~\cite{Gleisberg:2008fv} and \OPENLOOPS~\cite{Cascioli:2011va}, and merged with the \SHERPA parton shower based on Catani--Seymour dipole factorisation~\cite{Schumann:2007mg} according to the MEPS@NLO prescription~\cite{Hoeche:2011fd,Hoeche:2012yf,Catani:2001cc,Hoeche:2009rj}. The \NNPDF[3.0nnlo] PDF set is used. The EW production of $W^\pm Z/\gamma^* jj$ is simulated using \MGNLO{2.6.2}~\cite{Alwall:2014hca} at LO in pQCD, using the \NNPDF[3.0nlo] PDF set, and interfaced to \PYTHIA{8.235} for modelling the parton shower in the dipole recoil scheme. The EW production of $Z/\gamma^* Z/\gamma^*jj$ is modelled using \SHERPA{2.2.2} at LO.

Samples for $V\gamma$~($V=W,Z/\gamma^*$) processes are produced using \SHERPA{2.2.11}. All off-shell contributions are taken into account. The NLO matrix elements with up to one additional parton and LO matrix elements with up to three partons are merged with the parton shower using an MEPS@NLO merging scale of $Q=20$~GeV. The \NNPDF[3.0nnlo] PDF set is used. The photon is required to have $\pt>7$~GeV and be isolated from leptons~\cite{Frixione:1998jh}. The EW production of $V\gamma$ is modelled using \MGNLO{2.6.5}+\PYTHIA{8.240} at LO in pQCD.
 
Triboson processes $VVV$ ($V=W,Z$), which lead to two-jets plus four-lepton (charged or neutrinos) final states, and do not contain $W$ boson pairs with the same charge, are considered background and are simulated at LO with up to one extra parton in \SHERPA{2.2.2} using the \NNPDF[3.0nnlo] PDF set.
 
The $t\bar{t}V$~($V=W,Z$) samples are generated at NLO in pQCD using \MGNLO{2.3.3.p0}~\cite{Alwall:2014hca} and showered by \PYTHIA{8.210}, with the cross section normalised to the prediction including NLO EW and NLO QCD corrections~\cite{deFlorian:2016spz}. The \NNPDF[3.0nlo] PDF is used in the matrix element. The $tZq$ contributions were modelled at LO accuracy in pQCD using \MGNLO{2.3.3.p1} interfaced to \PYTHIA{8.212} for parton showering.
 
Several MC samples are used to subtract prompt lepton contributions in the region used to derive scale factors for the data-driven non-prompt lepton background estimate described in Section~\ref{subsec:bkg_nonprompt}.
Events containing a $W$ or $Z$ boson with associated jets are simulated using \MGNLO{2.3.2.p1}~\cite{Alwall:2014hca} at LO, interfaced to the \PYTHIA{8.210}~\cite{Sjostrand:2014zea} parton shower model. The $W/Z$ samples are normalised using next-to-next-to-leading-order (NNLO) cross sections~\cite{Hamberg:1990np,Gavin:2012sy}.
 
The \POWHEGBOX{v2}~\cite{Alioli:2010xd} program is used with the \NNPDF[3.0nlo] PDF set to generate $t\bar{t}$ and single top-quark events in both the $Wt$- and $s$-channels. The parton shower, hadronisation, and the underlying event are simulated using \PYTHIA{8.230} with the \NNPDF[2.3lo] PDF set~\cite{Pumplin:2002vw} and the corresponding A14~\cite{ATL-PHYS-PUB-2014-021} set of tuned parameters. The top quark mass is set to 172.5~GeV. Single-top events in the $t$-channel are generated with \POWHEGBOX{v2} using the 4-flavour scheme for the NLO matrix element calculations together with the fixed four-flavour PDF set \NNPDF[3.0nlo4f]. \PYTHIA{8} with the A14 set of tuned parameters is used for the parton shower. For all top processes, top-quark spin correlations are preserved (for $t$-channel, top quarks are decayed using MadSpin~\cite{Artoisenet:2012st}). The cross sections of these processes involving top quarks are normalised to the NLO (NNLO for $t\bar{t}$) pQCD predictions including next-to-next-to-leading logarithmic soft gluon terms~\cite{Beneke:2011mq,Cacciari:2011hy,Baernreuther:2012ws,Czakon:2012zr,Czakon:2012pz,Czakon:2013goa,Czakon:2011xx,Wt_NLO_NNLL}.
 
Contributions from prompt photon + jet production, where the converted photon is misreconstructed as an electron, are subtracted with other prompt electron contributions, and are simulated using \SHERPA{2.2.2} with up to two additional jets at NLO and with up to four jets at LO in pQCD, using the \NNPDF[3.0nnlo] PDF set.
 
\subsection{\sswwjj\ EFT samples}
\label{sec:sswwjj_EFT_samples}
 
For the EFT interpretation presented in Section~\ref{sec:EFT}, individual samples that vary only one term at a time (SM, interference, quadratic or cross, see~\cite{ATL-PHYS-PUB-2023-002manual} for details) are generated. This amplitude decomposition technique is used to avoid generating multiple samples with different Wilson coefficient values or relying on an event-by-event matrix element reweighting.
The \sswwjj\ EFT samples are simulated using \MGNLO[2.6.5]~\cite{Alwall:2014hca} at LO interfaced to \PYTHIA{8.235}~\cite{Sjostrand:2007gs} for the modelling of the parton shower (using the dipole recoil scheme), hadronisation and underlying event. The PDF set used was \NNPDF[3.0nlo]~\cite{Ball:2014uwa}, with a strong coupling constant $\as(M_Z) = 0.118$.
Contrary to the samples described in Section~\ref{sec:sswwjj_samples}, where a custom scale was used as described, the samples that include the effect of EFT operators use half of the sum of the transverse masses of all objects in the event~\cite{Hirschi:2015iia}.
 
\subsection{$\mathrm{H}^{\pm\pm}_5$ samples}
\label{sec:H5pp_samples}
 
The signal samples used to set limits on doubly-charged Higgs boson production (see Section~\ref{sec:GM}) are simulated with \MGNLO[2.9.5]~\cite{Alwall:2014hca} at LO interfaced to \PYTHIA{8.245}~\cite{Sjostrand:2007gs} for the modelling of the parton shower in the dipole recoil scheme, hadronisation and underlying event. The PDF set used was \NNPDF[3.0nlo]~\cite{Ball:2014uwa}, implemented at NLO in pQCD and with a strong coupling constant $\as(M_Z) = 0.118$. The signal simulation is produced for 23 mass points from $200$~GeV to $3$~TeV using the \texttt{H5plane} benchmark~\cite{Logan:2017jpr} assuming a narrow-width signal with the width-to-mass ratio of $H5$ states below 5\%. The $\sin\theta_{\mathrm{H}}$ values are set to $0.5$ for masses up to $800$~GeV and $0.25$ for higher-mass samples to be compatible with present constraints.


\section{Event reconstruction and selection}
\label{sec:event_selection}

 
The signature of $W^\pm W^\pm jj$ events is the presence of two high-energy forward jets (tagging jets) in opposite hemispheres and the presence of a same-sign charged-lepton pair ($e$ or $\mu$) and missing transverse momentum (\MET). Due to the emission of a vector boson from each initial quark line, the final-state quarks tend to be at high absolute values of rapidity, and have rather large momenta. The two jets therefore have a large absolute rapidity difference, $\dyjj = \left| y_{\mathrm{j}1} - y_{\mathrm{j}2} \right |$,  and a large invariant mass \mjj. In addition, the leptons from the decays of the two $W$ bosons tend to lie between the tagging jets in rapidity.
 
Lepton candidates are first preselected using baseline criteria.
Electron candidates are reconstructed~\cite{EGAM-2018-01} from electromagnetic calorimeter energy clusters and matched to a track reconstructed in the ID\@. Baseline electrons are required to have $|\eta|<2.47$ and transverse momentum $\pT > 4.5~\mathrm{GeV}$. They must be outside the barrel/endcap transition region ($1.37 < |\eta| < 1.52$) of the calorimeter. Muons are reconstructed~\cite{MUON-2018-03} from tracks in the MS, matched to a corresponding track in the ID where possible. Baseline muons are required to have $\pT>3~\mathrm{GeV}$ and $|\eta|<2.7$. Electron and muon tracks are required to originate from the primary vertex.\footnote{The primary vertex is identified as the vertex in the event with the highest scalar sum of the squared transverse momenta of its associated tracks.} The transverse impact parameter significance\footnote{The transverse impact parameter significance is defined as $|d_0|/\sigma(d_0)$, where $d_0$ is the distance of closest approach of $e$ or $\mu$ to the primary vertex in the transverse plane and $\sigma(d_0)$ is its uncertainty.} is required to be less than 5 for electrons and 15 for muons. The longitudinal impact parameter\footnote{The longitudinal impact parameter is equal to $|z_0 \cdot \sin \theta|$, where $z_0$ is the difference between the value of the $z$ coordinate of the point on the track at which $d_0$ is defined and the longitudinal position of the primary vertex.} must be less than 0.5 (1.5)~mm for electrons (muons).
Baseline electrons and muons are required to satisfy respective loose identification criteria~\cite{EGAM-2018-01,MUON-2018-03}.
 
Jets are reconstructed using the \antikt algorithm~\cite{Cacciari:2008gp}, with a radius parameter of \(R=0.4\), using particle-flow objects~\cite{PERF-2015-09} as input. The jets
are required to have $\pt > 25~\GeV$ and $|\eta| \le 4.5$.
Contamination from jets originating in pile-up collisions is reduced by using the jet-vertex-tagger algorithm~\cite{PERF-2014-03}. Jets with $\pT > 20~\GeV$ and $|\eta| < 2.5$ that contain a $b$-hadron are identified using the 85\% efficiency working point of the DL1r heavy-flavour tagger~\cite{ATL-PHYS-PUB-2017-013}.
 
To avoid the double counting of physics objects, an overlap removal procedure is applied to baseline leptons and jets. First, non-$b$-jets are removed if they overlap with an electron or muon
within $\Delta R < 0.2$
and have less than three associated tracks with $\pT > 500~\MeV$. At the next step, electrons or muons that overlap with jets, including $b$-jets, within $\Delta R < 0.4$ are removed. Finally, electrons that share an ID track with a muon are removed.
 
Two further categories of leptons are defined, which are mutually exclusive subsets of baseline leptons: signal leptons for the signal extraction and background leptons for the estimate of the non-prompt lepton background.
 
Signal electrons must satisfy the Tight likelihood-based identification criteria~\cite{PERF-2017-01}, and signal muons the Medium identification criteria~\cite{MUON-2018-03}. Electrons (muons) are required to satisfy the Gradient~\cite{EGAM-2018-01} (PflowTight~\cite{MUON-2018-03}) isolation criteria.
The longitudinal impact parameter of muons is required to be less than 0.5~mm, while the transverse impact parameter significance must be less than 3. A charge selector tool based on boosted decision trees uses shower shape and track-to-cluster matching variables~\cite{EGAM-2018-01} to reject electron candidates where the charge is likely misidentified. The efficiency of the charge selector tool for selecting signal electrons with correctly identified charge is reaching 99.4\% (95\%), in the same time achieving a rejection of 83\% (92\%) of electrons with wrongly identified charge in the barrel (endcap) region.
 
Background lepton candidates are used to estimate the non-prompt lepton background. Background electrons are required to satisfy the Medium likelihood identification~\cite{PERF-2017-01}, while an isolation requirement is dropped. The isolation requirement for background muons is changed to PflowLoose~\cite{MUON-2018-03} and the transverse impact parameter significance is required to be less than 10. All background leptons are required to fail the signal lepton selection to ensure that the samples of signal and background leptons are statistically independent.
 
Two signal leptons with the same charge are required in each event. Each lepton must have transverse momentum $\pt>27~\GeV$. Muon pseudorapidity is restricted to the range $|\eta| < 2.5$, where a matching to the ID track is possible. One of the leptons must be matched to the lepton that fired a single lepton trigger mentioned in Section~\ref{sec:data_simulation}.
The dilepton invariant mass $m_{\ell\ell}$ is required to be above $20$~GeV.
 
Due to a non-negligible charge misidentification rate for electrons (see Section~\ref{subsec:bkg_chflip}), $Z/\gamma* \to ee$ events can pass the same-sign selection. To reduce the Drell--Yan background,
the pseudorapidities $\eta$ of each electron must satisfy $\left|\eta\right|<1.37$.
This requirement restricts electrons to the detector region with a lower material budget, where the production of secondary particles (photons and electron-positron pairs) is less probable and allows a more accurate charge identification.
In addition,
events with two signal electrons must survive a $Z$ veto, $|m_{ee}-m_Z| > 15~\GeV$, where $m_Z$ is the mass of the $Z$ boson and $m_{ee}$ is the dielectron invariant mass.
 
To suppress background from processes with more than two leptons in the final state (e.g., $WZ$ and $ZZ$ background), a third lepton veto is applied in the following way: events in which a third baseline lepton survives the overlap removal are rejected. If a third baseline lepton does not survive the overlap removal, but the dilepton invariant mass of a same-flavour opposite-charge signal lepton and the third lepton is compatible with the $Z$ boson mass, $|m_{\ell\ell} - m_{Z}| < 15~\GeV$, the event is also rejected.
 
Due to the presence of two neutrinos in the final state, signal events usually
have a large $\MET$.
The latter is reconstructed from the \pt imbalance of baseline objects that survive the overlap removal and the soft track term~\cite{PERF-2016-07}, and must satisfy $\MET > 30~\GeV$. 
 
At least two jets are required to be present in each event, with transverse momentum $\pt$ that exceeds $65~\GeV$ for the leading jet and $35~\GeV$ for the subleading jet. The two jets with the highest $\pt$ in the event are selected as tagging jets, i.e.\ they are used to calculate $\mjj$
and $|\dyjj|$ used in the further selection.
To suppress background contributions from top processes, events are vetoed if any jet
with $\pT > 20~\GeV$ and $|\eta| < 2.5$ is identified as a $b$-jet.
 
The VBS signature of two jets with high invariant mass and large angular separation is used to further purify the events and in particular to separate the signal from QCD-induced processes that lead to the same final state. The dijet invariant mass and the difference in jet rapidities must satisfy $\mjj \ge 500~\GeV$ and $|\dyjj| > 2$, respectively.
The expected purity of the EW (QCD, Int) \sswwjj process in the signal region (SR) is 52\% (5.4\%, 1.7\%).
 
\begin{table}[htb]
\caption{Summary of the event selection for the signal and control regions.}
\centering
\begin{tabular}{r c c c}
\toprule
Requirement & SR & Low-\mjj CR & WZ CR \\
\midrule
Leading and subleading lepton \pT & \multicolumn{3}{c}{\(>27~\GeV\)} \\
Electron \(|\eta|\) & \multicolumn{3}{c}{\(<2.47\) (1.37 in \(ee\)), excluding \(1.37 \le |\eta| \le 1.52\)} \\
Muon \(|\eta|\) & \multicolumn{3}{c}{\(<2.5\)} \\
\midrule
Leading (subleading) jet \pT & \multicolumn{3}{c}{\(>65\,(35)~\GeV\)} \\
Additional jet \pT & \multicolumn{3}{c}{\(>25~\GeV\)} \\
Jet \(|\eta|\)  & \multicolumn{3}{c}{\(<4.5\)} \\
\midrule
$m_{\ell\ell}$ & \multicolumn{3}{c}{\(>20~\GeV\)} \\
$\met$ & \multicolumn{3}{c}{\(>30~\GeV\)} \\
Charge misid. $Z\rightarrow ee$ veto & \multicolumn{2}{c}{$|m_{ee}-m_Z| > 15~\GeV$} & \(-\) \\
\(b\)-jet veto & \multicolumn{3}{c}{\(N_{\bjet}=0\), \(\pT^{\bjet}>20~\GeV\), \(|\eta^{\bjet}|<2.5\)} \\
\(N_\mathrm{veto\,leptons}\) & \(=0\) & \(=0\) & \(= 1\,,\pT>15~\GeV\) \\
\(\mlll\) & \(-\) & \(-\) & \(>106~\GeV\) \\
\midrule
\mjj & \(>500~\GeV\) & \(200<\mjj<500~\GeV\) & \(>200~\GeV\) \\
\(|\dyjj|\) & \multicolumn{3}{c}{\(>2\)} \\
\bottomrule
\end{tabular}
\label{tab:event_selection}
\end{table}
 
In addition to the SR, the measurement makes use of two control regions (CR), referred to in the following as the ``\(WZ\) CR'' and ``low-\mjj CR''. The \(WZ\) CR follows closely the SR selection with the exception that the third-lepton veto requirement is dropped, and exactly one more signal lepton with \(\pT > 15~\GeV\) is required. In addition, the \mjj condition is loosened to \(\mjj > 200~\GeV\) and the trilepton invariant mass is required to satisfy \( \mlll > 106~\GeV \) to suppress contamination from radiative $Z$ boson decays. The \(WZ\) CR is used to improve the modelling of the dominant background coming from QCD-induced \wzjj events.
 
The low-\mjj CR and the SR use the same selection criteria except on the dijet invariant mass,
where the CR selection requires \(200 < \mjj < 500~\GeV\).
As a result, this CR has a similar background composition to the SR but the contribution from signal events is reduced to 11\%.
The low-\mjj CR is used in the signal extraction fit to control the uncertainties of major background contributions, described in Sections~\ref{subsec:bkg_WZ}-\ref{subsec:bkg_chflip}.
 
The SR and CR event selections are summarised in Table~\ref{tab:event_selection}.


\section{Background estimation}

 
The dominant background, which comes from $WZ/\gamma^* jj$, is estimated using MC and includes a data-driven correction to the shape of the \mjj distribution and an adjustment to the normalisation based on a dedicated CR\@. The non-prompt lepton and electron charge misidentification backgrounds are estimated using data-driven methods. The remaining background sources are modelled using MC simulations. The contributions from the hard double parton scattering process, where two $W$ bosons are produced in two separate partonic interactions from the same $pp$ collision, was checked using MC and found to be negligible. The fraction of signal events where at least one of the two signal jets originates from pile-up $pp$ interactions was estimated using MC simulations to be about 0.5\%.
 
\subsection{$WZ/\gamma^*$ background}
\label{subsec:bkg_WZ}
The $WZ/\gamma^*$ production, dubbed $WZ$ in the following, where both the $W$ and $Z/\gamma^*$ bosons decay leptonically,
comprises the dominant background in this measurement and contributes 22\%\ of the overall expected event yield in the SR\@. It contributes when one lepton escapes the third lepton veto requirement, typically because it was outside of the geometrical acceptance of the detector.
 
The $\wzjj$ final states are modelled using MC simulations and are dominated by the QCD-induced production mode. The cross section extraction of the
$\sswwjj$ signal relies on the proper modelling of $\mjj$ templates, and to avoid a bias in the signal extraction induced by a possible $\mjj$ mismodelling in background, the shape of this variable given by the QCD $\wzjj$ model is reweighted to data~\cite{STDM-2016-09} in a dedicated
region. First, the selection of the $WZ$ CR described in Section~\ref{sec:event_selection} is applied. In addition, to reduce the contamination from EW $\wzjj$ events, a requirement on the $WZ$ system centrality, $\xi_{WZ} > 0.4$, is imposed, which suppresses events where the $WZ$ system is located between the two tagging jets in rapidity --- a characteristic topology for EW production mode.
The centrality is defined as $\xi_{WZ} = \left|\frac{y_{WZ}-(y_{\mathrm{j}1}+y_{\mathrm{j}2})/2}{y_{\mathrm{j}1}-y_{\mathrm{j}2}}\right|$, where
$y_{WZ} = (y_{\ell(W)} + y_{\ell\ell(Z)})/2$ is the average of the rapidity $y_{\ell(W)}$ of the charged lepton coming from the $W$
and the rapidity $y_{\ell\ell(Z)}$ of the dilepton coming from the $Z$ boson, and $y_{\mathrm{j}1}$ ($y_{\mathrm{j}2}$) is the rapidity of the leading (subleading) jet. This requirement reduces the contribution of EW \wzjj events to the total \wzjj yield from 13\% to 5\%. The $\mjj$ reweighting, which was derived in this special data region, is applied to the \wzjj model in the SR and all CRs and is constructed to preserve the MC normalisation of the QCD $\wzjj$ background for $\mjj > 200~\GeV$. To test the impact of the selection on the reweighting function, the function was also derived in the $WZ$ CR without applying the $WZ$ centrality selection, and the result was found to be consistent with the nominal reweighting function within the uncertainty. The weight values range from 1.2 at $\mjj = 200~\GeV$ to approximately 0.3 at $\mjj = 3~\TeV$.  The uncertainty in the reweighting function, dominated by the limited data in the derivation region, is \mjj dependent, ranging between 10\% at $\mjj=200~\GeV$ and 70\% at $\mjj = 3~\TeV$. Additional sources of uncertainty in the reweighting function include renormalisation and factorisation scale variations, as well as PDF and \as variations, are negligible compared with the statistical component.
 
\subsection{Non-prompt lepton background}
\label{subsec:bkg_nonprompt}
Leptons from hadron decays and jets misidentified as leptons are referred to as non-prompt leptons. They constitute the second-largest background source,
corresponding to 12\% of the expected event yield in the signal region, and arise mainly from $W$+jets and semileptonic $t\bar{t}$ processes. This background is estimated using a data-driven method involving the computation of
scale factors defined as the ratio of the probability for a non-prompt lepton to pass the signal lepton requirements to the
probability for it to pass the background lepton selection, as defined in Section~\ref{sec:event_selection}. This method is similar to the one used in Ref.~\cite{ATLAS:2019cbr}.
 
The scale factors are derived from a data sample
containing events with a single non-prompt lepton candidate balanced by a jet.
Jets with transverse momenta $\pT > 25\,(30)~\GeV$ in the pseudorapidity range $|\eta| < 2.5$ $(2.5 < |\eta| < 4.5)$ are selected if the jet and lepton candidates are opposite in azimuthal angle and satisfy $\left|\Delta \phi(\ell, j) \right| > 2.8$.
The prompt lepton contributions are reduced by adding a requirement on the sum of \MET and the transverse mass $m_\mathrm{T}$ (see definition in Section~\ref{subsec:diff_cross_sec}) of the lepton + \MET system, $\MET + m_\mathrm{T} < 50~\GeV$.
Residual prompt lepton and prompt photon conversion contributions are modelled using MC\@.
The scale factors are computed in bins of lepton $\pT$ and, for electrons, in two bins of $\eta$ corresponding to the barrel and endcap LAr calorimeters.
The events in this sample are triggered by single-electron or single-muon triggers with low-\pt thresholds of 12 and 14~GeV, respectively, which require loosely identified leptons without any isolation requirement.
 
The uncertainty in the scale factors ranges from 10\%\ to 40\%. The dominant contributions to the uncertainty come from varying the prompt lepton contributions by $\pm5\%$ and from the variation of the $b$-tagged jet selection ($b$-veto versus at least one $b$-jet), which is done to vary the amount of
non-prompt leptons from light-flavour hadrons versus heavy-flavour hadrons. Additional uncertainties in the scale factors are estimated by varying the $\MET + m_\mathrm{T}$ requirement, and by varying the level of prompt electron and muon contributions, related to a potential mismodelling of the identification and isolation efficiencies for background leptons (see Section~\ref{sec:event_selection}) that satisfy corresponding looser requirements but fail tighter requirements. Statistical uncertainties arising from the limited number of data and MC events in the region used to derive the scale factors are also included.
 
The estimate of the non-prompt background in the signal region is obtained by applying these scale factors to data events in the kinematic region defined by the SR selection, with the exception that instead of having two signal leptons, events are required to have one signal and one background lepton. In this region, contributions from processes with two prompt leptons or one prompt lepton plus a prompt photon, where the latter converts and is misreconstructed as an electron, are subtracted using MC. If a signal lepton plus background lepton event is triggered by the background lepton, a dedicated set of scale factors is applied, derived with the analysis triggers mentioned in Section~\ref{sec:data_simulation} instead of low-\pt triggers. This helps to avoid a bias due to having the lepton identification and isolation requirements of the analysis triggers be tighter than the selection of background leptons. Systematic uncertainties in these dedicated scale factors are obtained, considering the same sources used for the scale factors obtained using low-\pT single-lepton triggers.
 
\subsection{Electron charge misidentification and photon conversion background}
\label{subsec:bkg_chflip}
Background from $Z$ and dileptonic $t\bar{t}$ events involving electrons, where the charge of an electron is misidentified in the detector,
is estimated by weighting opposite-charge dilepton data events with charge misidentification probabilities obtained from MC\@. After applying the electron charge-selector tool~\cite{EGAM-2018-01} described in Section~\ref{sec:event_selection}, the residual contribution of the overall dilepton background in the signal region
is 2.3\%, and is 1.4\% for events with one signal electron and one signal muon. The uncertainty in this background contribution reaches 40\% and is dominated by the uncertainties in the scale factors that calibrate the charge-selector tool, which were obtained using  $Z \to e^+e^-$ events~\cite{EGAM-2018-01}.
 
The background contribution from $\ell\gamma jj$ events, where the prompt photon $\gamma$
is misreconstructed as an electron, is simulated using $V\gamma$ MC (see Table~\ref{tab:mc_samples}). An overall normalisation uncertainty of 40\% is assigned to this
background, motivated by the scale factor uncertainties of the electron charge-selector tool. The expected contribution of this process to the SR event yield is 2.4\%.
 
\subsection{Other background}
\label{subsec:bkg_other}
The rest of the prompt-lepton background arises from $ZZ$, triboson $VVV$ (excluding triboson final states
involving two same-sign $W$ bosons), $t\bar{t}V$, and $tZq$ processes, which together correspond to 1.6\% of the expected total SR yield, and are modelled using MC (see Table~\ref{tab:mc_samples}). A normalisation uncertainty of
20\%~\cite{ATL-PHYS-PUB-2017-005} (30\%~\cite{ATL-PHYS-PUB-2017-005,TOPQ-2018-08}) is applied on the cross section of the $ZZ$ ($VVV$ and top quark processes) production.
 
The expected signal and background yields in the SR are given in Table~\ref{tab:SR_yields_prefit}.
 
\begin{table}
\caption{Expected signal and background yields in the SR\@. The yields are shown for different dilepton final states where the first lepton has the highest \pT. The ``Other prompt'' category combines $ZZ$, $VVV$, $t\bar{t}V$, and $tZq$ background processes. The sum of all contributions may differ from the total value due to rounding. The uncertainty includes both the statistical and systematic components. }
\label{tab:SR_yields_prefit}
\centering
\begin{tabular}{lS@{\(\,\pm\,\)}SS@{\(\,\pm\,\)}SS@{\(\,\pm\,\)}SS@{\(\,\pm\,\)}SS@{\(\,\pm\,\)}l}
\bottomrule
Process & \multicolumn{2}{c}{$ee$} & \multicolumn{2}{c}{$e\mu$} & \multicolumn{2}{c}{$\mu e$} & \multicolumn{2}{c}{$\mu\mu$} & \multicolumn{2}{c}{Combined} \\
\midrule
$W^{\pm}W^{\pm}jj$ EW    &    27.6  &     0.9  &    68.2  &     1.6  &    61.3  &     1.5  &    77.8  &     1.7  &     235  &       5   \\
$W^{\pm}W^{\pm}jj$ QCD   &     1.6  &     0.5  &     7.3  &     2.2  &     6.4  &     1.9  &     8.8  &     2.5  &      24  &       7   \\
$W^{\pm}W^{\pm}jj$ Int   &    0.93  &    0.20  &     2.2  &     0.5  &     2.0  &     0.4  &     2.5  &     0.5  &     7.6  &     1.6   \\
$W^{\pm}Zjj$ QCD                & 8.4& 1.0   & 26.8& 3.0   & 26.7& 3.0   & 20.9& 2.2   & 83& 9 \\
$W^{\pm}Zjj$ EW                 & 1.71& 0.14   & 4.9& 0.4   & 4.1& 0.4   & 4.2& 0.4   & 14.9& 1.2 \\
Non-prompt               &     8.9  &     2.6  &      15  &       4  &    10.2  &     3.2  &      21  &       7  &      56  &      12   \\
$V\gamma$                &     1.3  &     0.8  &     5.1  &     2.2  &     4.6  &     2.6  &   \multicolumn{2}{c}{---} &      11  &       5   \\
Charge misid.            &     3.8  &     2.0  &     5.0  &     1.3  &     1.2  &     0.4  &     \multicolumn{2}{c}{---}  &      10  &       4   \\
Other prompt             &    1.02  &    0.29  &     2.5  &     0.6  &     1.8  &     0.5  &     1.7  &     2.2  &     7.1  &     2.8   \\
\midrule
Total expected           &      55  &       4  &     137  &       7  &     118  &       6  &     137  &       8  &     448  &      20   \\
 
\midrule
Data                  &    \multicolumn{2}{c}{52}  &   \multicolumn{2}{c}{149}  &   \multicolumn{2}{c}{127}  &   \multicolumn{2}{c}{147}  &   \multicolumn{2}{c}{475}\\
\bottomrule
\end{tabular}
\end{table}


\section{Systematic uncertainties}
\label{sec:systematic_uncertainties}

Systematic uncertainties affecting the measurement arise from experimental sources and signal and background modelling uncertainties.
 
\textbf{Experimental uncertainties.} Experimental sources include uncertainty in the calibration of the objects and algorithms used in the measurement. In particular, uncertainties in the
lepton reconstruction, identification, and isolation efficiencies~\cite{EGAM-2018-01,MUON-2018-03} and in the lepton trigger efficiency~\cite{TRIG-2018-05,TRIG-2018-01} are taken into account. Uncertainties in the calibration of jet energy scale and resolution~\cite{PERF-2015-09}, the jet vertex tagger~\cite{PERF-2014-03},  and in the calibration of the $b$-tagger~\cite{ATL-PHYS-PUB-2017-013} used for applying a $b$-jet veto are included. The uncertainties in the \MET calibration~\cite{PERF-2016-07} and MC pile-up reweighting are considered as well. A luminosity uncertainty of 1.7\%~\cite{ATLAS-CONF-2019-021} is assigned to all MC based predictions.
 
One of the dominant systematic uncertainty sources in this measurement is related to the non-prompt background estimate. The uncertainty estimates are obtained by varying $\pT$-dependent (and $\eta$-dependent for electrons) scale factors associated with the data-driven method used. More details about the types of systematic variations can be found in Section~\ref{subsec:bkg_nonprompt}.
 
Uncertainties in the data-driven electron charge misidentification background arise from the calibration uncertainties in the charge-selector tool~\cite{EGAM-2018-01}, as described in Section~\ref{subsec:bkg_chflip}.
 
\textbf{Signal and background modelling uncertainties.}
Theoretical signal modelling uncertainties are dominated by the effects of missing higher-order pQCD contributions in the nominal LO \MGHWshort signal model. To account for them, two separate variations are studied. First, the \(\mu_{\mathrm{R}}\) and \(\mu_{\mathrm{F}}\) are varied up and down by a factor two, excluding variations in opposite directions~\cite{ATL-PHYS-PUB-2016-002}, and
applied to EW, Int, and QCD \sswwjj samples coherently. It is important to note that the \(\mu_{\mathrm{R}}\) variation is zero for the EW \sswwjj LO model. Therefore, as a second variation, the difference with respect to the \SHERPA~EW (\SHERPA~QCD) calculation is included as an uncertainty in the EW (QCD) \sswwjj prediction. The NLO EW effects on the EW \sswwjj prediction were derived by the authors of Ref.~\cite{Biedermann:2017bss} for the fiducial region of this measurement and are taken as an uncertainty in the nominal model as well, using an \mjj-dependent correction factor.
The parton shower uncertainty is included by considering a difference between the nominal prediction using \PYTHIA and an alternative prediction using the \HERWIG parton shower model.
Additionally, PDF and $\as$ variations are included, using standard prescriptions from Ref.~\cite{Butterworth:2015oua}.
For the cross section extraction, the theory variations of the respective signal process (which can be EW \sswwjj or inclusive \sswwjj production) are normalised to the nominal prediction in the fiducial region, defined in Section~\ref{subsec:fid_cross_sec}.
 
The QCD-induced $\wzjj$ background MC model suffers from \mjj mismodelling, which is corrected for by reweighting the \mjj distribution to data with other processes subtracted. The uncertainty of the reweighting function is dominated by limited data in the reweighting region and is represented by corresponding parameter variations of the exponential reweighting fit function. Further details about the systematic uncertainties from the \mjj reweighting can be found in Section~\ref{subsec:bkg_WZ}.
Additionally, uncertainties in the theoretical modelling of the QCD- and EW-induced \wzjj processes are taken into account, including \(\mu_{\mathrm{R}}\) and \(\mu_{\mathrm{F}}\) variations~\cite{ATL-PHYS-PUB-2016-002}, PDF, and \as~\cite{Butterworth:2015oua} variations.
 
Normalisation uncertainties of minor prompt lepton background sources are included, as quantified in Section~\ref{subsec:bkg_other}.
 
Statistical uncertainties of all signal and background predictions are considered and are subdominant.
 
Systematic uncertainties are included in the cross section extraction fits, described in Section~\ref{subsec:fid_cross_sec}, as nuisance parameters (NP) constrained by Gaussian functions. The statistical uncertainties in the model are included as NPs constrained by Poisson distributions. Table~\ref{tab:EW_cross_sec_syst} shows the impact of these uncertainties on the fitted signal strength of the EW \sswwjj production.
 
\begin{table}
\caption{Impact of the uncertainty on the EW \sswwjj cross section measurement. The contribution of a systematic uncertainty (uncertainty group) to the total uncertainty is evaluated by fixing the respective NP (NPs) to its (their) best-fit value(s), redoing the fit, and subtracting the uncertainties of the cross section in quadrature. The procedure is implemented incrementally such that the grouped systematic and statistical uncertainties added in quadrature correspond to the total cross section uncertainty by construction. Lepton calibration uncertainties encompass the effects of calibration of lepton energy or momentum scale and resolution, as well as lepton trigger, reconstruction, identification, and isolation efficiencies. ``Background, other'' includes normalisation uncertainties of background samples modelled with MC where their normalisation is not obtained in a dedicated CR. The ``Model statistical'' category is related to the finite number of MC events and data events used for data-driven background estimates. }
\label{tab:EW_cross_sec_syst}
\centering
\begin{tabular}{p{2ex}lcc} 
\toprule 
\multicolumn{2}{l}{Source} & Impact [\%] \\ 
\midrule 
\multicolumn{2}{l}{Experimental} & 4.6\\ 
\midrule 
& Electron calibration & ${\hphantom{0}0.4}$ \\ 
& Muon calibration & ${\hphantom{0}0.5}$ \\ 
& Jet energy scale and resolution & ${\hphantom{0}1.9}$ \\ 
& $E_{\mathrm{T}}^{\mathrm{miss}}$ scale and resolution & ${\hphantom{0}0.2}$ \\ 
& $b$-tagging inefficiency & ${\hphantom{0}0.7}$ \\ 
& Background, misid. leptons & ${\hphantom{0}3.4}$ \\ 
& Background, charge misrec. & ${\hphantom{0}1.0}$ \\ 
& Pile-up modelling & ${\hphantom{0}0.1}$ \\ 
& Luminosity & ${\hphantom{0}1.9}$ \\ 
\midrule 
\multicolumn{2}{l}{Modelling} & 4.5\\ 
\midrule 
& EW \sswwjj, shower, scale, PDF \& $\alpha_s$ & ${\hphantom{0}0.7}$ \\ 
& EW \sswwjj, QCD corrections & ${\hphantom{0}1.9}$ \\ 
& EW \sswwjj, EW corrections & ${\hphantom{0}0.9}$ \\ 
& Int \sswwjj, shower, scale, PDF \& $\alpha_s$ & ${\hphantom{0}0.6}$ \\ 
& QCD \sswwjj, shower, scale, PDF \& $\alpha_s$ & ${\hphantom{0}2.6}$ \\ 
& QCD \sswwjj, QCD corrections & ${\hphantom{0}0.8}$ \\ 
& Background, WZ scale, PDF \& $\alpha_s$ & ${\hphantom{0}0.3}$ \\ 
& Background, WZ reweighting & ${\hphantom{0}1.5}$ \\ 
& Background, other & ${\hphantom{0}1.3}$ \\ 
& Model statistical & ${\hphantom{0}1.8}$ \\ 
\midrule 
\multicolumn{2}{l}{ Experimental and modelling} & ${\hphantom{0}6.4}$ \\ 
\multicolumn{2}{l}{ Data statistical} & ${\hphantom{0}7.4}$ \\ 
\midrule 
\multicolumn{2}{l}{ Total} & ${\hphantom{0}9.8}$ \\ 
\bottomrule 
\end{tabular} 
\end{table}


\FloatBarrier
\section{Cross section extraction}
\label{sec:cross_section}

 
\subsection{Fiducial cross section extraction}
\label{subsec:fid_cross_sec}
The fiducial phase space is chosen to conform as closely as possible to the detector acceptance and to the analysis selection, described in Section~\ref{sec:event_selection}. The analysis fiducial region is defined at the particle level using the collection of stable\footnote{A particle is considered stable if its lifetime is greater than $3\times 10^{-11}$~s.} particles from matrix-element plus parton-shower generators.
Events are required to have two prompt leptons ($e$ or $\mu$) with $\pT > 27~\GeV$ and $|\eta| < 2.5$ dressed with prompt photons that lie within $\Delta R < 0.1$. Events with \(\tau\)-leptons from $W$ decays in the matrix element calculation are vetoed.
The two leptons must have the same sign of electric charge and have an invariant mass $\mll > 20~\GeV$.
At least two jets reconstructed with the \antikt algorithm with a radius parameter of \(R=0.4\) are required, with $|\eta|<4.5$ and $\pT > 65~\GeV\ (35~\GeV)$ for the leading (sub-leading) jet. In cases where a third jet is required, its \pT requirement is lowered to 25~GeV. The jets are clustered including electrons among the clustering inputs, but excluding muons and neutrinos, so an overlap removal step between electrons and jets is applied. If an electron and a jet overlap with $\Delta R(e,\mathrm{jet}) < 0.4$, and the ratio $p_{\mathrm{T,e}}/p_{\mathrm{T,jet}} < 0.5$, the electron is removed, otherwise the jet is removed.
No overlap removal between muons and jets is applied.
Events with \bjets within $|\eta|<2.5$ and $\pT > 20~\GeV$, that are identified via ghost matching~\cite{Cacciari:2008gp} to weakly decaying $b$-hadrons, are vetoed. Also events are vetoed when there is a third prompt lepton with $\pT > 3~\GeV$ and within $|\eta|<2.5$.
Events with the dielectron invariant mass in the range $|m_{ee}-m_Z| < 15~\mathrm{GeV}$ are rejected. The particle-level \MET is reconstructed from the visible final-state objects that meet these criteria and must satisfy the condition $\met>30~\GeV$.
Finally, the invariant mass of the two highest-\pT jets must satisfy $\mjj>500~\GeV$ and their separation in rapidity, $\left|\dyjj\right|$, must exceed 2.

Fiducial cross sections of the EW and inclusive \sswwjj production are measured by performing separate maximum likelihood fits.
The EW (inclusive) \sswwjj signal strength, \(\mu_\mathrm{sig}^{\mathrm{EW}}\) (\(\mu_\mathrm{sig}^{\mathrm{EW+Int+QCD}}\)), defined as the ratio of the observed to expected signal cross section, is a free parameter in the fit. The second free parameter in the fit is the normalisation coefficient of the dominant background, QCD \wzjj, \(\mu^{\mathrm{QCD~WZ}}\).
To constrain the latter parameter, the $WZ$ CR, as defined in Section~\ref{sec:event_selection}, is included in the fit as a single bin.
Systematic uncertainties are included as NPs with Gaussian priors, correlated across all fit regions.
Event yields in the low-\mjj CR, described in Section~\ref{sec:event_selection}, are also included in the fit.
The SR and low-\mjj CR are each split into four subregions depending on the flavours of the leading- and subleading-\pT leptons:
\(ee\), \(e\mu\), \(\mu e\), \(\mu\mu\),
to enable a better constraint on
\pT and flavour-dependent non-prompt and charge misidentification background uncertainties. The \mjj distribution, which provides good discrimination between the \sswwjj signal and backgrounds,
is used for the fit in the SR, with the binning optimised to maximise the expected signal sensitivity.
In the fit for extracting the EW \sswwjj cross section, the histogram corresponding to the interference term between the EW and QCD \sswwjj production modes is
scaled by \(\sqrt{\mu_\mathrm{sig}^{\mathrm{EW}}}\).
 
The measured fiducial cross section, \(\sigma^{\mathrm{EW(+Int+QCD)}}_\mathrm{fid,meas}\), is obtained by multiplying the fiducial cross section predicted by MC, \(\sigma^{\mathrm{EW(+Int+QCD)}}_\mathrm{fid,pred}\) with the fitted signal strength \(\mu_\mathrm{sig}^{\mathrm{EW(+Int+QCD)}}\),
\(\sigma^{\mathrm{EW(+Int+QCD)}}_\mathrm{fid,meas} = \mu_\mathrm{sig}^{\mathrm{EW(+Int+QCD)}} \cdot \sigma^{\mathrm{EW(+Int+QCD)}}_\mathrm{fid,pred}\).
 
Even though the contributions from $W$ bosons that decay to \(\tau\)-leptons with a subsequent decay into electrons or muons are excluded
in the fiducial region definition, they are treated as signal in the fit, assuming lepton universality in $W$ decays,
and correspond to around 11\% of the total signal yield in the SR. The fraction of EW \sswwjj signal events in the SR which do not pass the fiducial region selection corresponds to 31\%. Such events are also treated as signal events in the fit, i.e. are scaled by the same signal strength parameter.
 
\subsection{Differential cross section extraction}
\label{subsec:diff_cross_sec}
A fit procedure similar to the one described in Section~\ref{subsec:fid_cross_sec} is used for extracting
single-differential cross sections, with the exception that two-dimensional distributions are used in the SR and
low-\mjj CR. The variables of interest include the dilepton, \mll, and dijet, \mjj, invariant masses, and the transverse
mass, \mt, of the dilepton--\MET system, defined as
\begin{equation*}
\mt = \sqrt{\left(\ET^{\ell\ell} + \MET \right)^2 - \left|\vec{p}_\mathrm{T}^{\ell\ell} + \vec{E}^{\mathrm{miss}}_{\mathrm{T}}  \right|^2},
\end{equation*}
where \( \ET^{\ell\ell} \) is the transverse energy of the dilepton system, \( \vec{p}_\mathrm{T}^{\ell\ell} \)
is the vectorial sum of the lepton transverse momenta, and \( \vec{E}^{\mathrm{miss}}_{\mathrm{T}} \) is the missing transverse momentum vector.
In addition, the cross sections are measured as a function of the
number of jets between the two signal jets in rapidity, \Ngap, and
the Zeppenfeld variable of the third jet, \Zep~\cite{Ballestrero:2018anz}:
\begin{equation*}
\Zep = \left| \frac{\eta_{\mathrm{j}_3} - \frac{1}{2}(\eta_{\mathrm{j}_1}+\eta_{\mathrm{j}_2})}{\eta_{\mathrm{j}_2}-\eta_{\mathrm{j}_1}}\right|.
\end{equation*}
The latter cross section is measured in the subset of the SR where a third jet is present.
 
One fit per variable of interest is performed to obtain the respective differential cross section.
Separate fits are made to extract the EW \sswwjj and inclusive \sswwjj cross sections.
 
In every bin of the variable of interest, the \mjj distribution is fitted to obtain a better constraint on the signal
strength. For the differential cross section extraction as a function of \mjj,
the \mll shape is used for the signal fit in every \mjj bin, and the low-\mjj CR is dropped.
No division according to lepton flavour is performed in the SR and the low-\mjj CR in these $\mjj$ fits to simplify the fit model.
 
The cross section unfolding is based on a maximum-likelihood fit following the method of Ref.~\cite{TOPQ-2020-27}.
The unfolding procedure is applied to the SR and CR distributions at the detector level. The detector-level signal distribution consists of a sum of subsamples, where each subsample contains signal events with the particle level value of the variable of interest in a specified range (``cross section bin''). The binning of the variable of interest in the SR and low-\mjj CR detector-level
distributions used in the fit matches the cross section binning. Signal strength parameters
associated with the particle-level signal predictions in the respective cross section bin are determined in the fit.  The normalisation of the QCD \wzjj background is also a free parameter.
No regularisation
is applied in the unfolding.
Signal events that fail the fiducial region selection but pass the SR selection are scaled by the same signal strength parameter as the events that pass the fiducial region selection in the respective cross section bin.
The signal strength parameters obtained in the fit are directly used to scale the MC particle-level cross sections,
bin-wise, to obtain the unfolded measured cross sections.

The differential cross section binning for the \mll, \mt, and \mjj variables is optimised to have similar expected EW \sswwjj
signal significances per bin. For the \Ngap variable, only two bins are possible due to the small number of events at high
gap jet multiplicities. The binning of \Zep is chosen to
display a specific feature of the EW \sswwjj cross section --- the suppressed third jet production in the central region,
\(0<\Zep<0.5\).
 
\subsection{Results}
\label{cross_sec_results}
 
Post-fit yields for signal and background processes in the SR are shown in Table~\ref{tab:SR_yields}, corresponding to the fit
used for the EW \sswwjj fiducial cross section extraction. The fitted value of the QCD \wzjj background
normalisation coefficient is found to be \(\mu_{WZ}^{\mathrm{QCD}} = \muwzewfit\), where the uncertainties correspond to statistical, modelling systematic, experimental systematic,  and
luminosity uncertainties, respectively.
The value well below unity is in agreement with the one found in the ATLAS observation of electroweak \wzjj production~\cite{STDM-2017-23}, and with the inclusive $WZ$ cross section measurement in the two-jet bin at high \mjj values~\cite{STDM-2018-03}. The EW \sswwjj signal strength is found to be \(\mu_\mathrm{sig}^{\mathrm{EW}}=\musigewfit\) and corresponds
to the measured value of the fiducial cross section of EW \sswwjj production of \(\sigma^{\mathrm{EW}}_\mathrm{fid} = \sigmaew~\fb\).
 
\begin{table}
\caption{Signal and background yields in the SR after the fit for the EW \sswwjj fiducial cross section. The yields are shown for different dilepton final states where the first lepton has the highest \pT. The ``Other prompt'' category combines $ZZ$, $VVV$, $t\bar{t}V$, and $tZq$ background processes. The sum of all contributions may differ from the total value due to rounding. The uncertainty includes both the statistical and systematic components.}
\label{tab:SR_yields}
\centering
\begin{tabular}{lS@{\(\,\pm\,\)}SS@{\(\,\pm\,\)}SS@{\(\,\pm\,\)}SS@{\(\,\pm\,\)}SS@{\(\,\pm\,\)}l}
\bottomrule
Process & \multicolumn{2}{c}{$ee$} & \multicolumn{2}{c}{$e\mu$} & \multicolumn{2}{c}{$\mu e$} & \multicolumn{2}{c}{$\mu\mu$} & \multicolumn{2}{c}{Combined} \\
\midrule
$W^{\pm}W^{\pm}jj$ EW    &     32.9  &       3.4  &      81  &       8  &      73  &       7  &      90  &      9  &     277  &      26\\
$W^{\pm}W^{\pm}jj$ QCD   &     1.7  &     0.5  &     8.0  &     2.4  &     7.1  &     2.1  &     9.7  &     2.9  &      27  &       8\\
$W^{\pm}W^{\pm}jj$ Int   &    1.00  &    0.22  &    2.4  &    0.5  &    2.1  &    0.4  &    2.7  &    0.6  &     8.2  &     1.7\\
$W^{\pm}Zjj$ QCD                & 5.5& 0.7   & 18.2& 2.1   & 18.2& 2.2   & 14.0& 1.7   & 56& 6 \\
$W^{\pm}Zjj$ EW                 & 1.69& 0.14   & 4.9& 0.4   & 4.1& 0.4   & 4.2& 0.4   & 14.9& 1.2 \\
Non-prompt               &     8.4  &     1.6  &    14.9  &     2.4  &    10.2  &     1.6  &      21  &       5  &      55  &      9\\
$V\gamma$                   &     1.5  &     0.7  &     6.1  &     2.4  &     5.5  &     2.8  &   \multicolumn{2}{c}{---}  &      13  &       5\\
Charge misid.               &     4.3  &     2.0  &     5.4  &     1.2  &     1.4  &     0.4  &     \multicolumn{2}{c}{---}  &      11  &       4\\
Other prompt             &    0.99  &    0.25  &     2.5  &     0.5  &     1.9  &     0.5  &     1.4  &     1.4  &     6.8  &     2.1\\
\midrule
Total                    &      58  &       4  &     143  &      7  &     123  &      6  &     143  &      8  & 468  & 21\\
\midrule
Data                  &    \multicolumn{2}{c}{52}  &   \multicolumn{2}{c}{149}  &   \multicolumn{2}{c}{127}  &   \multicolumn{2}{c}{147}  &   \multicolumn{2}{c}{475}\\
\bottomrule
\end{tabular}
\end{table}
 
The fitted value of the signal strength scaling the total \sswwjj production, as obtained in a dedicated fit,
is \(\mu_\mathrm{sig}^{\mathrm{EW+Int+QCD}}=\musigewqcdfit\) and corresponds to the measured value of the fiducial cross section of the \sswwjj production of
\(\sigma^{\mathrm{EW+Int+QCD}}_\mathrm{fid} = \sigmaewqcd~\fb\).
A summary of the measured and predicted EW and inclusive \sswwjj fiducial cross sections is presented in Table~\ref{tab:fid_cross_sec}. The MC predictions generally agree with the measurement within the experimental and theoretical uncertainty, although all of the predictions underestimate the measured cross section values.
The \SHERPA MC models partial NLO QCD corrections by including real emissions of up to one extra parton and gives a 2\% lower cross section of the EW \sswwjj production than the LO \MGHWshort\ and \MGPYshort models. The NLO EW corrections~\cite{Biedermann:2017bss} reduce the \SHERPA prediction by around 15\%. Both these observations are consistent with the effects of NLO corrections described in Ref.~\cite{Biedermann:2017bss}. The scale uncertainties are lower for \MGHWshort\ and \MGPYshort than for \SHERPA, because renormalisation-scale variation of the LO EW \sswwjj prediction is exactly zero due to the absence of QCD interactions at the matrix element level.
The prediction from \POWHEGBOX{}+\PYTHIA MC, made at NLO in the VBS approximation, gives a 5\% higher cross section than the LO models. This might be related to the renormalisation and factorisation scales of \POWHEGBOX{}+\PYTHIA being fixed to a constant value of \(\mu_{\mathrm{R}} = \mu_{\mathrm{F}} = m_W\).
 
\begin{table}
\caption{Measured and predicted fiducial cross sections of the EW and inclusive \sswwjj production, quoted with their respective uncertainties. For \POWHEGBOX{}+\PYTHIA
only the central value 
is provided.}
\label{tab:fid_cross_sec}
\centering
\begin{tabular}{lcc}
\toprule
Description & $\sigma^{\mathrm{EW}}_\mathrm{fid}$ \([\fb]\) & $\sigma^{\mathrm{EW+Int+QCD}}_\mathrm{fid}$ \([\fb]\) \\
\midrule
Measured cross section & $\sigmaewStatSyst$ & $\sigmaewqcdStatSyst$ \\
\MGHWshort & \(2.53  \pm 0.04\,\text{(PDF)}  ~^{+\,0.22}_{-\,0.19}\,\text{(scale)}\) & \(2.92 \pm 0.05\,\text{(PDF)}  ~^{+\,0.34}_{-\,0.27}\,\text{(scale)}\)\\
\MGPYshort & \(2.53  \pm 0.04\,\text{(PDF)} ~ ^{+\,0.22}_{-\,0.19}\,\text{(scale)}\) & \(2.90 \pm 0.05\,\text{(PDF)}  ~^{+\,0.33}_{-\,0.26}\,\text{(scale)}\)\\
\SHERPA & \(2.48  \pm 0.04\,\text{(PDF)}  ~^{+\,0.40}_{-\,0.27}\,\text{(scale)}\) & \(2.92 \pm 0.03\,\text{(PDF)}  ~^{+\,0.60}_{-\,0.40}\,\text{(scale)}\)\\
\SHERPA \(\otimes\) NLO EW & \(2.10  \pm 0.03\,\text{(PDF)}  ~^{+\,0.34}_{-\,0.23}\,\text{(scale)}\) & \(2.54 \pm 0.03\,\text{(PDF)}  ~^{+\,0.50}_{-\,0.33}\,\text{(scale)}\)\\
\POWHEGBOX{}+\PYTHIA & 2.64 & --\\
\bottomrule
\end{tabular}
\end{table}
 
\begin{figure}
\centering
\subfloat[]{
\includegraphics[width=0.45\textwidth]{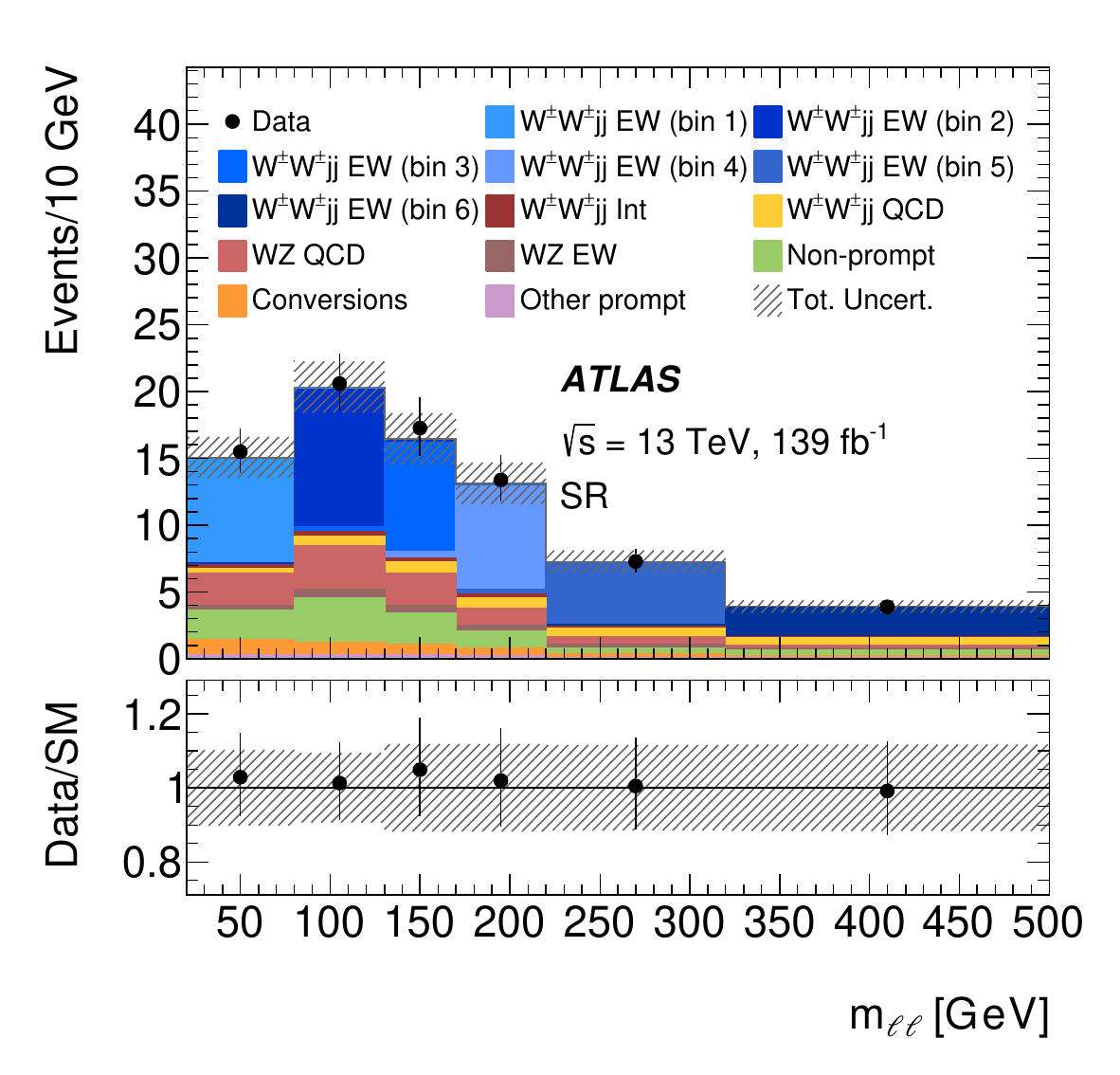}
}
\subfloat[]{
\includegraphics[width=0.45\textwidth]{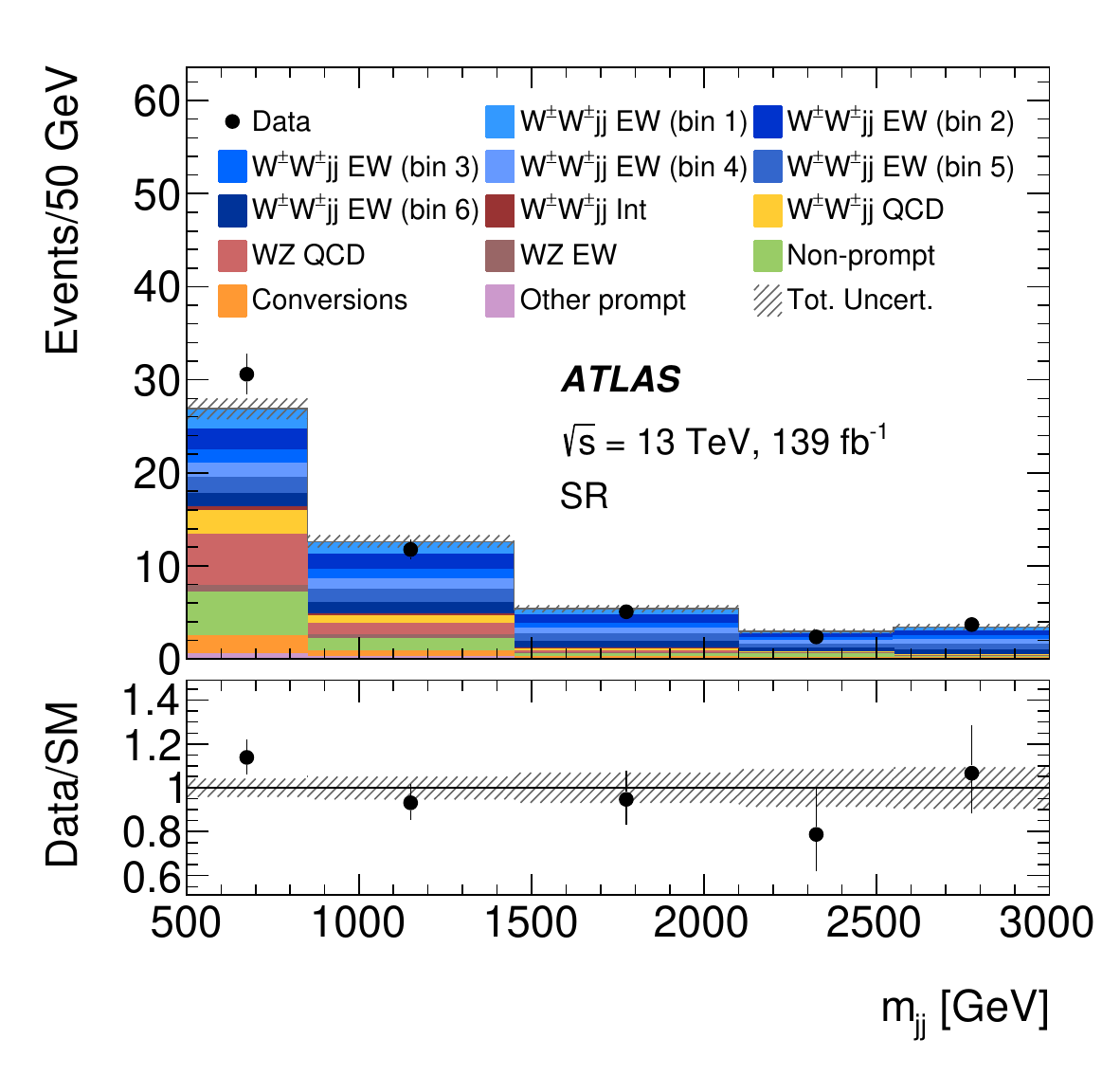}
}\\
\subfloat[]{
\includegraphics[width=.45\linewidth]{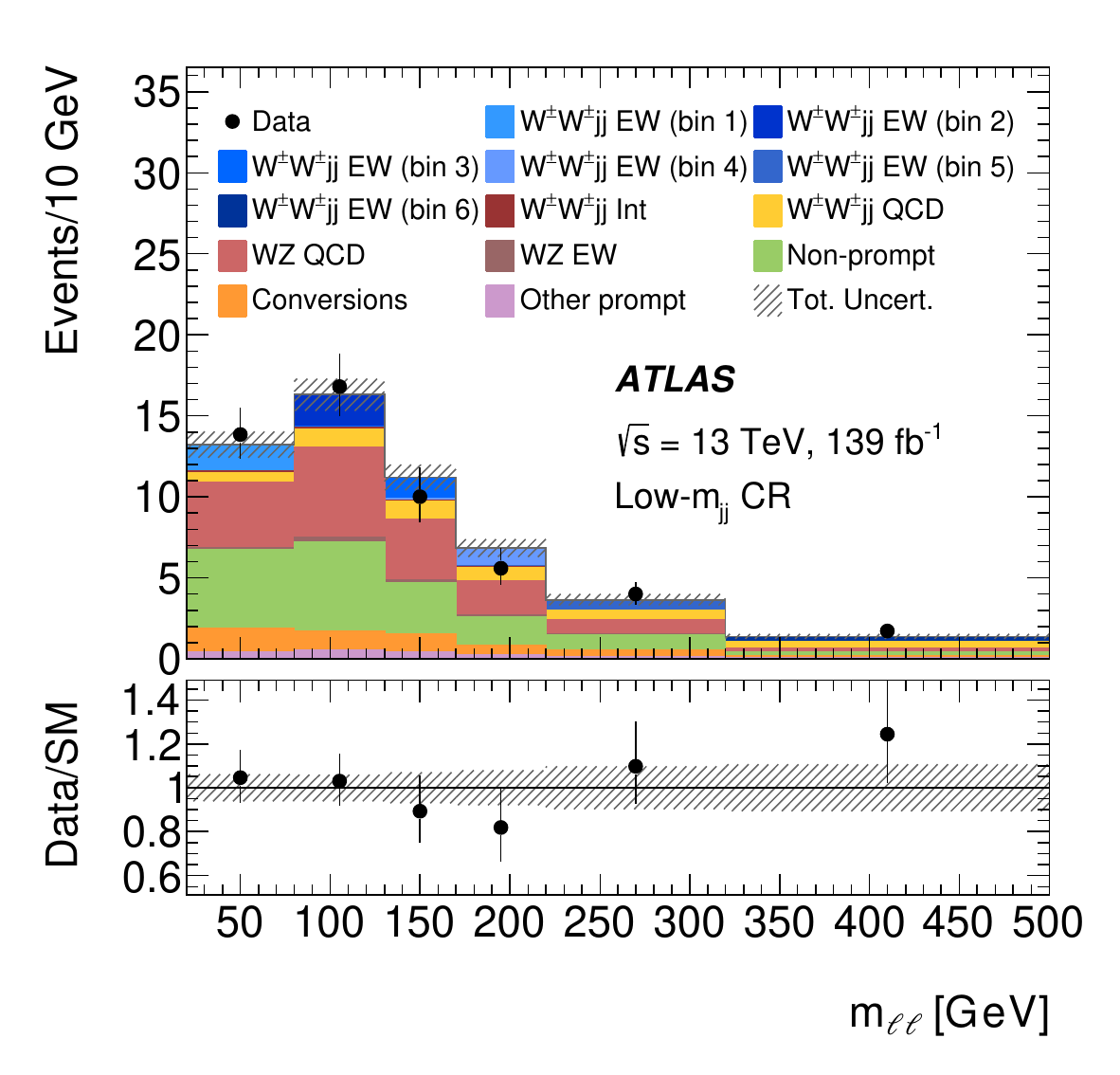}
}
\subfloat[]{
\includegraphics[width=.45\linewidth]{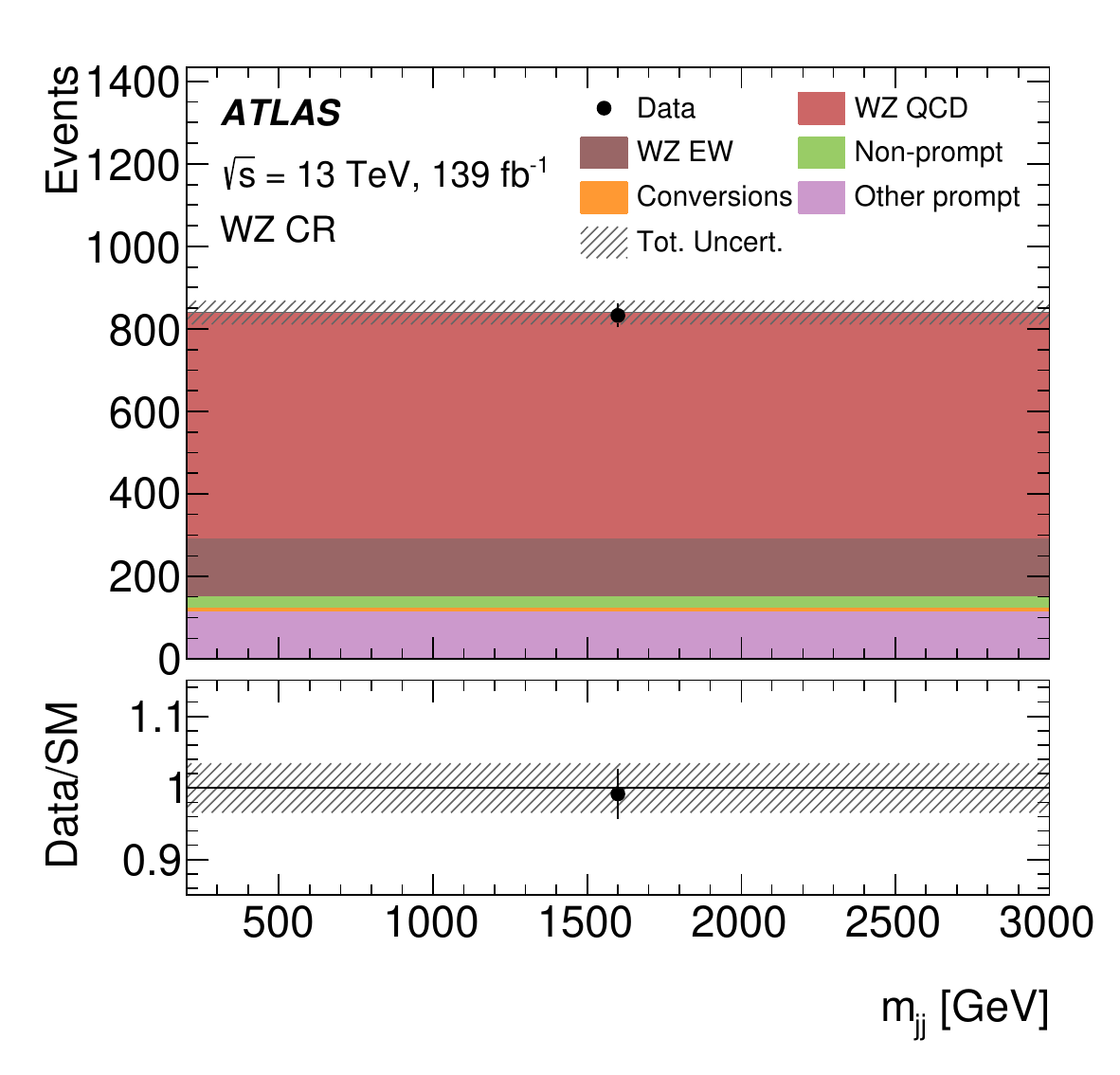}
}
\caption{Post-fit distributions of (a) \mll in SR, (b) \mjj in SR, (c) \mll in low-\mjj CR, and (d) event yield in the WZ CR, as obtained in the extraction of the differential cross section of EW \sswwjj production as a function of \mll. Data are shown as black markers with vertical error bars representing the statistical uncertainty. Contributions of EW \sswwjj events in SR and low-\mjj CR from different particle level \mll bins are presented in different shades of blue, denoted with ``bin $N$'' ($N=1,2,...,6$) in the legend. The last bin of each distribution includes overflow events. The hatched error band on the prediction corresponds to the total uncertainty. The lower panel of each plot shows the ratio of data to prediction.}
\label{fig:postfit_Mll_Mjj}
\end{figure}

Post-fit distributions of \mjj and \mll as obtained in the fit
for extracting the differential cross section as a function of \mll are presented in Figure~\ref{fig:postfit_Mll_Mjj}. Also shown in Figure~\ref{fig:postfit_Mll_Mjj}
are the corresponding post-fit \mll distribution in the low-\mjj CR and the WZ CR yield. The EW \sswwjj signal from different particle-level \mll slices is shown in different shades of blue and denoted by ``bin N'' in the legend, where \(N=1,...,6\) gives the number of the \mll bin. Overall, a very good agreement between the data and post-fit model is seen, the agreement is nearly perfect in the \mll distribution in SR and in the WZ CR, where by construction the free-floating fit parameters compensate for the differences between the pre-fit model and the data.
 
\begin{figure}
\centering
\subfloat[]{
\includegraphics[width=0.37\textwidth]{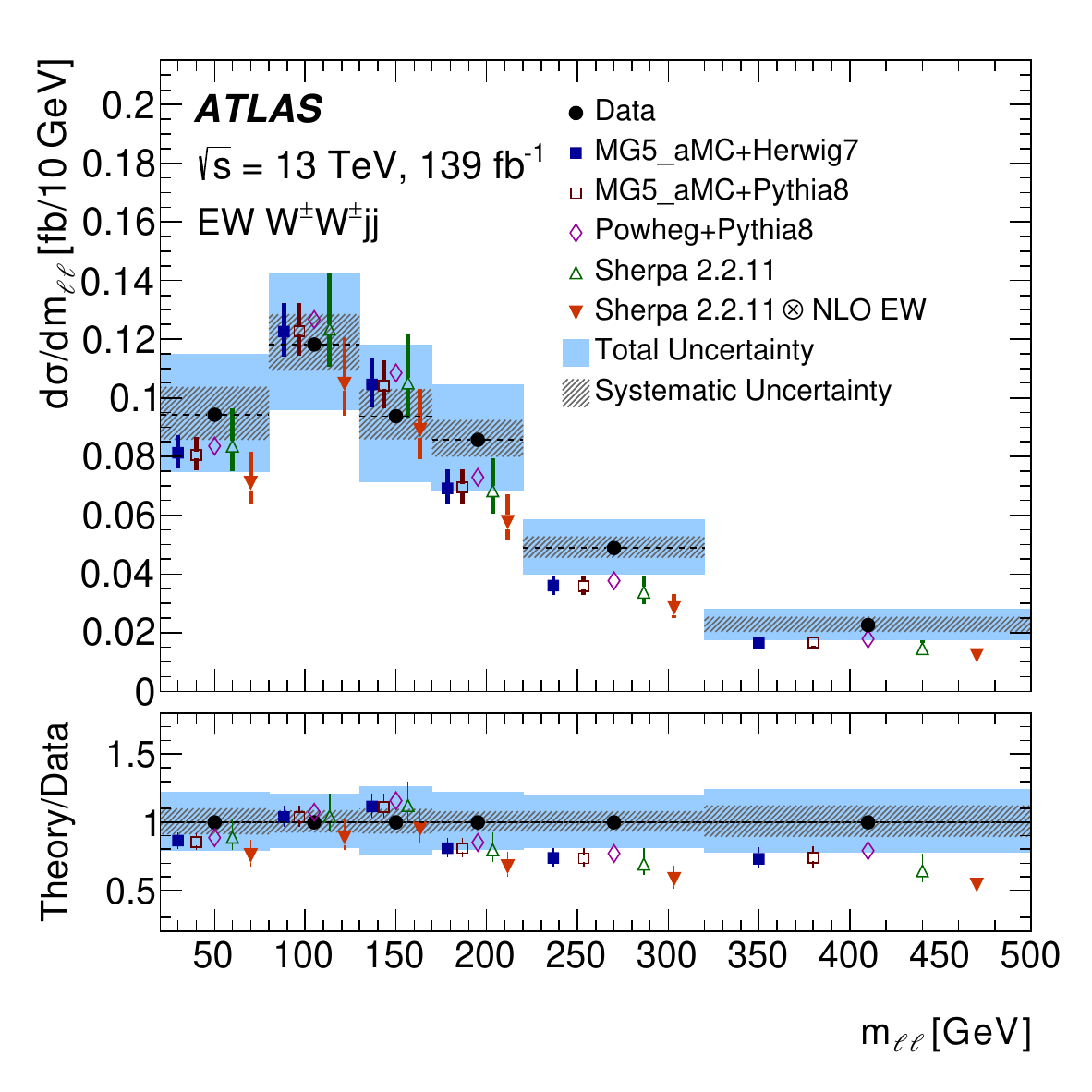}
}
\subfloat[]{
\includegraphics[width=0.37\textwidth]{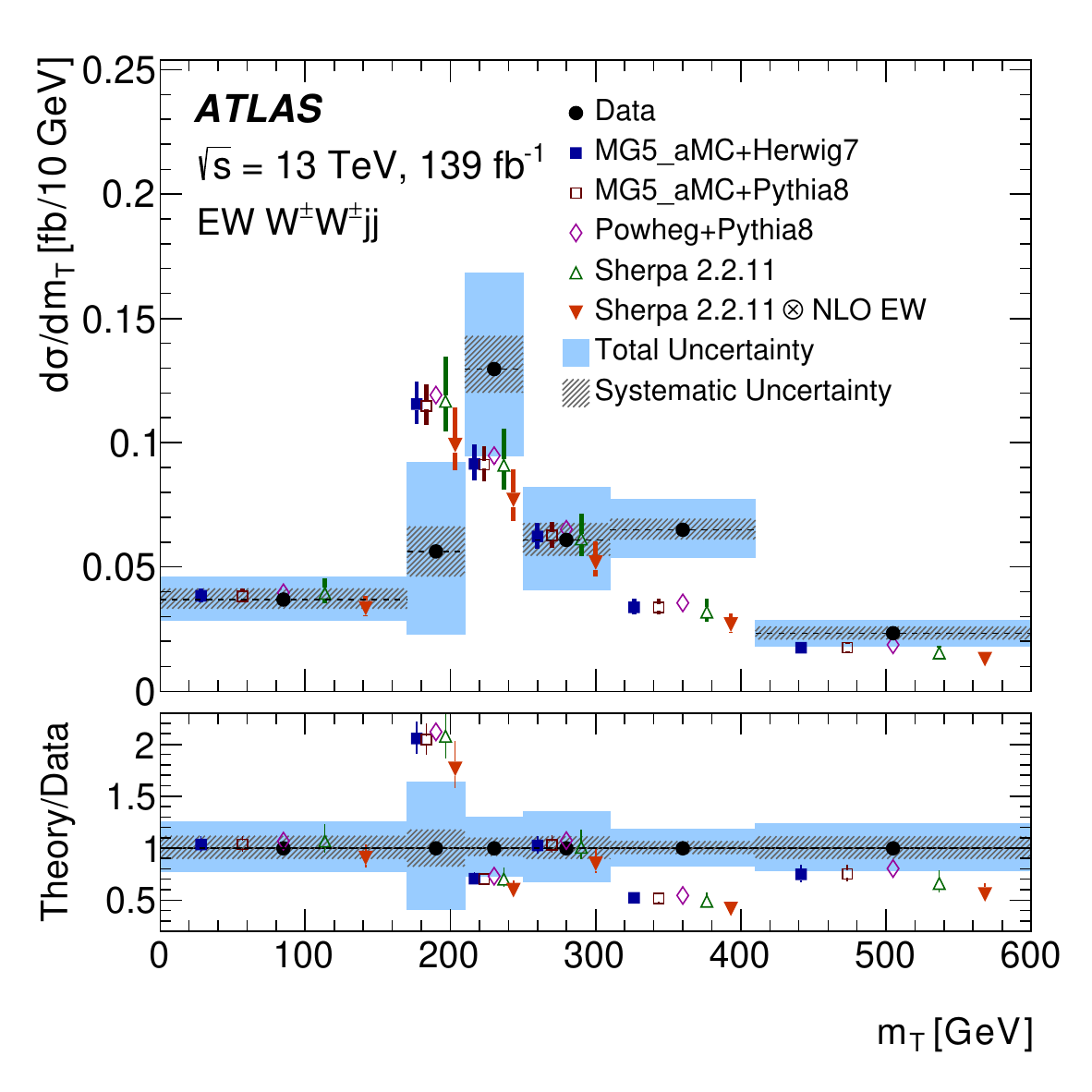}
}\\
\subfloat[]{
\includegraphics[width=0.37\textwidth]{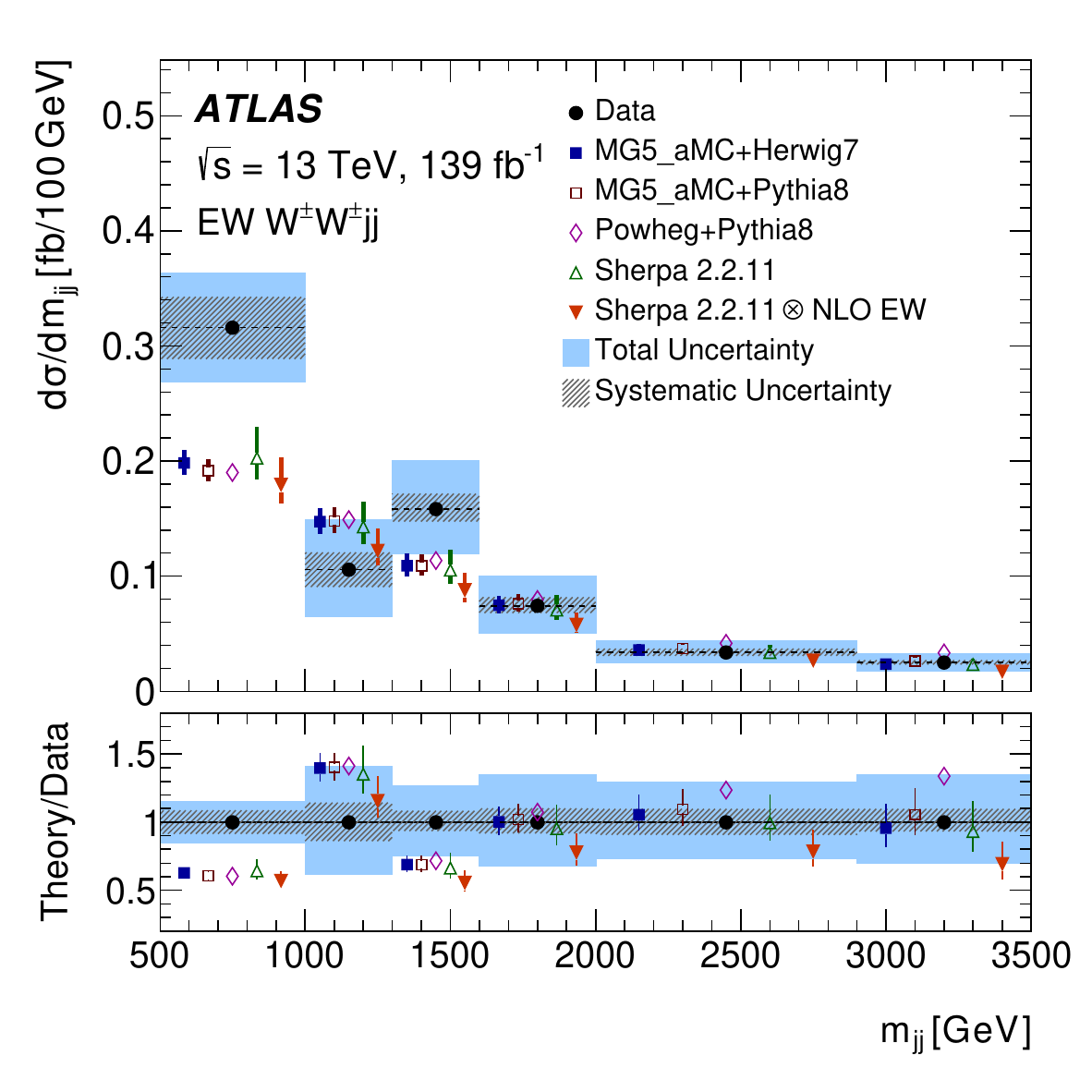}
}
\subfloat[]{
\includegraphics[width=0.37\textwidth]{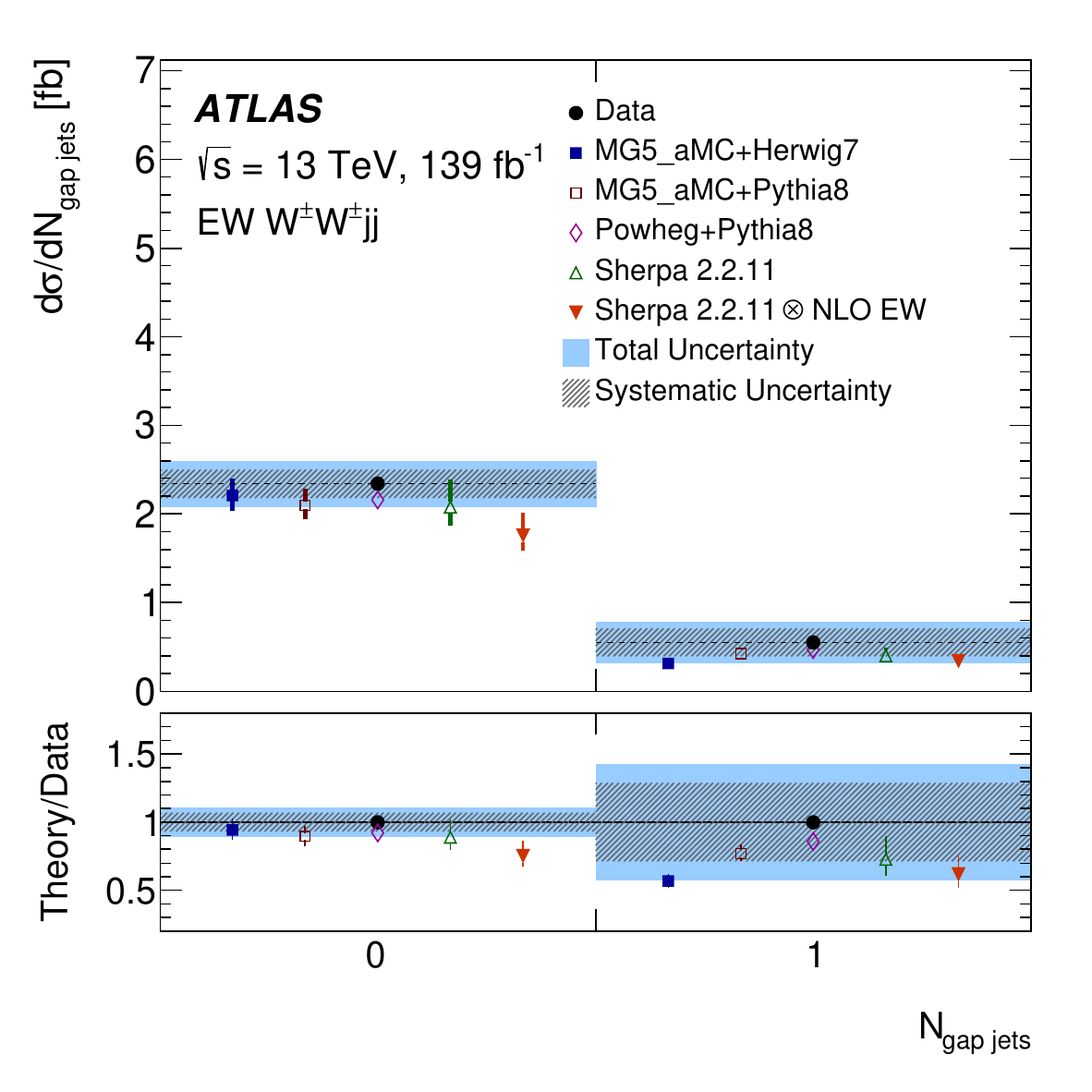}
}\\
\subfloat[]{
\includegraphics[width=0.37\textwidth]{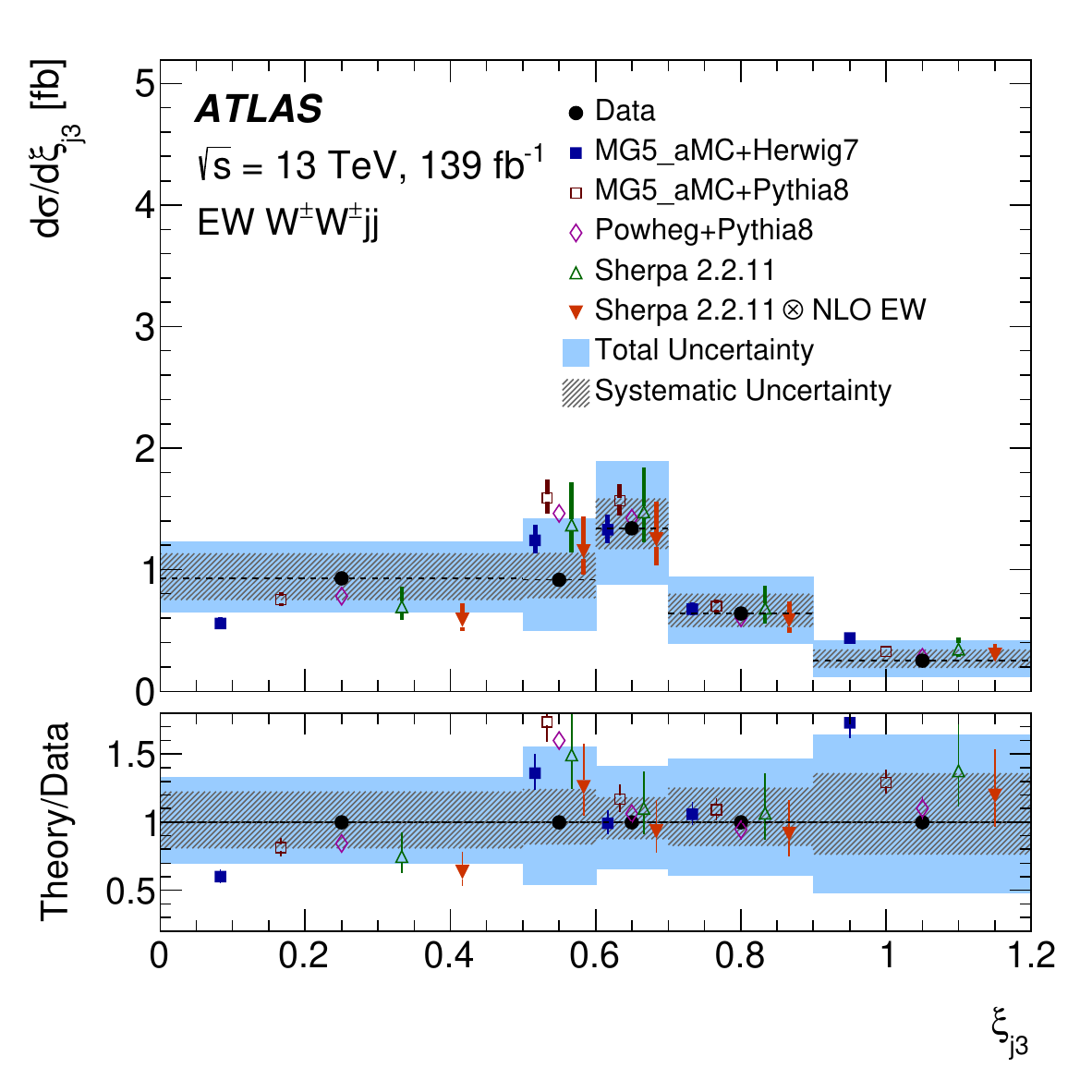}
}
\caption{Fiducial differential cross sections for the EW \sswwjj production as a function of
(a) \mll, (b) \mt, (c) \mjj, (d) \Ngap, and (e) \Zep. The measured data are shown as black points with horizontal bars indicating the bin range and hatched (filled) boxes representing the systematic (total) uncertainty. Different SM predictions as described in the text are compared to the data. The vertical error bars shown on the predictions correspond to the uncertainty coming from the variations of the renormalisation and factorisation scales, PDF and $\as$. Overflow events are included in the last bin. The lower panel of each plot shows the ratio of the predicted to measured cross sections.}
\label{fig:diff_cross_EW}
\end{figure}

Differential cross sections for the EW \sswwjj production as a function of \mll, \mt, \mjj, \Ngap, and \Zep are shown in Figure~\ref{fig:diff_cross_EW} and
compared with SM predictions from  \MGHWshort, \MGPYshort,
\SHERPA-EW, and \POWHEGBOX{}+\PYTHIA. A variant of the \SHERPA-EW prediction with the NLO EW corrections applied via \mjj-dependent factors~\cite{Biedermann:2017bss} is also shown. The NLO EW corrections move the prediction to lower values, especially at high \mjj. The uncertainties of the measured cross sections are dominated by the statistical component in all bins. In general, the SM predictions agree well with the measured cross sections within the quoted uncertainties, although all of the predictions tend to underestimate the data.
The agreement is worse for \mt where an overprediction of the data in the region \(170<\mt<210~\GeV\) and underprediction in the region \(310<\mt<410~\GeV\) are observed, which follow the behaviour at the reconstructed level. This is reflected in Table~\ref{tab:chi2cross_sec} which summarises the compatibility between the nominal predicted and measured differential cross sections in the form of $\chi^2$ and $p$-values. With the exception of \mt, the $p$-values for the EW \sswwjj differential cross sections range from 0.27 to 0.61, while for the differential cross section as a function of \mt the $p$-value is only 0.043.

\begin{figure}
\centering
\subfloat[]{
\includegraphics[width=0.37\textwidth]{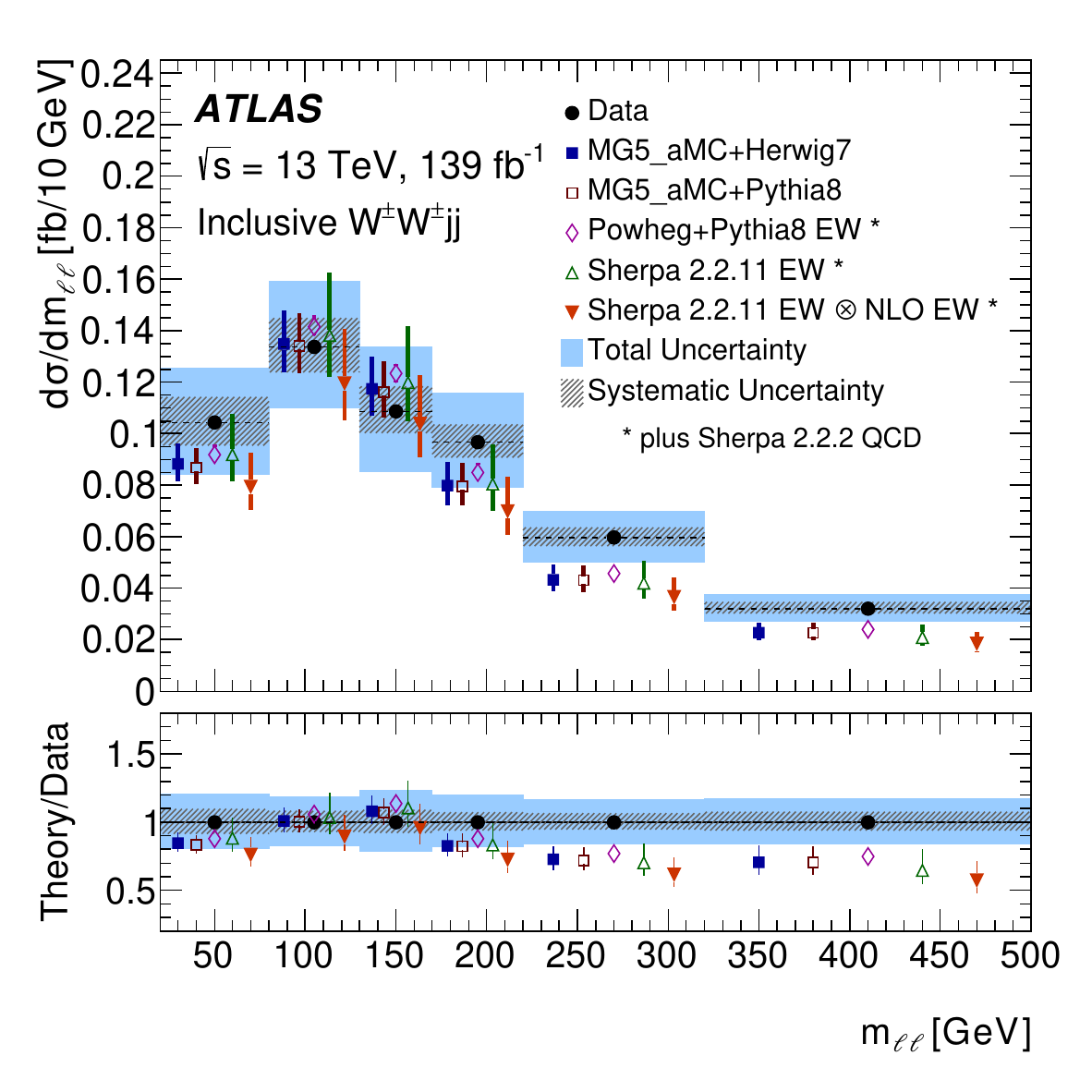}
}
\subfloat[]{
\includegraphics[width=0.37\textwidth]{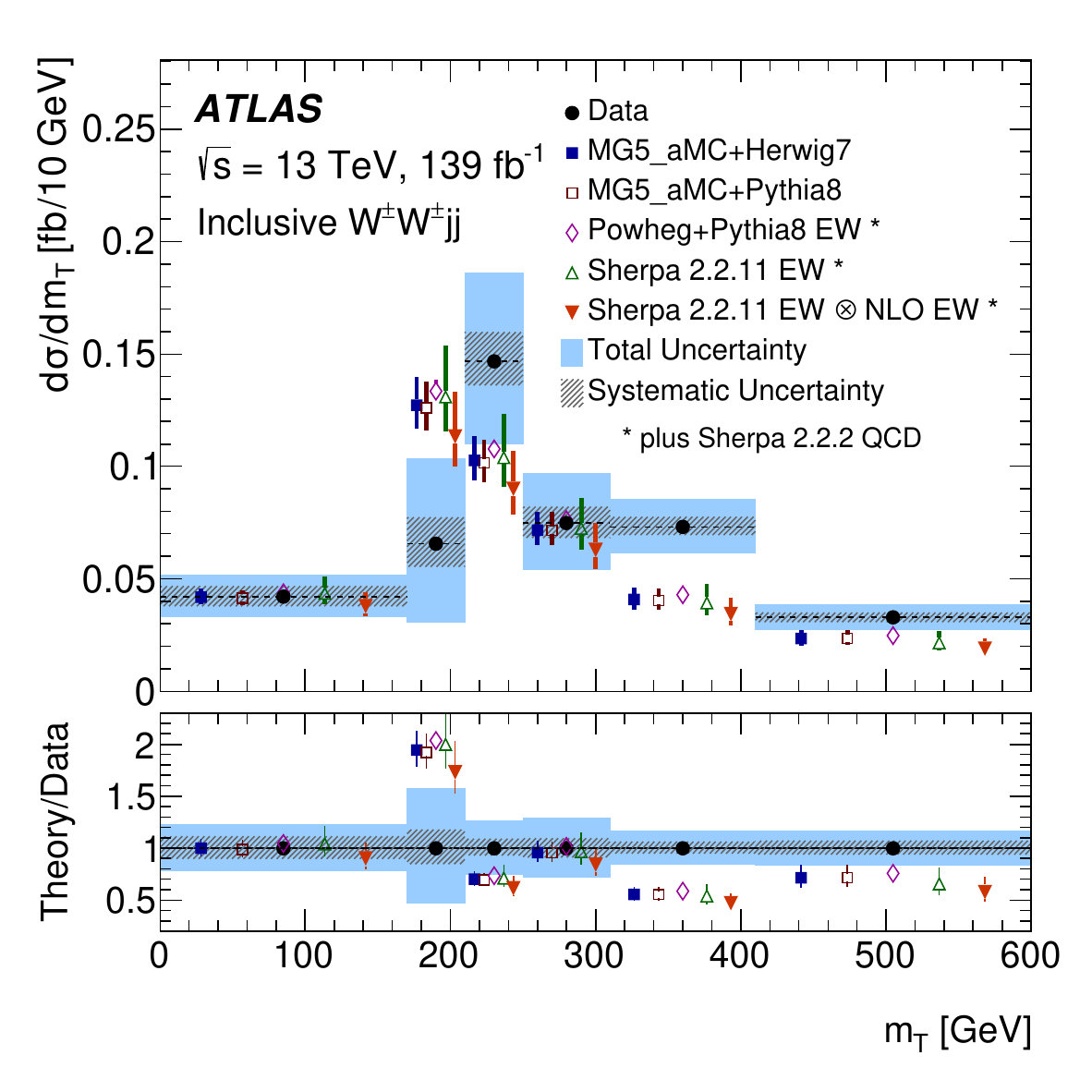}
}\\
\subfloat[]{
\includegraphics[width=0.37\textwidth]{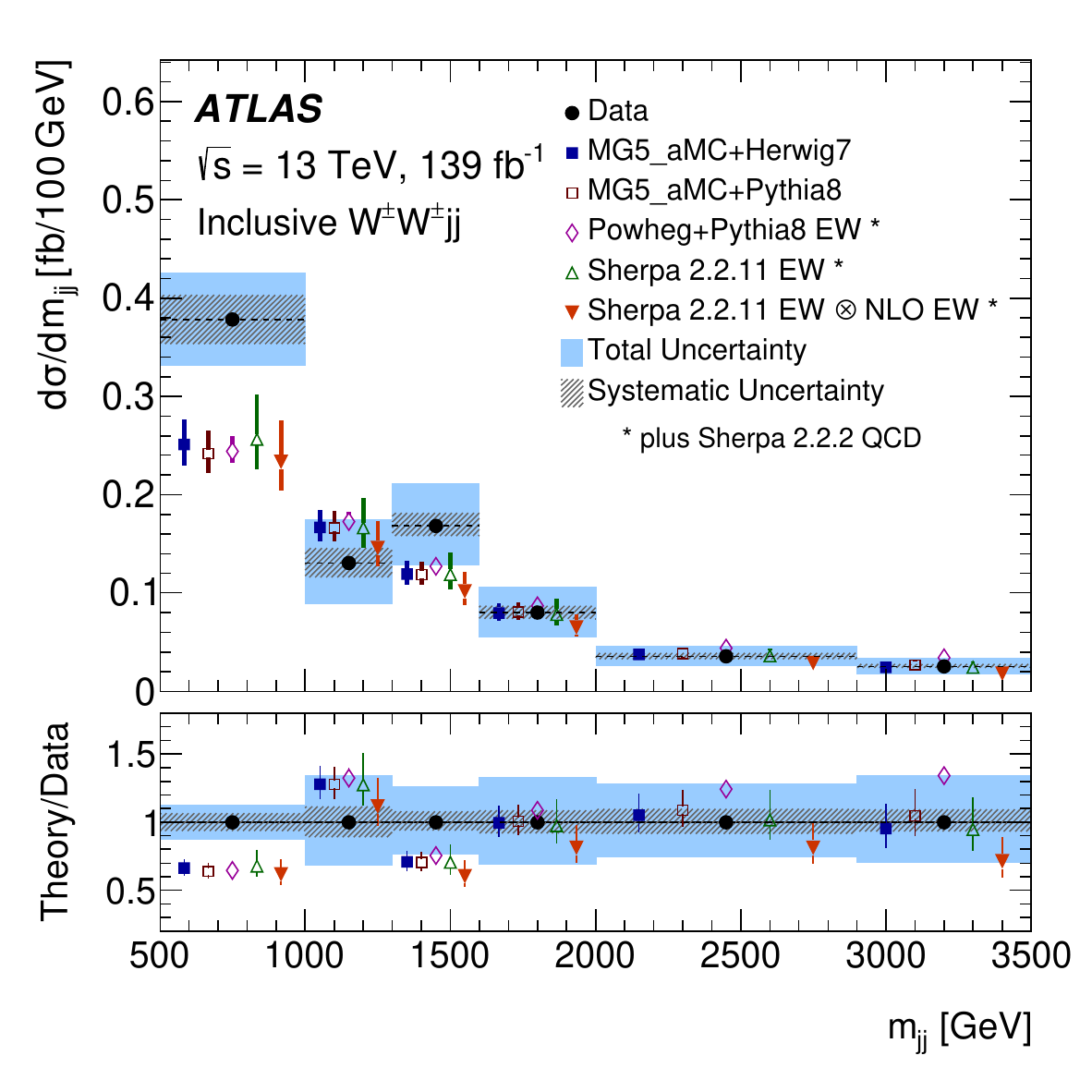}
}
\subfloat[]{
\includegraphics[width=0.37\textwidth]{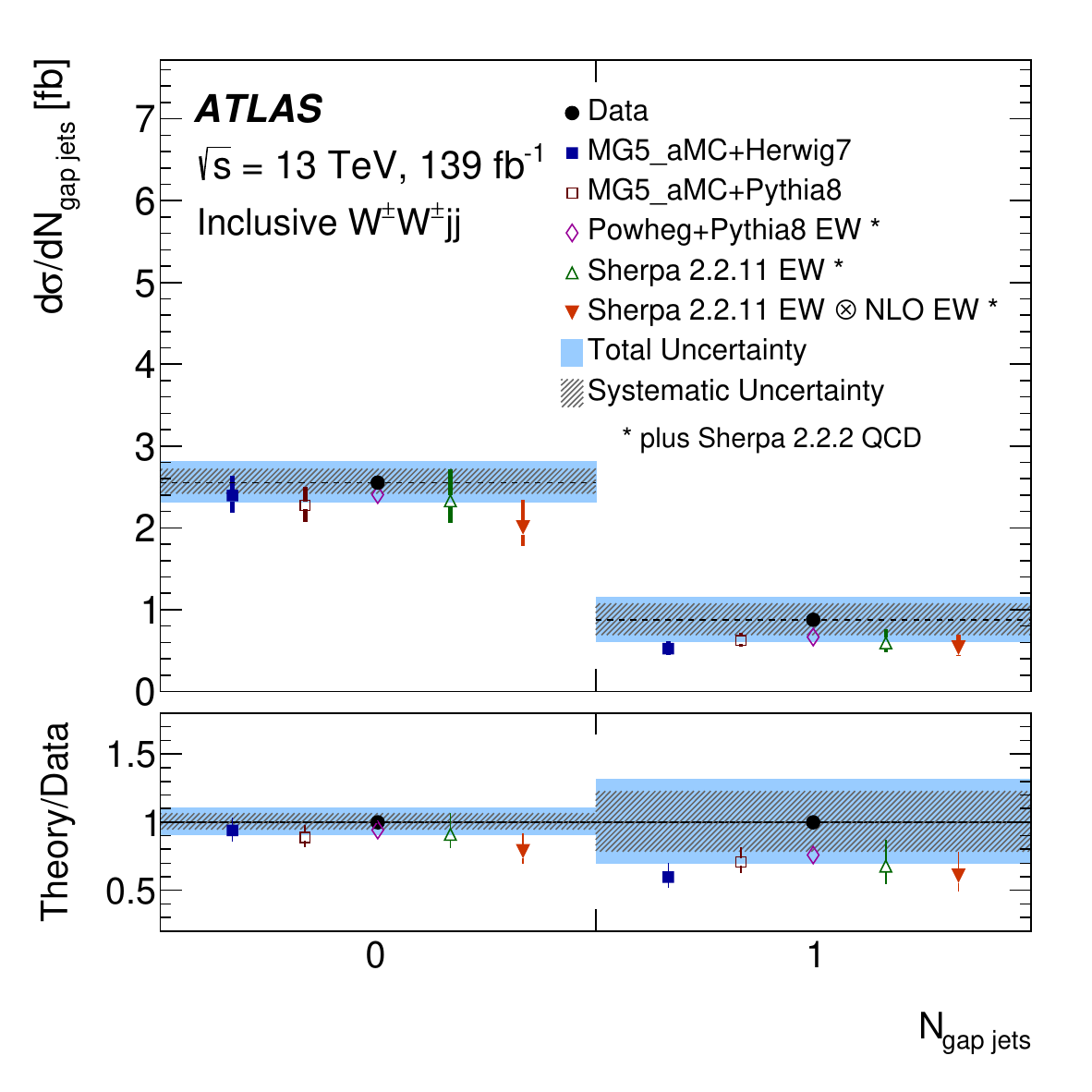}
}\\
\subfloat[]{
\includegraphics[width=0.37\textwidth]{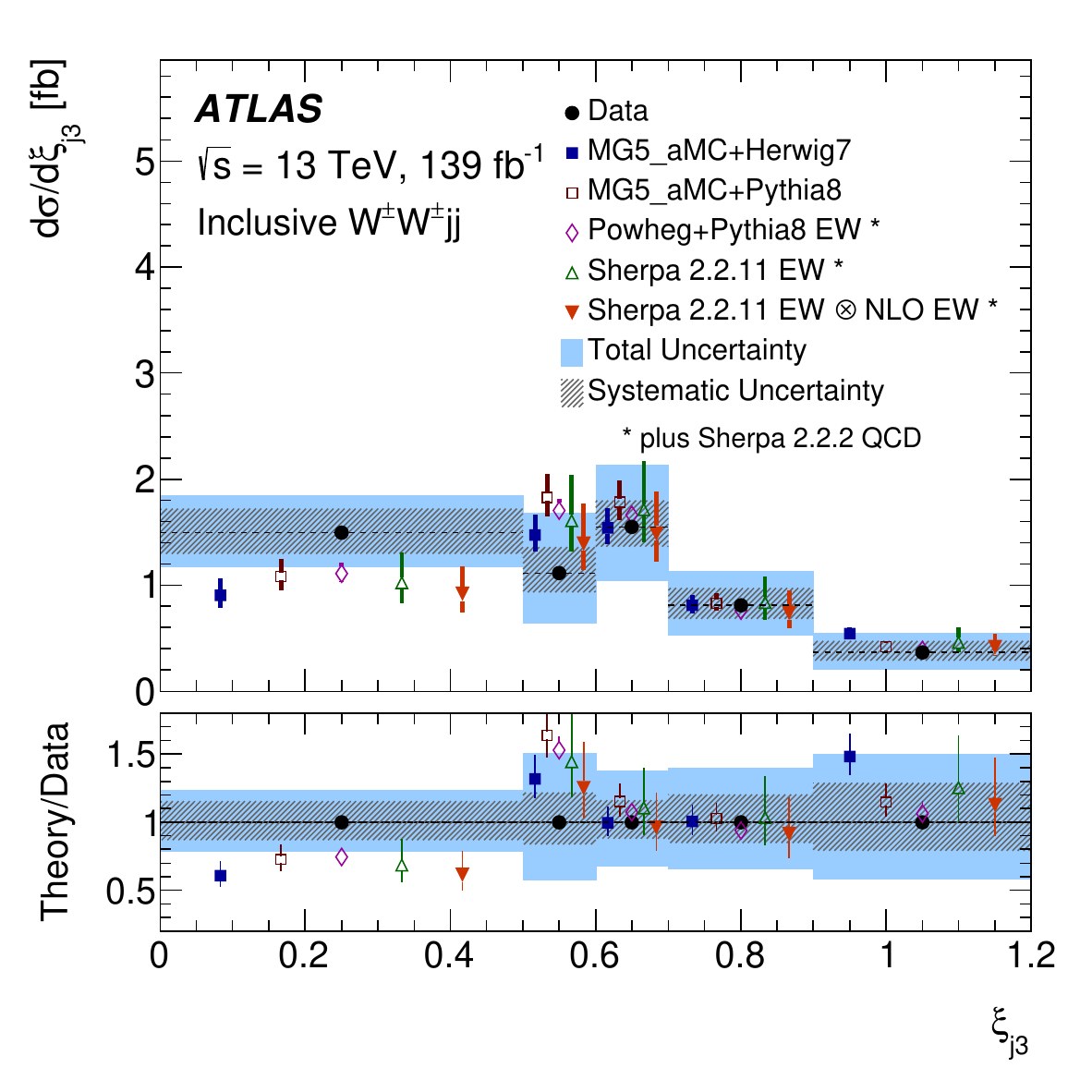}
}
\caption{Measured fiducial cross sections for inclusive \sswwjj production as a function of
(a) \mll, (b) \mt, (c) \mjj, (d) \Ngap, and (e) \Zep.
The measured data are shown as black points with horizontal bars indicating the bin range and hatched (filled) boxes representing the systematic (total) uncertainty. Different SM predictions as described in the text are compared to the data. The vertical error bars shown on the predictions correspond to the uncertainty coming from the variations of the renormalisation and factorisation scales, PDF and $\as$. Overflow events are included in the last bin. The lower panel of each plot shows the ratio of predicted to measured cross sections.}
\label{fig:diff_cross_EWQCD}
\end{figure}

Differential cross sections of the inclusive \sswwjj production as a function of \mll, \mt, \mjj, \Ngap, and \Zep are shown in Figure~\ref{fig:diff_cross_EWQCD} and
compared with SM predictions from
\MGHWshort and \MGPYshort, the sum of the
predictions by \SHERPA-EW and \SHERPA-QCD, and the sum of predictions from \POWHEGBOX{}+\PYTHIA and \SHERPA-QCD. In all cases the interference term between EW and QCD production modes is included. A similar or slightly worse level of agreement between data and SM is observed, compared with the case of EW \sswwjj cross sections (see Table~\ref{tab:chi2cross_sec}).
 
\begin{table}
\caption{\(\chi^2\) and \(p\)-values obtained from the measured differential cross sections and the nominal \MGHWshort prediction, computed using the covariance matrix of the measured differential cross section and the difference between data and model.  The number of degrees of freedom $N_\mathrm{dof}$ is equal to the number of the cross section bins. The uncertainties in the MC prediction are ignored when computing $\chi^2$ and $p$-values. The values are provided for both EW and inclusive differential \sswwjj cross sections. The last column shows the maximum value of the respective variable observed in data.}
\label{tab:chi2cross_sec}
\centering
\begin{tabular}{c|c c|c c|c}
\toprule
\multirow{2}{*}{Variable} & \multicolumn{2}{c|}{EW \sswwjj} & \multicolumn{2}{c|}{Inclusive \sswwjj} &  \multirow{2}{*}{Max. value in data}\\
& $\chi^2/N_{\mathrm{dof}}$ & $p$-value & $\chi^2/N_{\mathrm{dof}}$ & $p$-value & \\
\midrule
\mll & 4.5/6 & 0.605 & 7.34/6 & 0.291 & 1081 GeV \\
\mt & 13.0/6 & 0.043 & 16.33/6 & 0.012 & 1270 GeV \\
\mjj & 7.6/6 & 0.266 & 8.67/6 & 0.193 & 6328 GeV \\
\Ngap & 2.5/2 & 0.282 & 2.53/2 & 0.282 & 5 \\
\Zep & 4.2/5 & 0.517 & 4.93/5 & 0.424 & 1.74 \\
\bottomrule
\end{tabular}
\end{table}


\FloatBarrier
\section{Limits on anomalous quartic gauge couplings}
\label{sec:EFT}

 
Contributions from models beyond the SM can modify the quartic interactions of weak bosons (Figure~\ref{fig:diagrams_ewk_vbs}~(c)). The EW production of \sswwjj in particular is sensitive to the interaction of four $W$ bosons. This measurement can therefore be used to search for new physics that affects the $WWWW$ coupling. The sensitivity is quantified by setting limits on relevant D-8 EFT operators. The operators are expressed using a model~\cite{Eboli:2006wa,Eboli:2016kko} that provides nine independent charge-conjugate and parity conserving D-8 effective operators relevant to the $WWWW$ quartic couplings. The interpretation uses the reconstructed dilepton invariant mass \mll obtained in the current measurement. The approach used is similar to that of the re-interpretation of the \sswwjj and \wzjj measurements based on the $36~\mathrm{fb}^{-1}$ data set made by the ATLAS Collaboration~\cite{ATL-PHYS-PUB-2023-002manual}.

The EFT Lagrangian can be written as an expansion in inverse energy where the first terms that conserve baryon and lepton numbers have coefficients quadratic in energy. As a consequence, the corresponding field operators
relevant for the LHC
are dimension-6 (D-6) and D-8 operators.
Therefore, the effective Lagrangian can be written in terms of higher dimension operators and their respective Wilson coefficients as:
 
\begin{equation*}
\mathcal{L}_{\textrm{eff}} = \mathcal{L}_{\textrm{SM}} + \sum_{i} { \frac{f_i^{(6)}}{{\Lambda}^2}}  O^{(6)}_i+ \sum_{j} { \frac{f_j^{(8)}}{{\Lambda}^4}} O^{(8)}_j + ...
\end{equation*}
where $O^{(6),(8)}_{i, j}$ are the D-6 and D-8 operators and involve SM fields with respective dimensionless couplings
$f_{i}^{(6)}$ and $f_{j}^{(8)}$, and $\Lambda$ is the energy scale of the new processes.
 
The basic 
units needed
to construct the effective Lagrangian for VBS are genuine Quartic Gauge Coupling (QGC) vertices, which appear in lowest order as D-8 operators\footnote{D-8 operators are the lowest-dimension operators that affect QGC but do not affect triple gauge coupling vertices. The effect of D-6 operators in VBS processes and in QCD diboson production accompanied by jets is of interest on its own ~\cite{Kalinowski:2018oxd,Dedes:2020xmo,Bellan:2021dcy}, but is not studied here.} and can be classified into three groups~\cite{Eboli:2016kko}: operators that contain four covariant derivatives of the Higgs field ($O_\mathrm{S{0, 1, 2}}$ of scalar type);  those that contain two Higgs boson covariant derivatives and two field-strength tensors ($O_\mathrm{M{0, 1, 2, 3, 4, 5,7}}$ of mixed -- scalar and tensor -- type); and those with four field-strength tensors ($O_\mathrm{T{0,1,2,5,6,7,8,9}}$ of tensor type). Among these, the operators that affect $W^{\pm}W^{\pm}jj$ are $O_\mathrm{S{0,1,2}}$, $O_\mathrm{M{0,1,7}}$ and $O_\mathrm{T{0,1,2}}$. Consequently, the respective coefficients that are constrained from these measurements are: $f_\mathrm{S{02,1}}/\Lambda^4$, $f_\mathrm{M{0, 1, 7}}/\Lambda^4$ and
$f_\mathrm{T{0,1,2}}/\Lambda^4$. Because the operators $O_\mathrm{S0}$ and $O_\mathrm{S2}$ are Hermitian conjugates of each other, they are varied simultaneously, with equal coefficient values $f_\mathrm{S{0}}=f_\mathrm{S{1}}=f_\mathrm{S{02}}$.
For simplicity, the parameters $c_i = c^{(8)}_i = f_i^{(8)}/\Lambda^4$ are used in the following.

An individual sample for every term (SM, interference, quadratic or cross, see Ref.~\cite{ATL-PHYS-PUB-2023-002manual} for details) is generated as described in Section~\ref{sec:sswwjj_EFT_samples}. To generate events corresponding to a given value of the EFT coefficient $c_i$ (or $c_i$ and $c_j$ for cross terms), the respective sample is multiplied by the appropriate parameter(s) ($c_i$, $|c_i|^2$, $c_i c_j $).

The interpretation of the data is performed using a likelihood modelled as a Poisson distribution, with systematic uncertainties implemented via Gaussian constraints. It uses the $m_{ll}$ distribution in the SR and the low-\mjj CR, including also the \(WZ\) CR as previously defined.
The SM normalisation of EW \sswwjj is assumed, while the EFT signal strength parameter is parameterised in terms of real Wilson coefficients $c_i$, with a separate signal strength parameter for each $c_i$.
A profile-likelihood ratio test statistic is constructed to estimate the confidence intervals for a given $c_i$.
For each individual $c_i$ (or a pair $c_i$, $c_j$ for two-dimensional limits),  maximum likelihood fits are performed by setting other coefficients to zero and maximising the likelihood with respect to the NPs.
 
Confidence intervals are derived using Wilks' theorem~\cite{Wilks:1938dza}, namely, assuming that the profile likelihood test statistic is $\chi^2$ distributed~\cite{Cowan:2010js} (asymptotic formula). A cross-check is performed where the distribution of the profile likelihood test statistic was calculated in a set of pseudo-experiments that were generated by sampling in the neighbourhood of each scanned parameter point.  This confirmed the result obtained using the asymptotic formula.
 
In the EFT framework, the presence of non-zero anomalous QGCs will violate tree-level unitarity at sufficiently large energy. More physical limits are obtained by removing the EFT contributions above the unitarity limit and keeping the SM predictions for all $VV$ invariant masses, even above the unitarity limit.
The unitarity limits from Ref.~\cite{Almeida:2020ylr} are used with only one non-zero Wilson coefficient. For two-dimensional limits the eigenvalues from Ref.~\cite{Almeida:2020ylr} are used to calculate unitarity bounds where only two Wilson coefficients are non-zero.
In this paper, individual limits for the coefficients corresponding to the EFT operators are extracted by removing the EFT contributions that exceed a scale $M_\mathrm{cut}$. The
removal is performed using boson kinematics at parton level (i.e., before the parton shower) and applied on the invariant mass of the $VV$ system. This clipping (unitarisation) technique is described in Ref.~\cite{CMS-SMP-19-012}.
 
The systematic uncertainties are listed in Section~\ref{sec:systematic_uncertainties}. Theory uncertainties affecting the EFT model include the variation of \(\mu_{\mathrm{R}}\) and \(\mu_{\mathrm{F}}\) and the choice of functional form for these scales, along with PDF and $\alpha_S$ uncertainties. The SM EW \sswwjj MC sample used in the EFT interpretation corresponds to around 20\% lower yield in the SR, compared to the SM EW \sswwjj MC sample used for the cross section extraction in Section~\ref{sec:cross_section}, with the difference reaching 40\% in the highest-\mll bin. This difference stems from different choices of \(\mu_{\mathrm{R}}\) and \(\mu_{\mathrm{F}}\) for these two samples, as described in Section~\ref{sec:sswwjj_EFT_samples}. The difference was assigned as an uncertainty during limit setting. Systematic uncertainties weaken the derived limits by up to 10\%.

Using the samples described above, limits at 95\% CL for the parameters corresponding to the operators with label S02, S1, M0, M1, M7, T0, T1 and T2 are extracted, based on their impact on both EW \sswwjj and EW \wzjj. When including the effect of the EFT operators on EW \wzjj production, the limits typically improve by 1\% to 4\%.
 
\begin{figure}[!htb]
\begin{center}
\subfloat[]{
\includegraphics[width=0.49\textwidth]{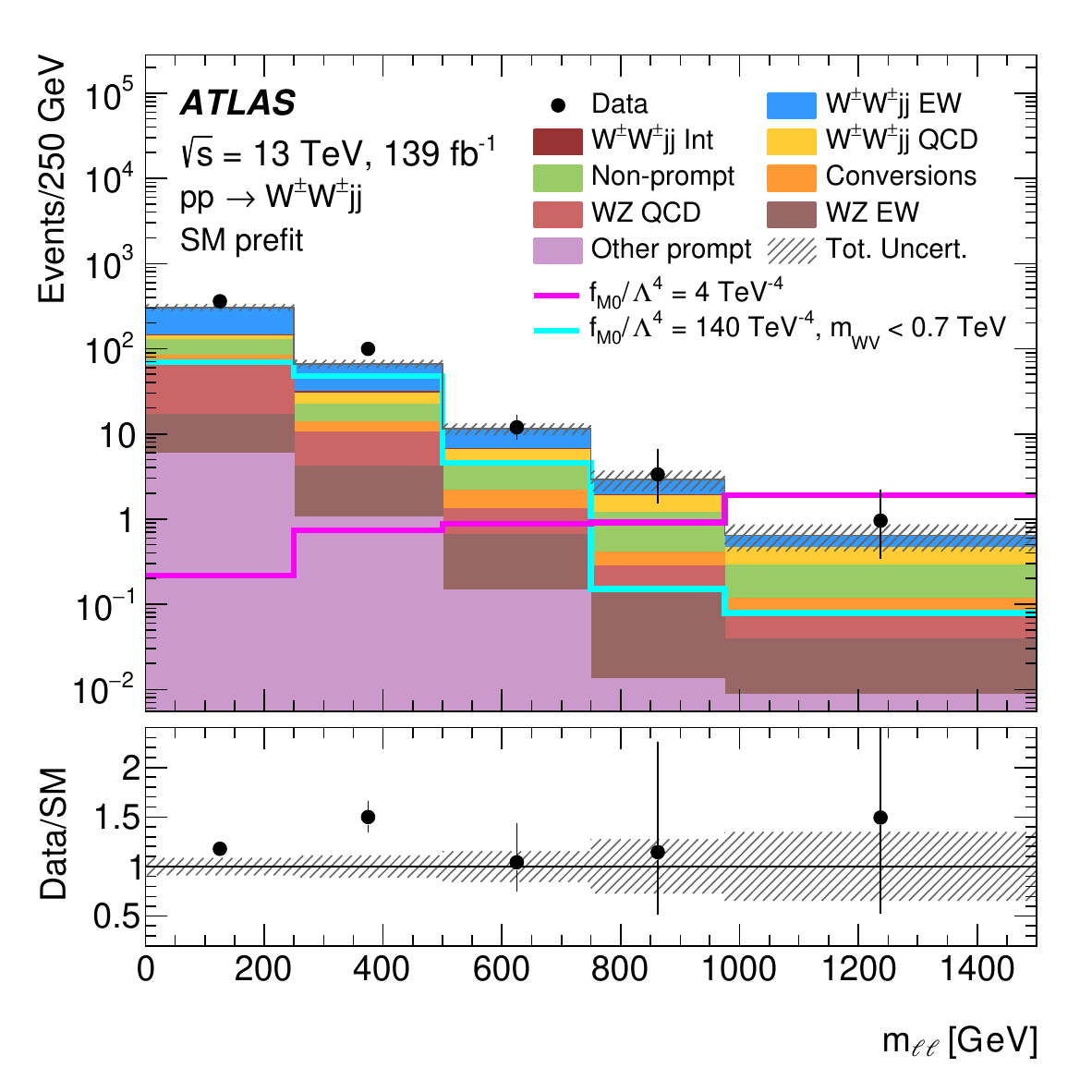}
\label{fig:EFT_prefit}
}
\subfloat[]{
\includegraphics[width=0.49\textwidth]{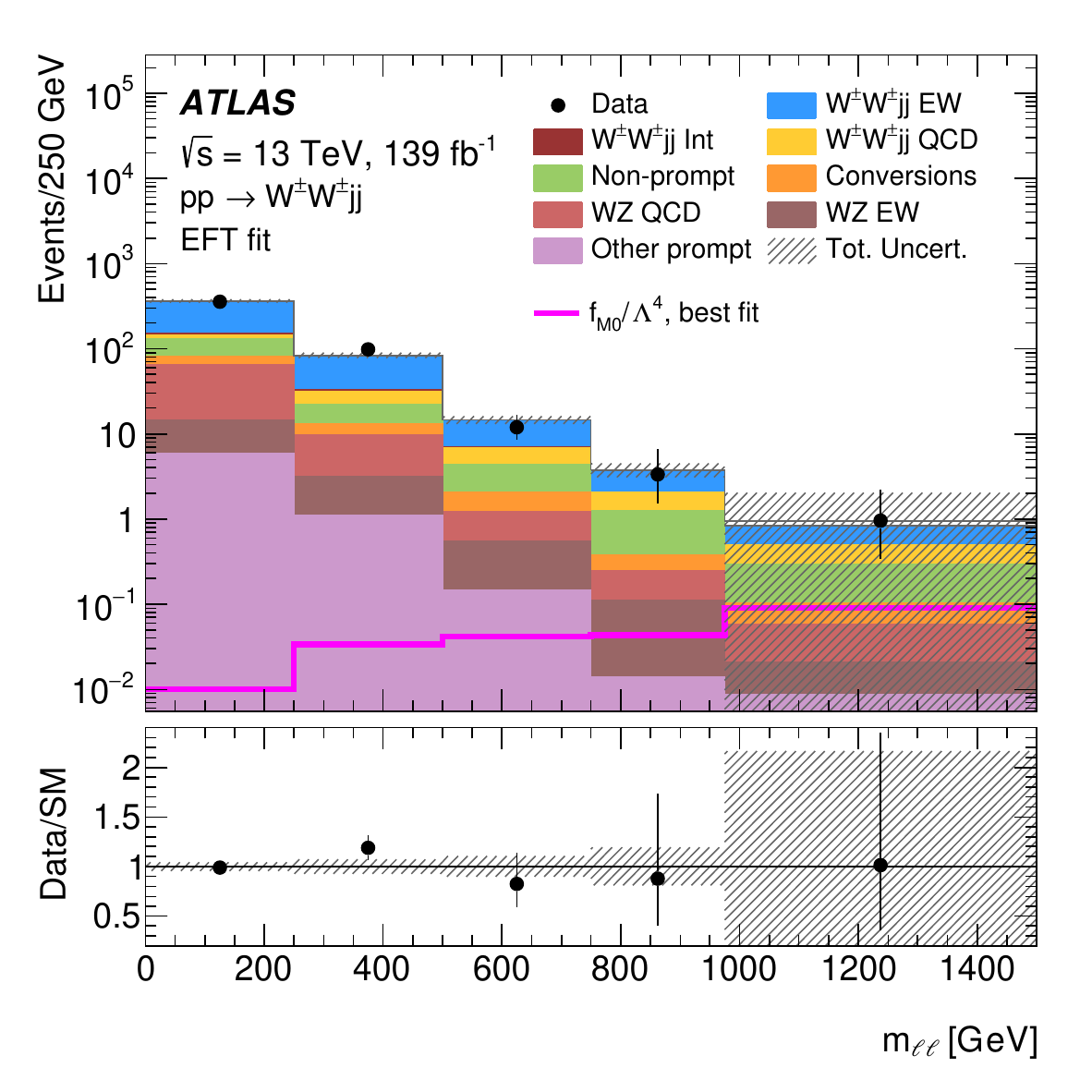}
\label{fig:EFT_postfit}
}
\end{center}
\caption{The observed \mll distribution and the SM distribution before (a) and after (b) the EFT fit. Data are shown as black markers with vertical error bars representing the statistical uncertainty. Filled histograms show contributions of SM processes, with the hatched error band corresponding to the total uncertainty. In (a), the sums of \sswwjj and \wzjj EFT contributions that correspond to the M0 operator with its Wilson coefficient set to its observed upper limit are shown as continuous lines for two cases, one where no unitarisation cut-off is applied, and another where the EFT contributions above $m_{WV} > 0.7~\TeV$ are removed. In (b), the continuous line presents the best-fit contribution of the M0 operator without the unitarisation cut-off. Overflow events are included in the last bin. The lower panel shows the ratio of data to SM prediction.}
\label{fig:Mll_M0}
\end{figure}
 
The pre-fit \mll distribution overlaid with the EFT M0 operator prediction are shown in Figure~\ref{fig:EFT_prefit} for cases when no unitarisation cut-off is applied, and for an EFT unitarisation cut-off of~700~\GeV. The boundaries of the last \mll bin are optimised to have best expected limits with no unitarisation cut-off, while other bins in the tail of the distribution are defined to improve the limits when a unitarisation cut-off is applied. Figure~\ref{fig:EFT_postfit} shows the \mll distribution after the fit of the sum of SM and M0 contributions to data, in case when no EFT unitarisation cut-off is applied.
 
\begin{figure}[!htb]
\begin{center}
\subfloat[]{
\includegraphics[width=0.32\textwidth]{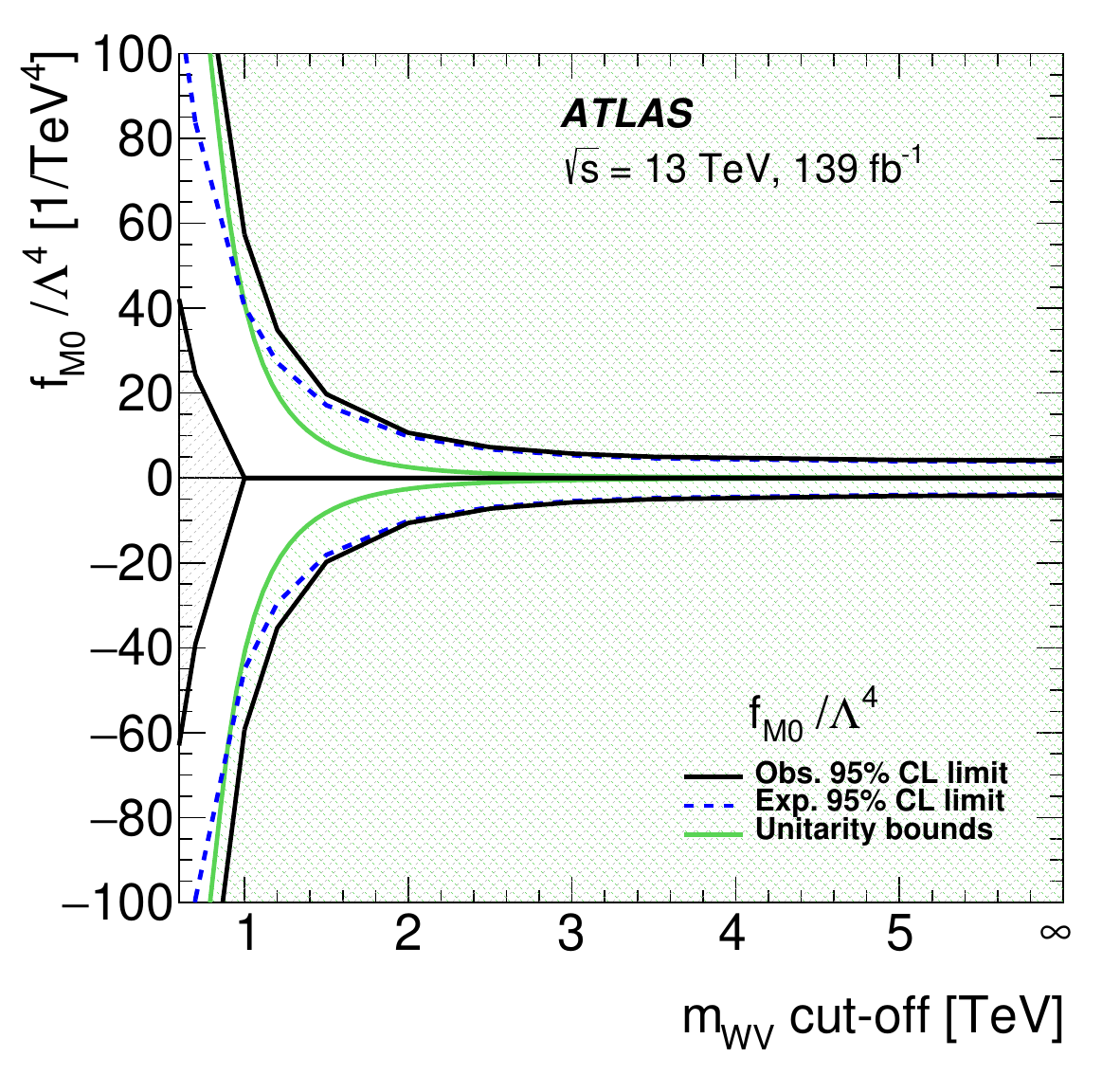}
}
\subfloat[]{
\includegraphics[width=0.32\textwidth]{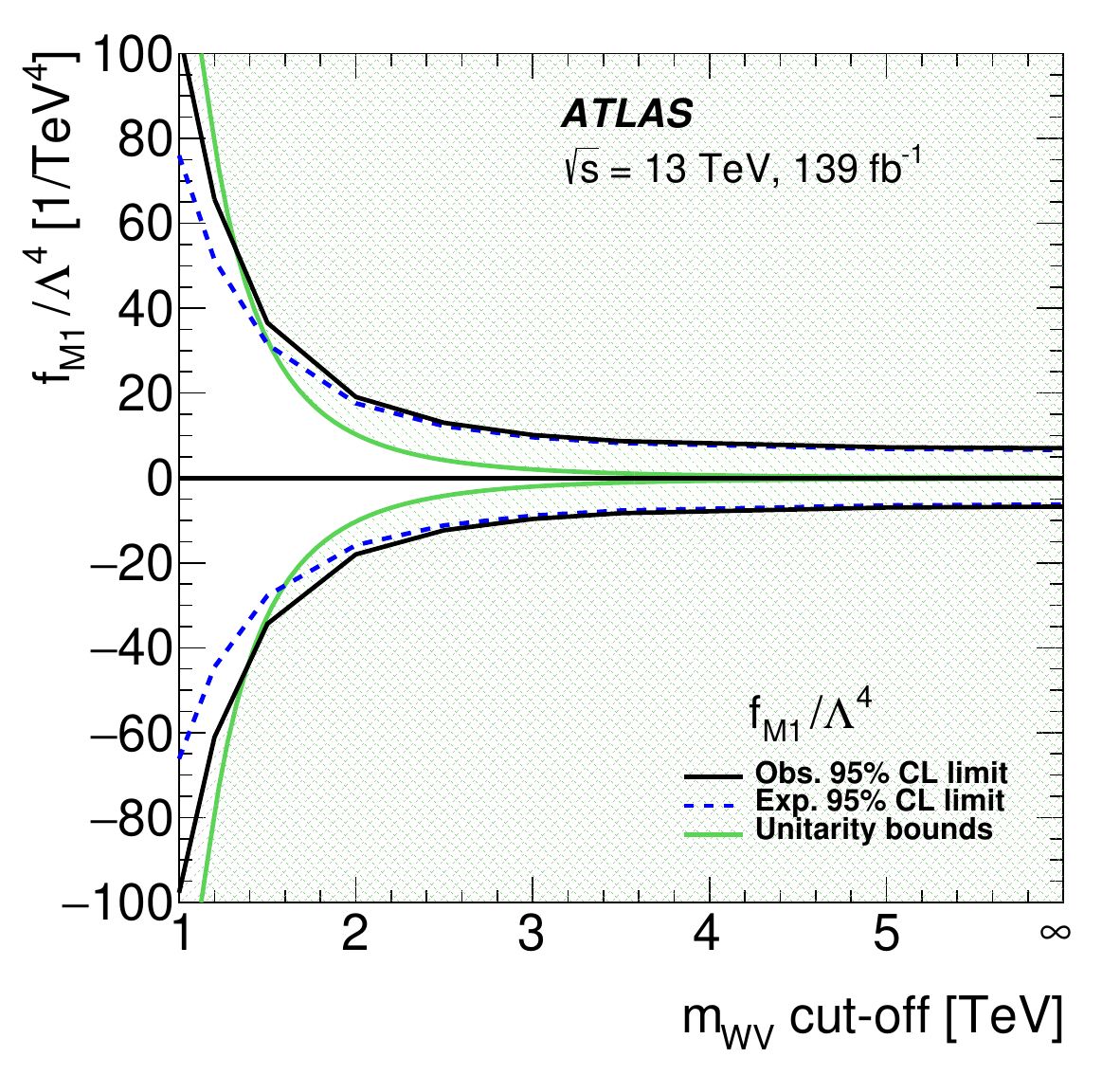}
}
\subfloat[]{
\includegraphics[width=0.32\textwidth]{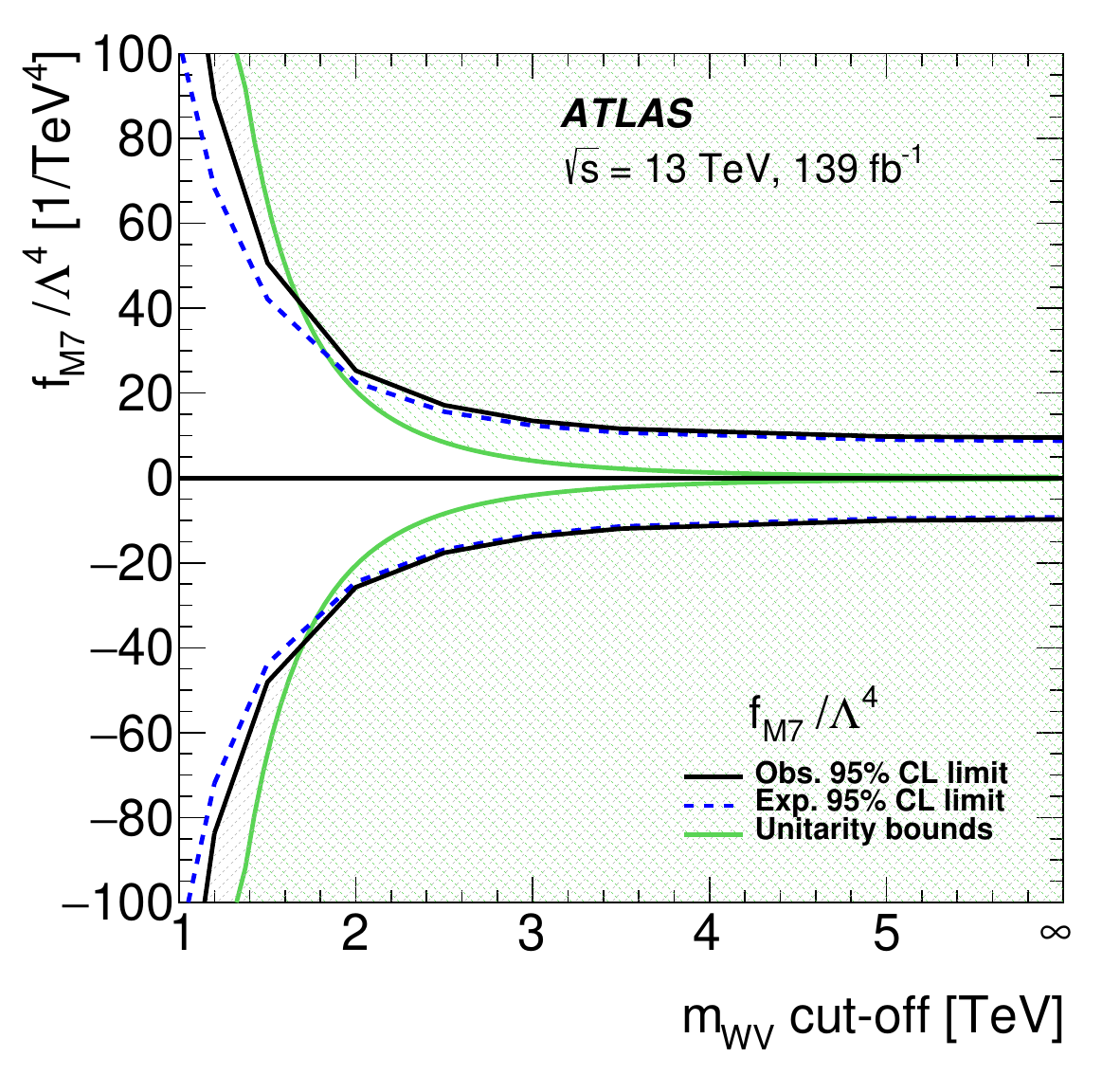}
}\\
\subfloat[]{
\includegraphics[width=0.32\textwidth]{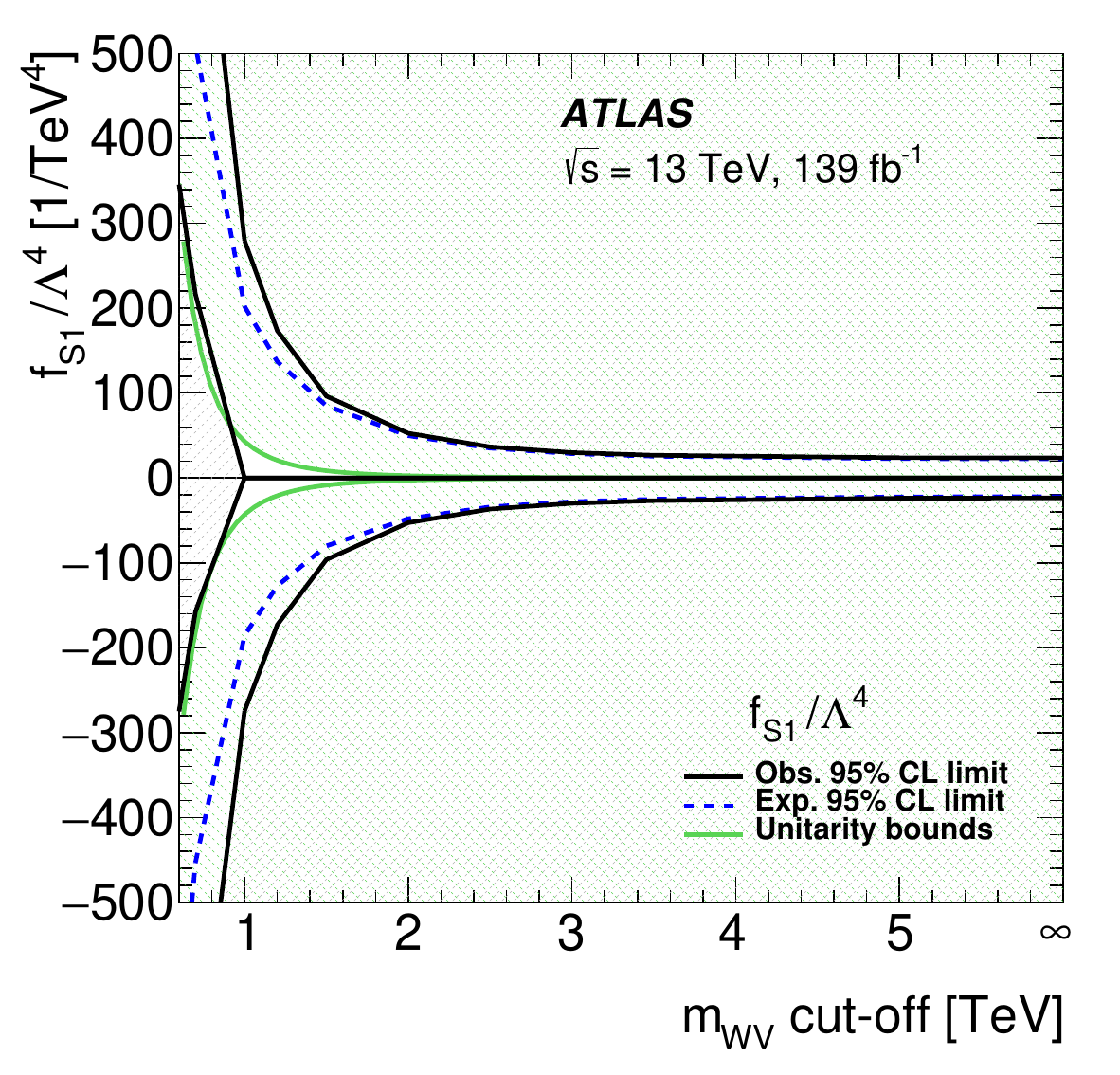}
}
\subfloat[]{
\includegraphics[width=0.32\textwidth]{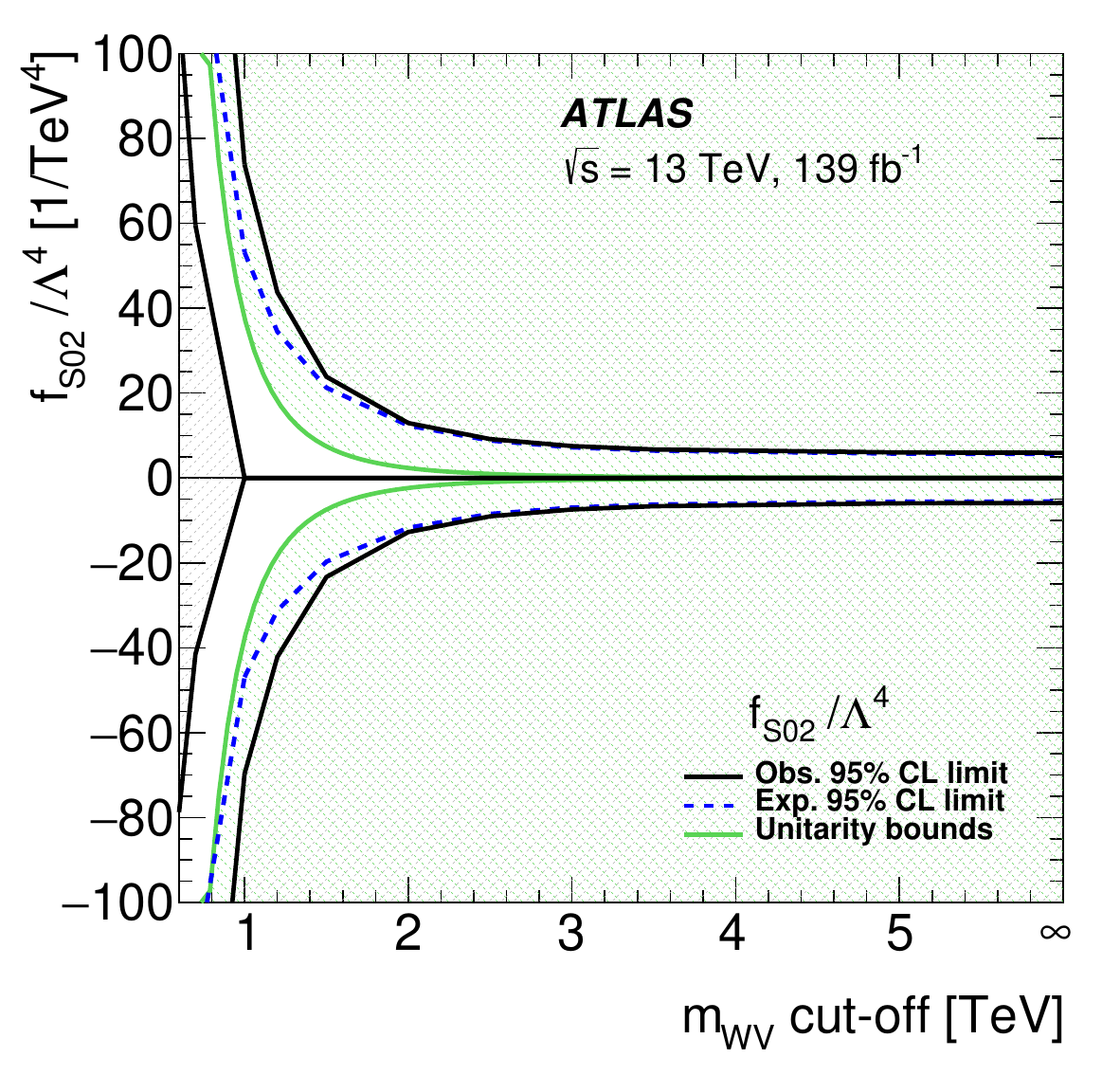}
}\\
\subfloat[]{
\includegraphics[width=0.32\textwidth]{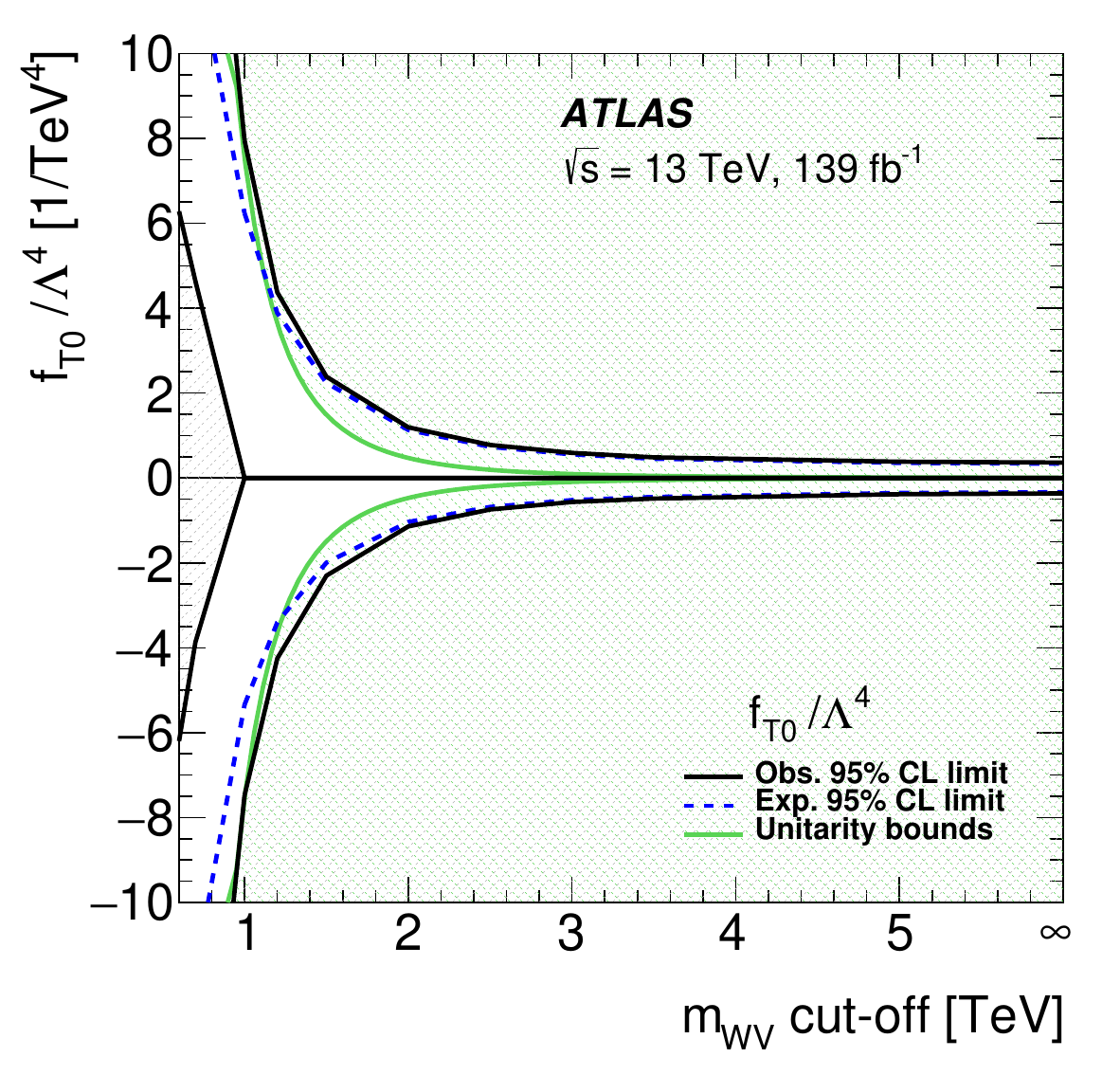}
}
\subfloat[]{
\includegraphics[width=0.32\textwidth]{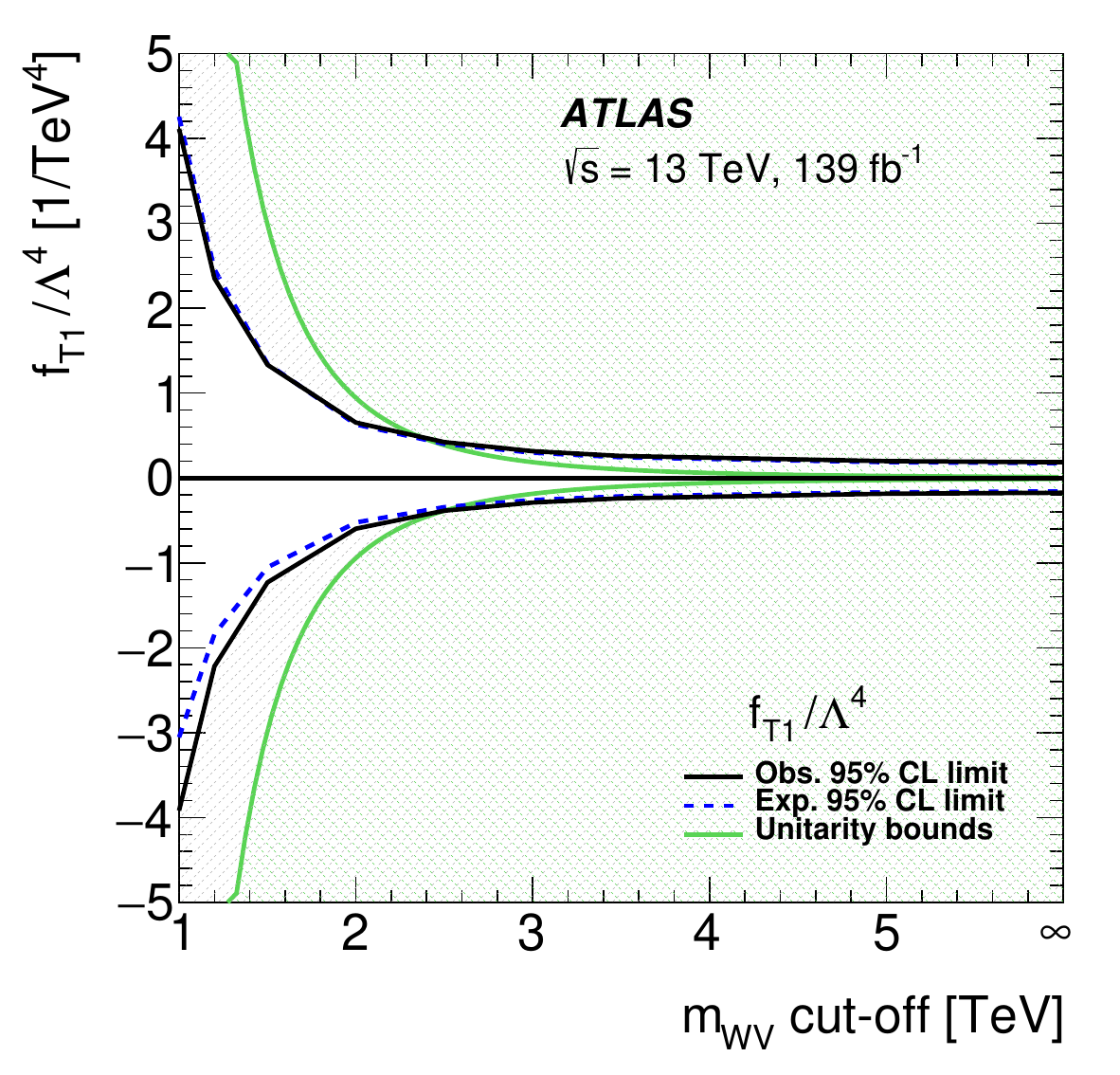}
}
\subfloat[]{
\includegraphics[width=0.32\textwidth]{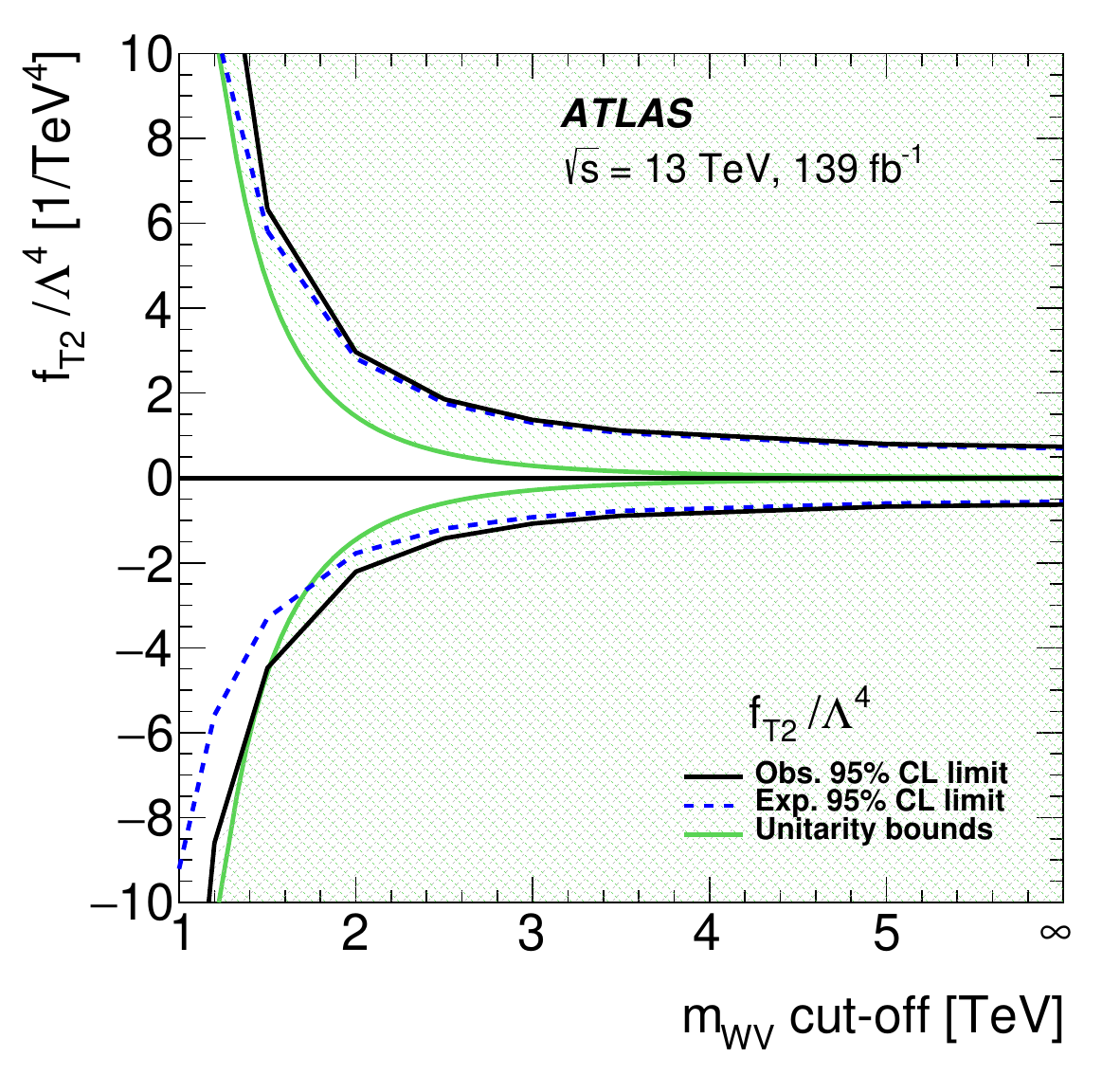}
}
\end{center}
\caption{Evolution of the one-dimensional expected (blue dashed line) and observed (black line) limits at 95\%~CL on the parameters corresponding to the quartic operators with label (a) M0, (b) M1, (c) M7, (d) S1, (e) S02, (f) T0, (g) T1, and (h) T2  as a function of the cut-off scale. The unitarity bounds (green line) for each operator as a function of the cut-off scale are defined for one non-zero Wilson coefficient following Ref.~\cite{Almeida:2020ylr}. The filled area corresponds to parameter values excluded either by the data at 95\%~CL or by the unitarity condition of the theory. The limits on M7 are obtained without taking into account the SM-EFT interference for the EW \wzjj final state.}
\label{fig:EFT_limits_clipping}
\end{figure}

Table~\ref{tab:EFT_limits} shows the expected and observed limits at 95\% CL  obtained without using any unitarisation procedure and with a unitarisation cut-off at the unitarity bound. The unitarity bounds for each operator as a function of the cut-off scale are defined for one non-zero Wilson coefficient following Ref.~\cite{Almeida:2020ylr}. Figure~\ref{fig:EFT_limits_clipping} presents the evolution of these limits as a function of the cut-off scale of the unitarisation procedure.
For clipping scales below approximately 1~\TeV, zero values of the coefficients $f_{M0}$, $f_{S1}$, $f_{S02}$, and $f_{T0}$ are excluded at 95\% CL as seen from Figure~\ref{fig:EFT_limits_clipping}. This is explained by the shape of the clipped EFT prediction, which exhibits a peak at \mll values around 400~\GeV, where an upward fluctuation is also present in data, as seen in Figure~\ref{fig:Mll_M0}, and leads to a double-minimum shape for the test statistic. The limits obtained are competitive with those from the CMS measurement of Ref.~\cite{CMS-SMP-19-012}, which is also based on the full Run~2 data set.
 
\begin{figure}[!htb]
\begin{center}
\subfloat[]{
\includegraphics[width=0.32\textwidth]{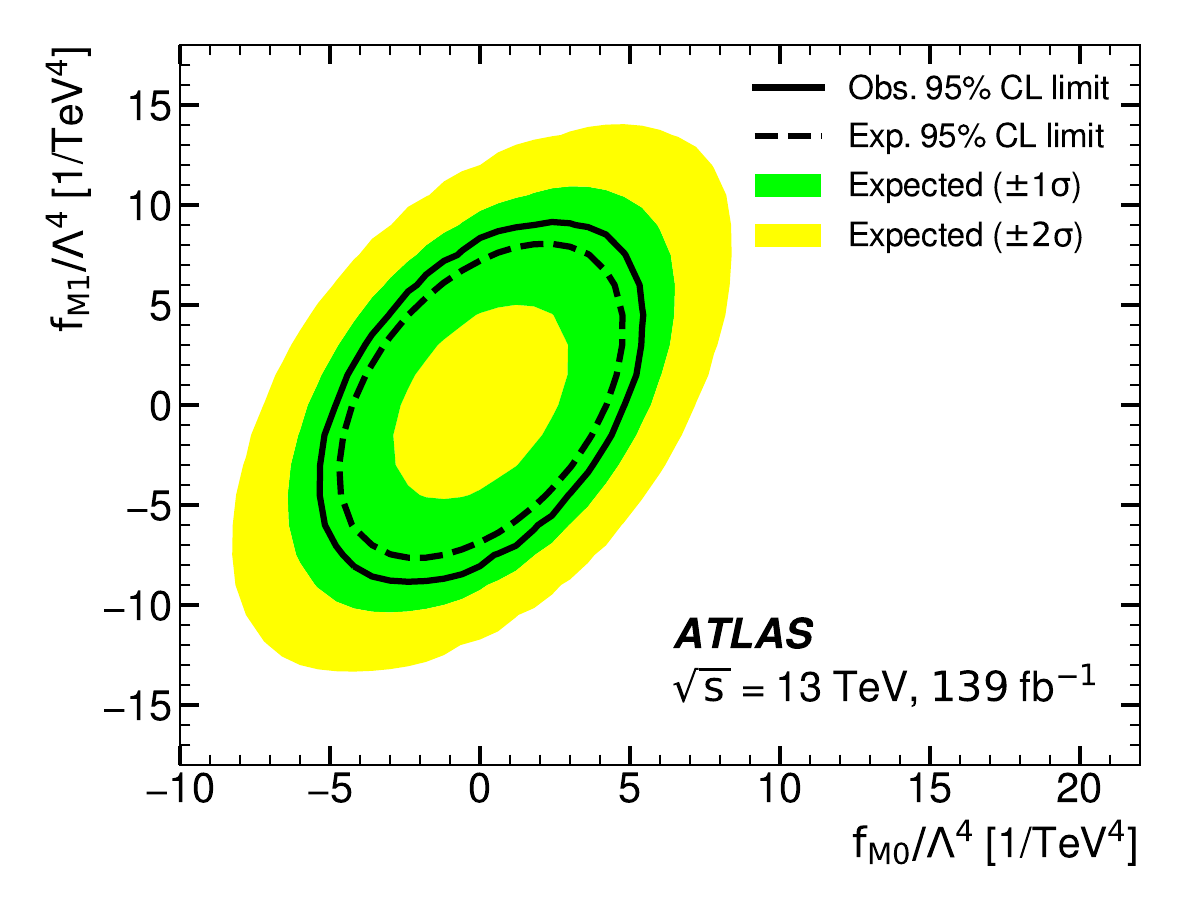}
}
\subfloat[]{
\includegraphics[width=0.32\textwidth]{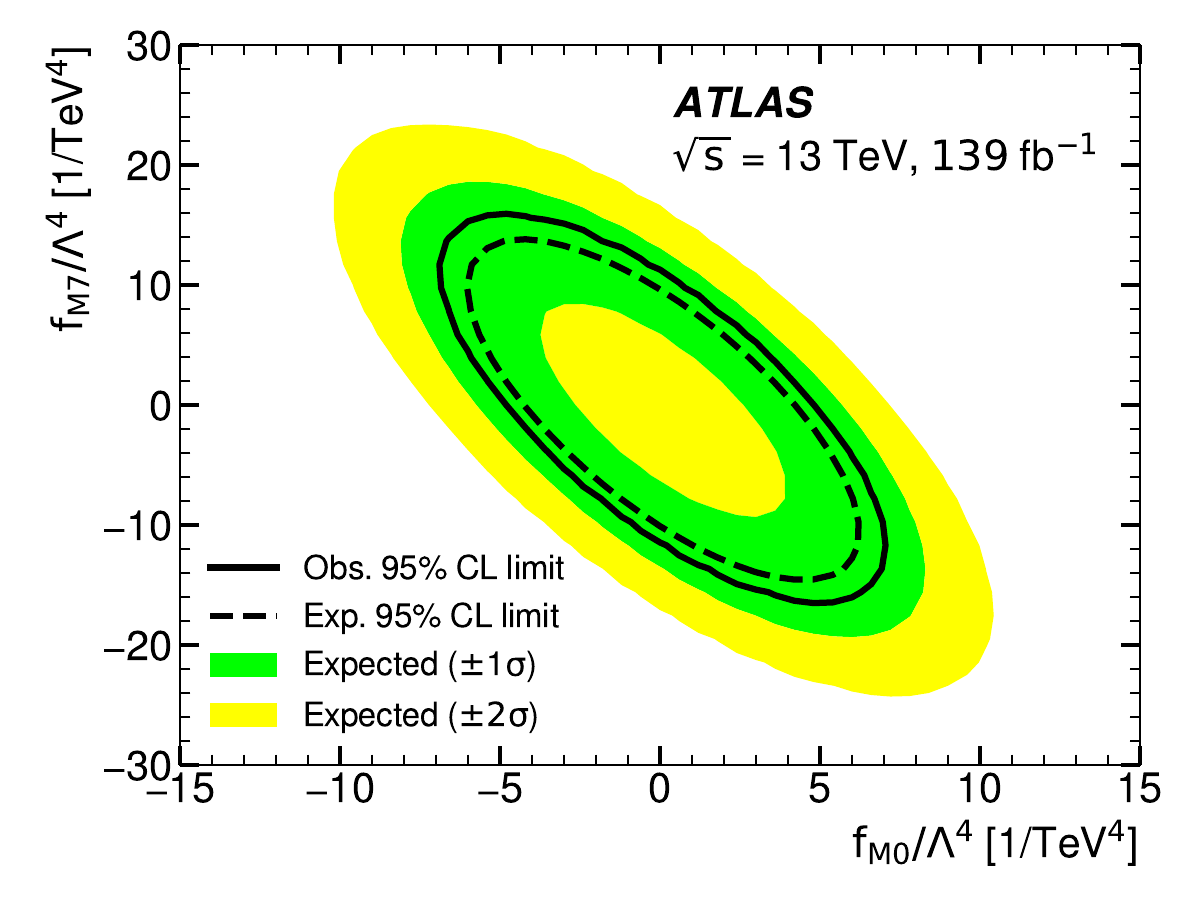}
}
\subfloat[]{
\includegraphics[width=0.32\textwidth]{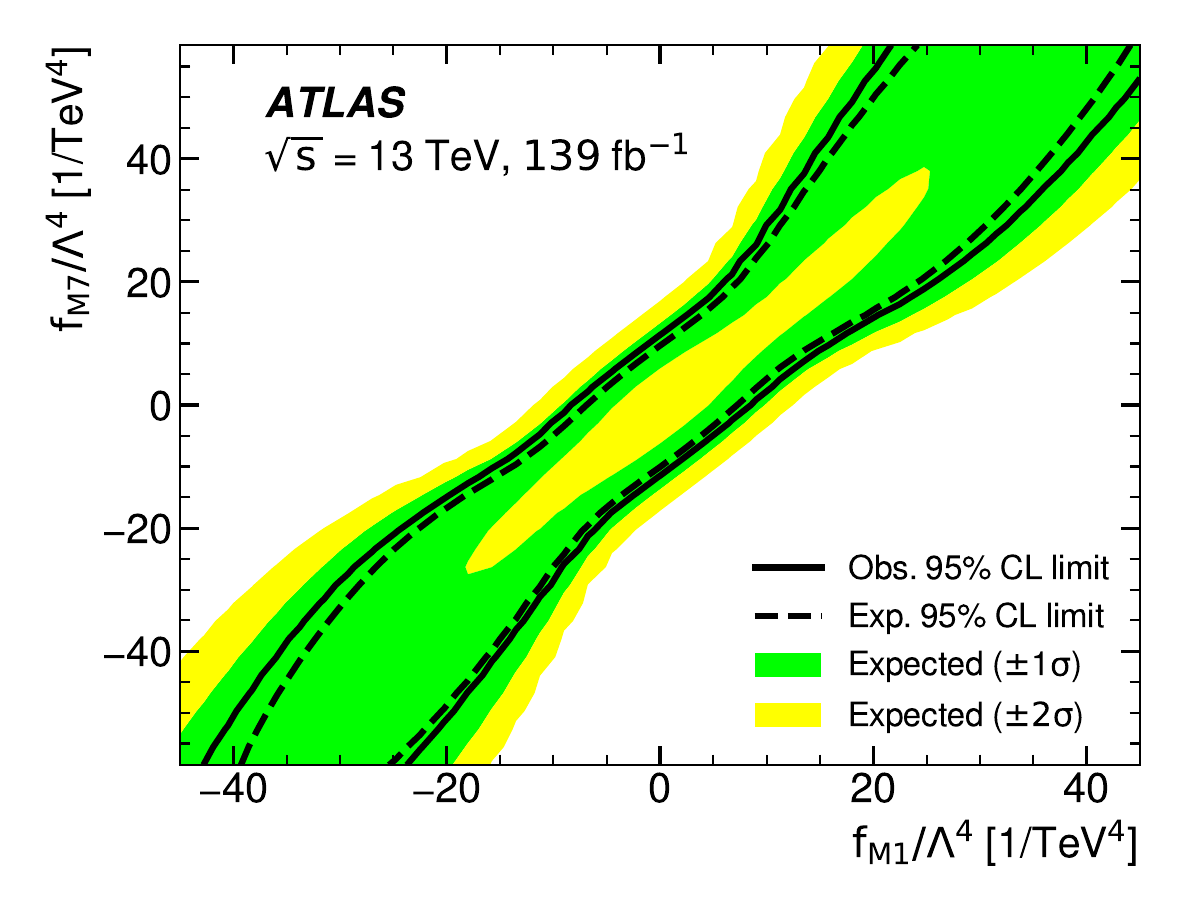}
}\\
\subfloat[]{
\includegraphics[width=0.32\textwidth]{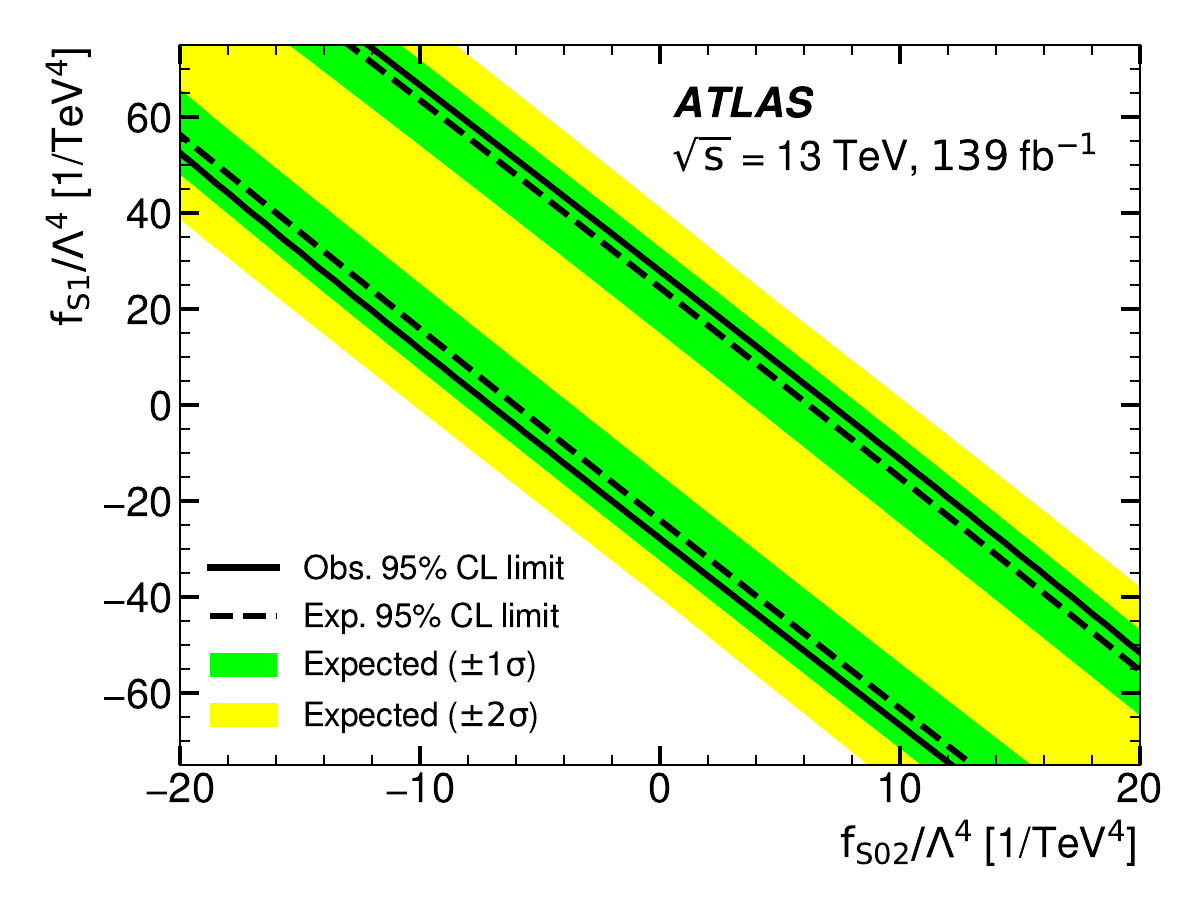}
}\\
\subfloat[]{
\includegraphics[width=0.32\textwidth]{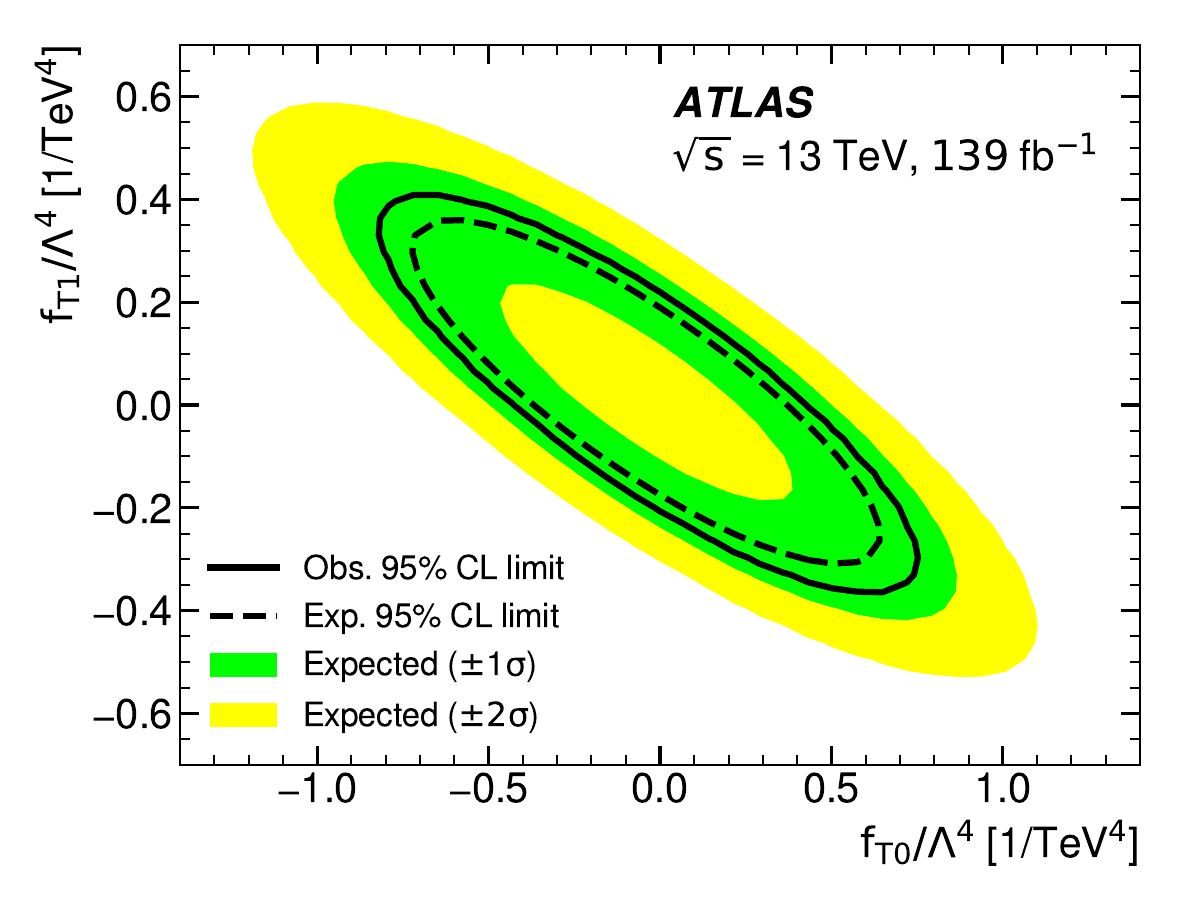}
}
\subfloat[]{
\includegraphics[width=0.32\textwidth]{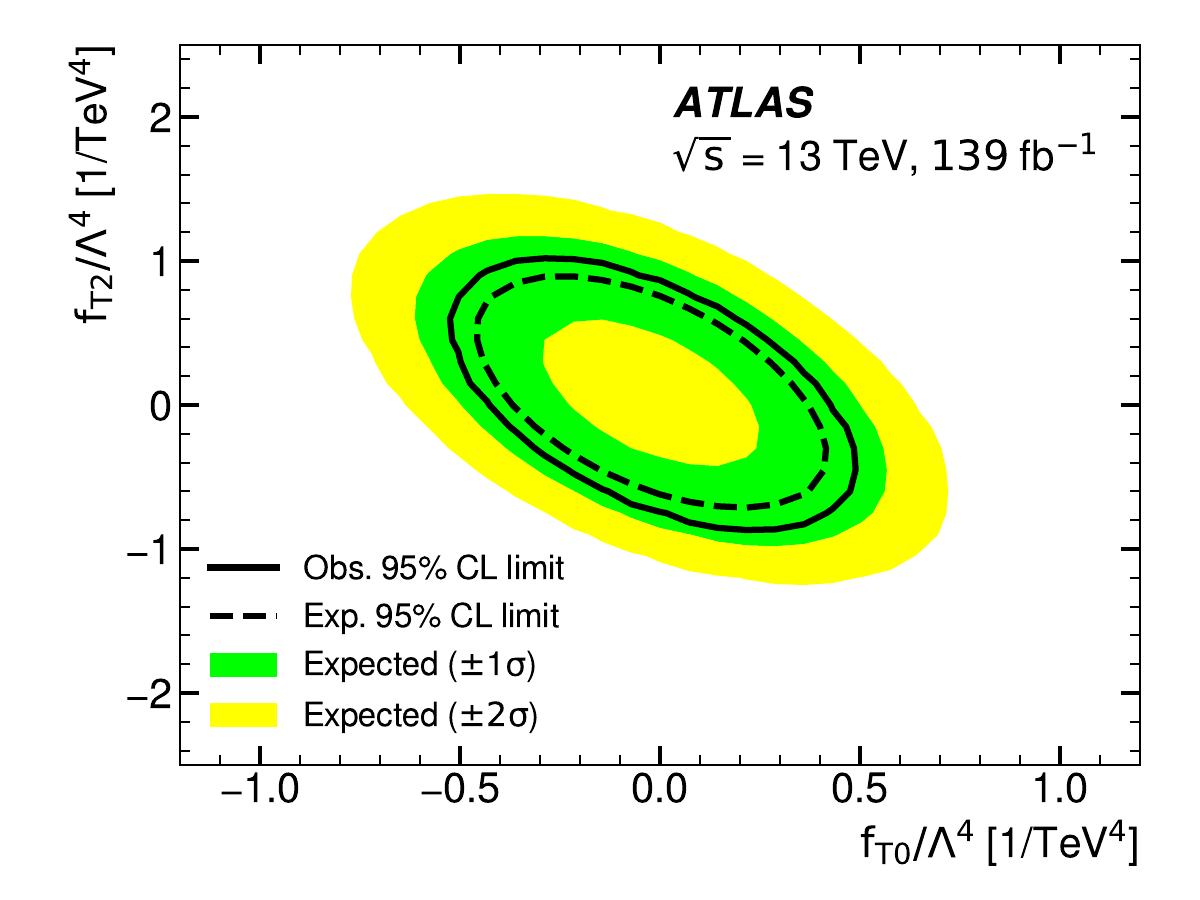}
}
\subfloat[]{
\includegraphics[width=0.32\textwidth]{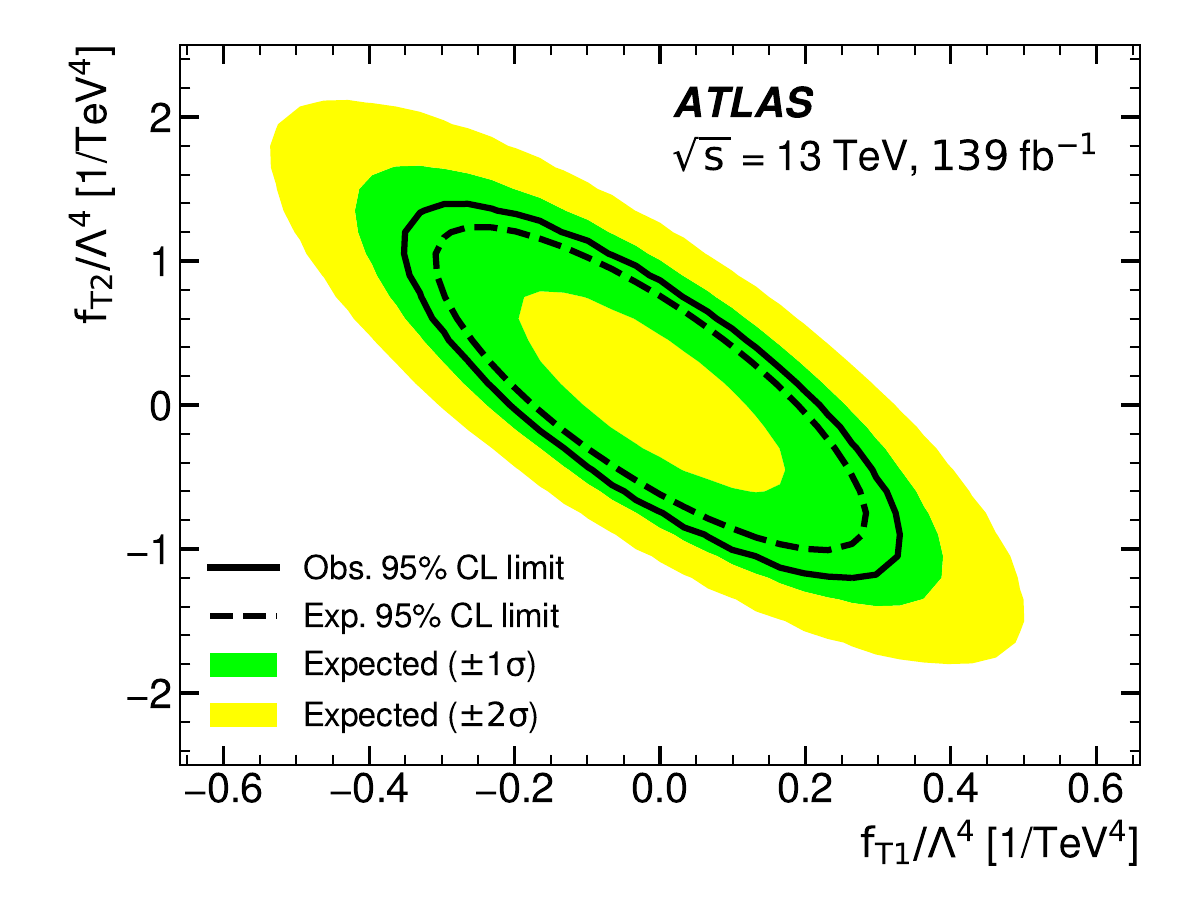}
}
\end{center}
\caption{Two-dimensional median expected (dashed line) and observed (solid line) 95\% CL intervals on parameters corresponding to the quartic operator combinations (a) M0-M1, (b) M0-M7, (c) M1-M7, (d) S1-S02, (e) T0-T1, (f) T0-T2 and (g) T1-T2 EFT parameters without any unitarisation procedure. The 1 (green) and 2 (yellow) sigma bands show the 68.3\% and 95.4\%\ CL regions for the expected limit curves, respectively. The limits on M7 are obtained without taking into account the SM-EFT interference term and EFT cross-term for the EW \wzjj final state.}
\label{fig:EFT_2Dlimits_noClipping}
\end{figure}

\begin{figure}[!htb]
\begin{center}
\subfloat[]{
\includegraphics[width=0.32\textwidth]{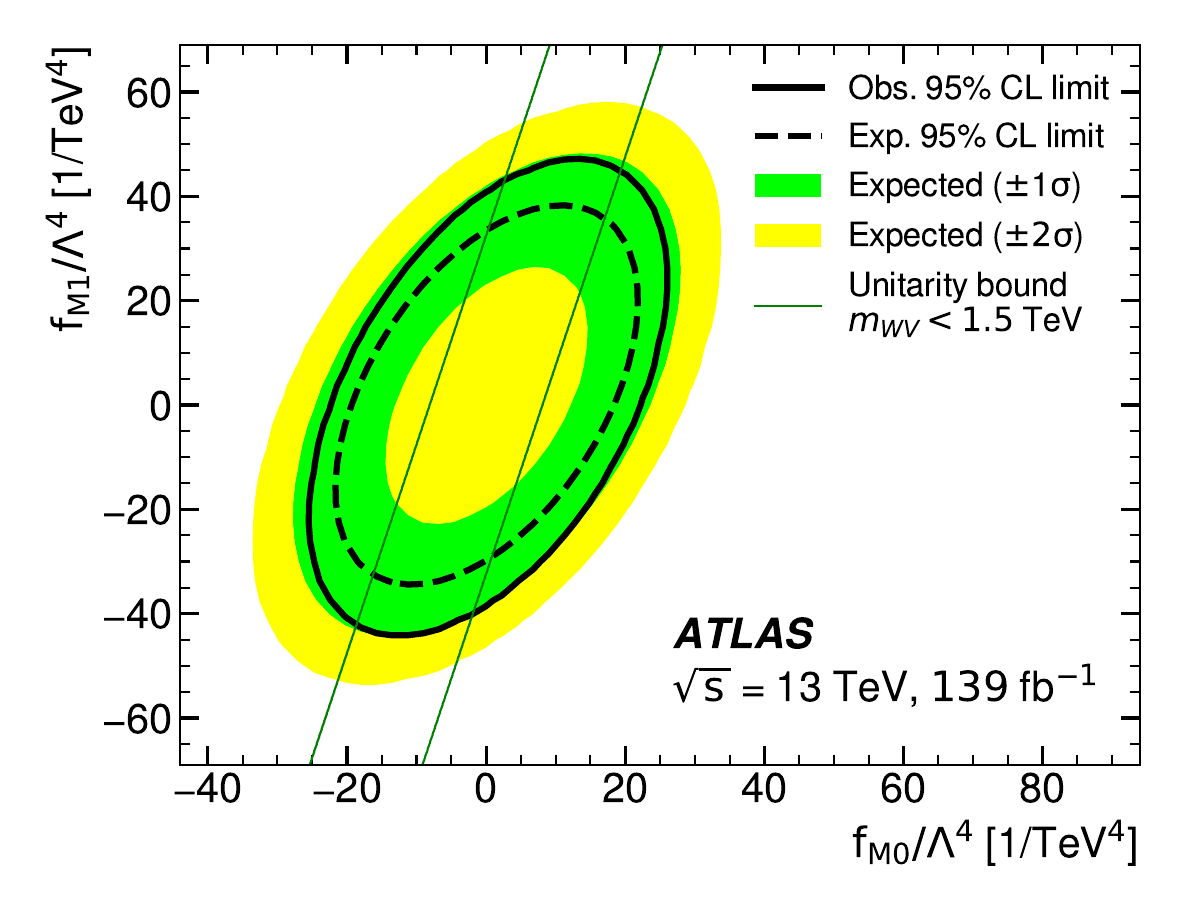}
}
\subfloat[]{
\includegraphics[width=0.32\textwidth]{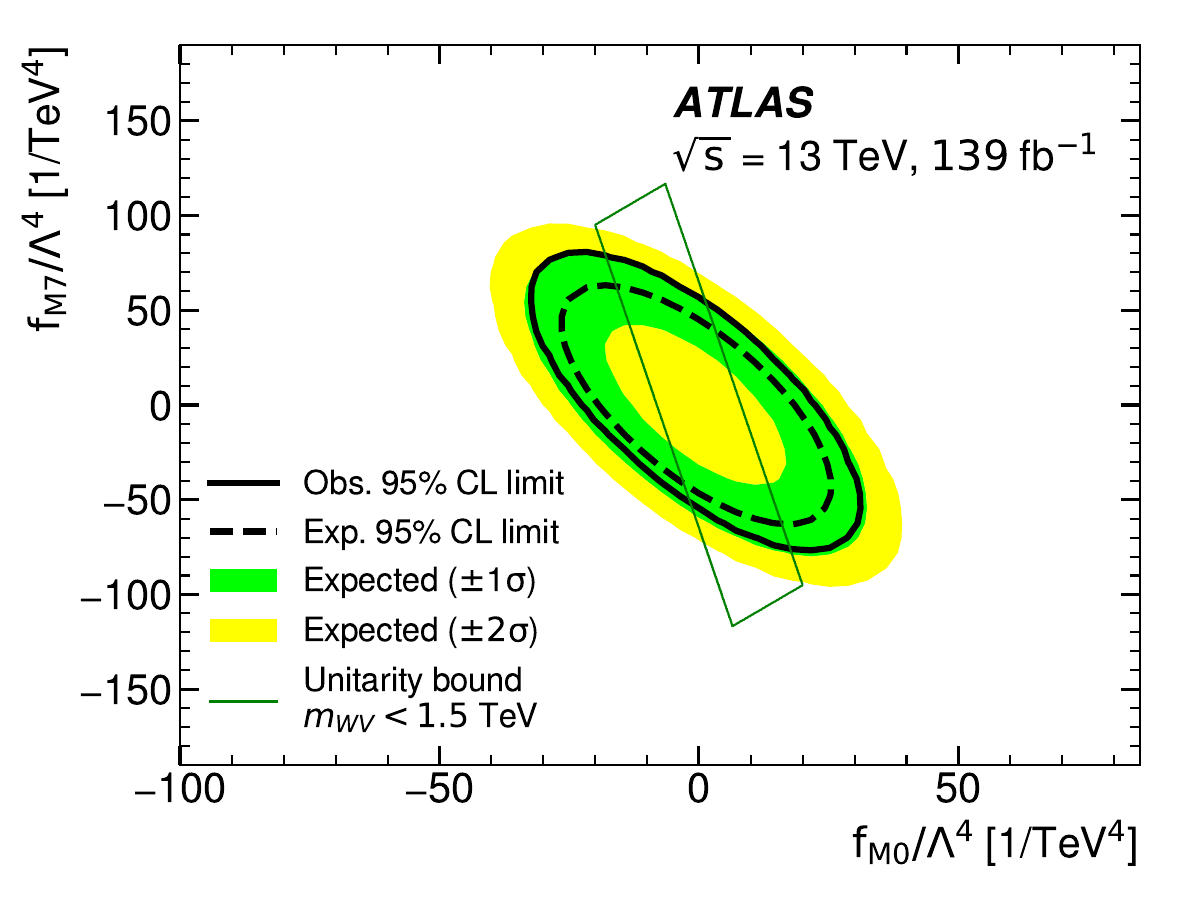}
}
\subfloat[]{
\includegraphics[width=0.32\textwidth]{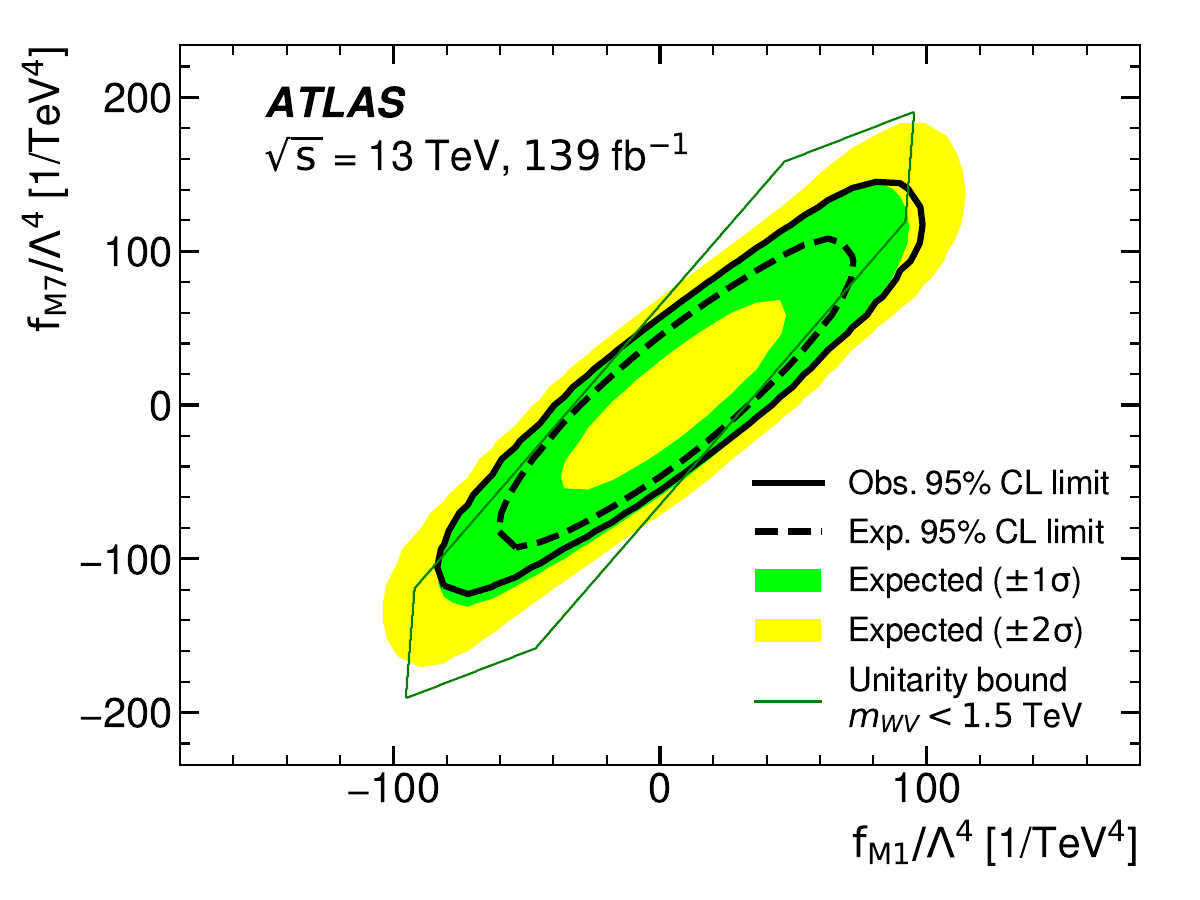}
}\\
\subfloat[]{
\includegraphics[width=0.32\textwidth]{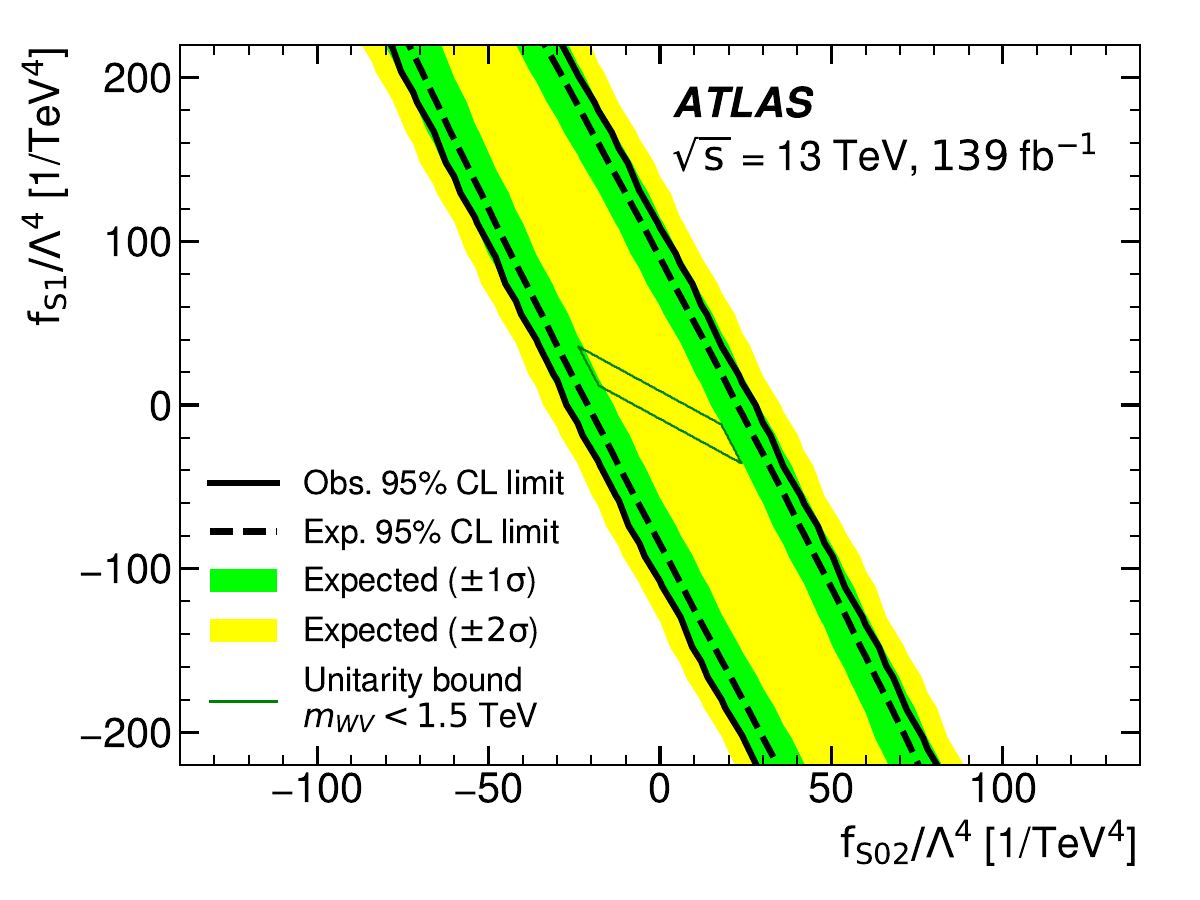}
}\\
\subfloat[]{
\includegraphics[width=0.32\textwidth]{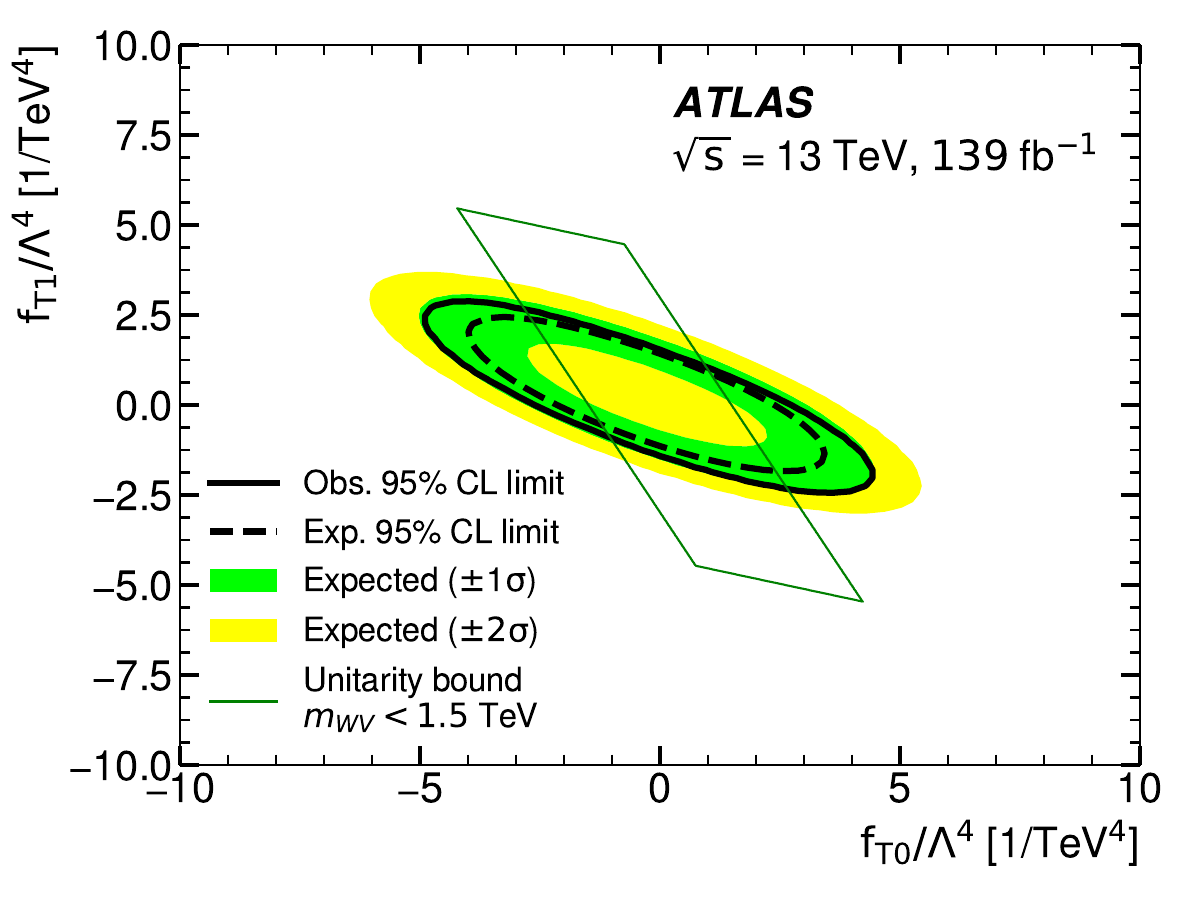}
}
\subfloat[]{
\includegraphics[width=0.32\textwidth]{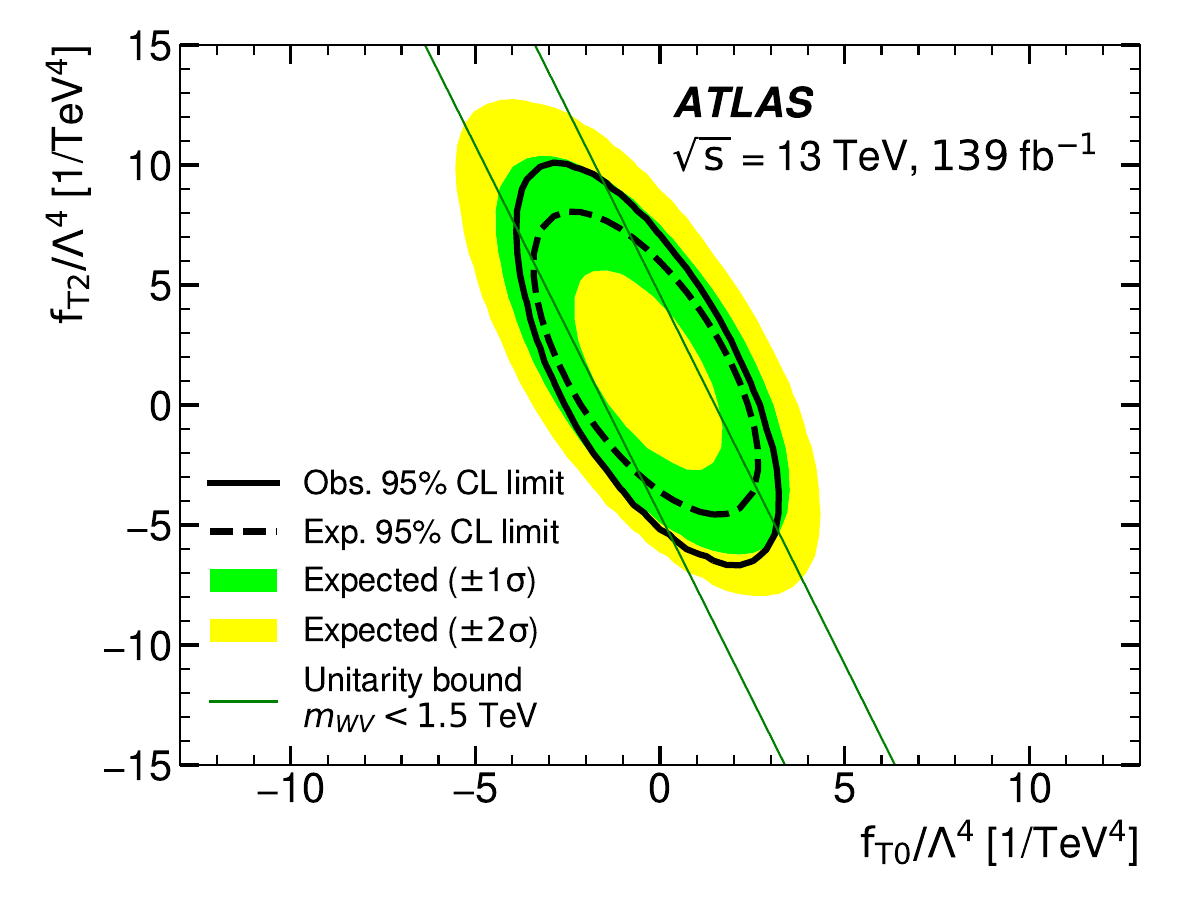}
}
\subfloat[]{
\includegraphics[width=0.32\textwidth]{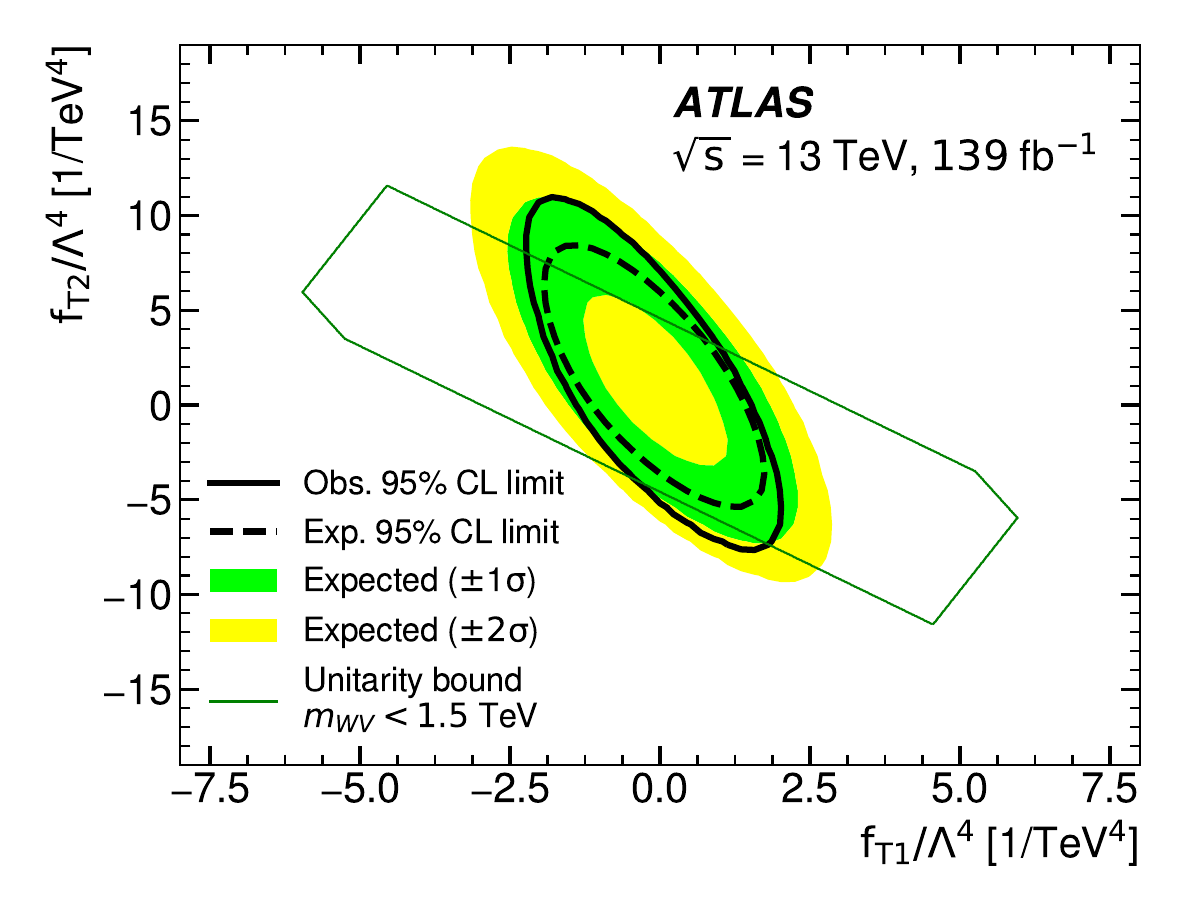}
}
\end{center}
\caption{Two-dimensional median expected (dashed line) and observed (solid line) 95\% CL intervals on parameters corresponding to the quartic operator combinations (a) M0-M1, (b) M0-M7, (c) M1-M7, (d) S1-S02, (e) T0-T1, (f) T0-T2 and (g) T1-T2 EFT parameters with a unitarisation cut-off scale of $1.5$~TeV and unitarity bounds (green line). The two-dimensional unitarity bounds for pairs of operators are obtained for the two non-zero Wilson coefficients from the eigenvalues from Ref.~\cite{Almeida:2020ylr}.
The 1 (green) and 2 (yellow) sigma bands show the 68.3\% and 95.4\%\ CL regions for the expected limit curves, respectively.
The limits on M7 are obtained without taking into account the SM-EFT interference term and EFT cross-term for the EW \wzjj final state.}
\label{fig:EFT_2Dlimits_Clipping1p5TeV_withUni}
\end{figure}

The expected and observed two-dimensional limits at 95\% CL on operator pairs belonging to the same group are presented in Figure~\ref{fig:EFT_2Dlimits_noClipping} where no unitarisation cut-off has been imposed.
Figure~\ref{fig:EFT_2Dlimits_Clipping1p5TeV_withUni} shows two-dimensional expected and observed limits at 95\% CL obtained with a unitarisation cut-off scale of $1.5$~TeV for all combinations of the operators studied. The predicted differential yields of nearly all operator pairs are positively correlated, except for the pairs with labels M0-M1 and M1-M7, which are anti-correlated. In the phase space studied, the predicted yields for $O_{S02}$ and $O_{S1}$ are nearly 100\% correlated, and thus a band cannot be excluded.
These two-dimensional limits improve upon the combined ATLAS interpretation of \sswwjj and \wzjj measurements based on 36~fb\(^{-1}\) data~\cite{ATL-PHYS-PUB-2023-002manual}, except for the M1-M7 combination, which is likely explained by the missing cross-term for the \wzjj final state in the present study.

\begin{table}[hbtp]
\caption{Expected and observed limits on the Wilson coefficients for various operators without any unitarisation procedure and with a unitarisation cut-off at the unitarity bound. The last column represents lower and upper limits at the respective cut-off value, where the unitarity bound and experimental bound cross. Cases where no crossing with the unitarity bound was found in the scanned region above $600$~GeV are labelled by "--". The notation S02 is used to indicate that the coefficients corresponding to the operators $O_{S0}$ and $O_{S2}$ are assigned the same value. The limits on M7 are obtained without taking into account the SM-EFT interference for the EW \wzjj final state.}
\label{tab:EFT_limits}
\begin{center}
\begin{tabular}{lccZc} 
\toprule 
Coefficient & Type & No unitarisation cut-off & At 1 TeV  & Lower, upper limit at the respective unitarity bound \\ 
& & [TeV$^{-4}$] & [TeV$^{-4}$] & [TeV$^{-4}$] \\
\midrule
\multirow{2}{*}{$f_{\mathrm{M}0}/\Lambda^4$} & Exp. &[-3.9, 3.8] & [-45, 40] & -64 at 0.9 TeV, 40 at 1.0 TeV\\
& Obs. &[-4.1, 4.1] & [-59, 57] & -140 at 0.7 TeV, 117 at 0.8 TeV\\
\multirow{2}{*}{$f_{\mathrm{M}1}/\Lambda^4$} & Exp. &[-6.3, 6.6] & [-66, 76] & -25.5 at 1.6 TeV, 31 at 1.5 TeV\\
& Obs. &[-6.8, 7.0] & [-98, 105] & -45 at 1.4 TeV, 54 at 1.3 TeV\\
\multirow{2}{*}{$f_{\mathrm{M}7}/\Lambda^4$} & Exp. &[-9.3, 8.8] & [-110, 103] & -33 at 1.8 TeV, 29.1 at 1.8 TeV\\
& Obs. &[-9.8, 9.5] & [-142, 146] & -39 at 1.7 TeV, 42 at 1.7 TeV\\
\multirow{2}{*}{$f_{\mathrm{S}02}/\Lambda^4$} & Exp. &[-5.5, 5.7] & [-47, 53] & -94 at 0.8 TeV, 122 at 0.7 TeV\\
& Obs. &[-5.9, 5.9] & [-70, 74] & --\\
\multirow{2}{*}{$f_{\mathrm{S}1}/\Lambda^4$} & Exp. &[-22.0, 22.5] & [-185, 202] & --\\
& Obs. &[-23.5, 23.6] & [-274, 280] & --\\
\multirow{2}{*}{$f_{\mathrm{T}0}/\Lambda^4$} & Exp. &[-0.34, 0.34] & [-5.3, 6.2] & -3.2 at 1.2 TeV, 4.9 at 1.1 TeV\\
& Obs. &[-0.36, 0.36] & [-7.5, 8.0] & -7.4 at 1.0 TeV, 12.4 at 0.9 TeV\\
\multirow{2}{*}{$f_{\mathrm{T}1}/\Lambda^4$} & Exp. &[-0.158, 0.174] & [-3.1, 4.3] & -0.32 at 2.6 TeV, 0.44 at 2.4 TeV\\
& Obs. &[-0.174, 0.186] & [-3.9, 4.1] & -0.38 at 2.5 TeV, 0.49 at 2.4 TeV\\
\multirow{2}{*}{$f_{\mathrm{T}2}/\Lambda^4$} & Exp. &[-0.56, 0.70] & [-9.2, 18.5] & -2.60 at 1.7 TeV, 10.3 at 1.2 TeV\\
& Obs. &[-0.63, 0.74] & [-17.3, 27.9] & --\\
\bottomrule 
\end{tabular} 
\end{center}
\end{table}

\FloatBarrier
\section{Limits on doubly-charged Higgs boson production}
\label{sec:GM}

 
In extended Higgs sectors with additional isotriplet scalar fields~\cite{Englert:2013wga}, charged Higgs bosons with couplings to $W$ and $Z$ bosons are present at tree level. The results from this measurement are interpreted to search for a doubly charged Higgs boson produced in VBF processes that decays into two same-sign $W$ bosons. Specifically, the GM model~\cite{Georgi:1985nv} with both real and complex triplets is considered. The scalar potential consists of a custodial quintuplet of fermiophobic Higgs bosons ($\mathrm{H}_5^{\pm\pm}$, $\mathrm{H}_5^{\pm}$, and $\mathrm{H}_5^0$) of common mass $m_{\mathrm{H}^{\pm\pm}_{5}}$ that couple to the $W$ and $Z$ bosons, and a custodial triplet of Higgs bosons ($\mathrm{H}_3^{\pm}$ and $\mathrm{H}_3^0$) of common mass $m_{\mathrm{H}_{3}}$ with only fermionic couplings. In addition, there are two CP-even scalars, $\mathrm{h}$ and $\mathrm{H}$ with masses $m_\mathrm{h} \approx 125$~GeV (identified as the SM-like Higgs boson) and $m_\mathrm{H}$ respectively, that are custodial singlets.
 
The VBF production and decays of the $\mathrm{H}^{\pm\pm}_5$ states depend on two parameters, $m_{\mathrm{H}^{\pm\pm}_{5}}$ and $\sin\theta_{\mathrm{H}}$, where $\sin\theta_{\mathrm{H}}$ characterises the contribution of the isotriplet scalar fields to the masses of the $W$ and $Z$ bosons. The \texttt{H5plane} benchmark is considered~\cite{deFlorian:2016spz}, where the triplet states are assumed to be heavier than the quintuplet states. Thus, in this benchmark the $\mathrm{H}^{\pm\pm}_5$ bosons decay to same-sign $W$ boson pairs with a branching fraction of 100\%~\cite{LHCHXSWG-2015-001}.
 
The signal extraction follows an approach similar to the one presented in Sections~\ref{sec:cross_section} and~\ref{sec:EFT} to extract limits from the samples corresponding to 23 mass points between $200$~GeV and $3$~TeV (see Section~\ref{sec:H5pp_samples}) using the transverse mass distribution $\mt$, which provides good discrimination between resonant signal and non-resonant background processes. The $\mjj$ binning in the signal region is identical to the binning used in the SM fits while the $\mt$ binning is modified to include an additional bin for the highest $\mt$ range.
The normalisations of the EW \sswwjj and QCD \wzjj background processes are freely floated in the fit. The \sswwjj EW-QCD interference contribution is normalised to the square root of the EW \sswwjj signal strength in the fit, similarly to the fits described in Section~\ref{subsec:fid_cross_sec}. The $\mathrm{H}^{\pm\pm}_5$ signal contribution is proportional to $\sin^2\theta_{\mathrm{H}}$. In addition to the sources of systematic uncertainty described in Section~\ref{sec:systematic_uncertainties}, theory uncertainties that affect the $\mathrm{H}^{\pm\pm}_5$ signal, namely variations of \(\mu_{\mathrm{R}}\) and \(\mu_{\mathrm{F}}\), $\as$, and PDFs, are considered. In addition, an uncertainty due to the missing NLO EW corrections is adopted, as recommended in Ref.~\cite{LHCHXSWG-2015-001}.
 
\begin{figure}[h]
\begin{center}
\subfloat[]{
\includegraphics[width=0.49\textwidth]{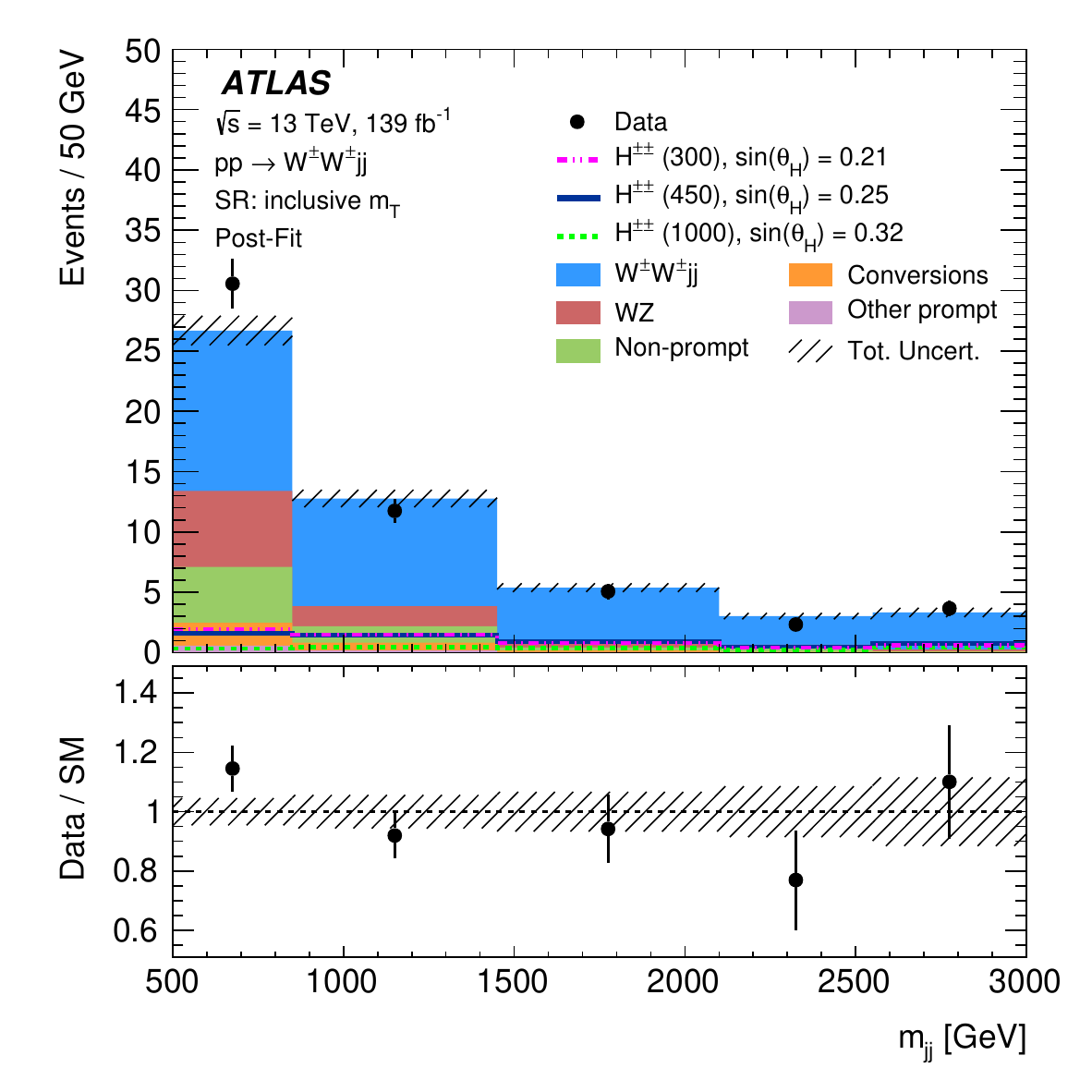}
\label{fig:hpp_mjj}
}
\subfloat[]{
\includegraphics[width=0.49\textwidth]{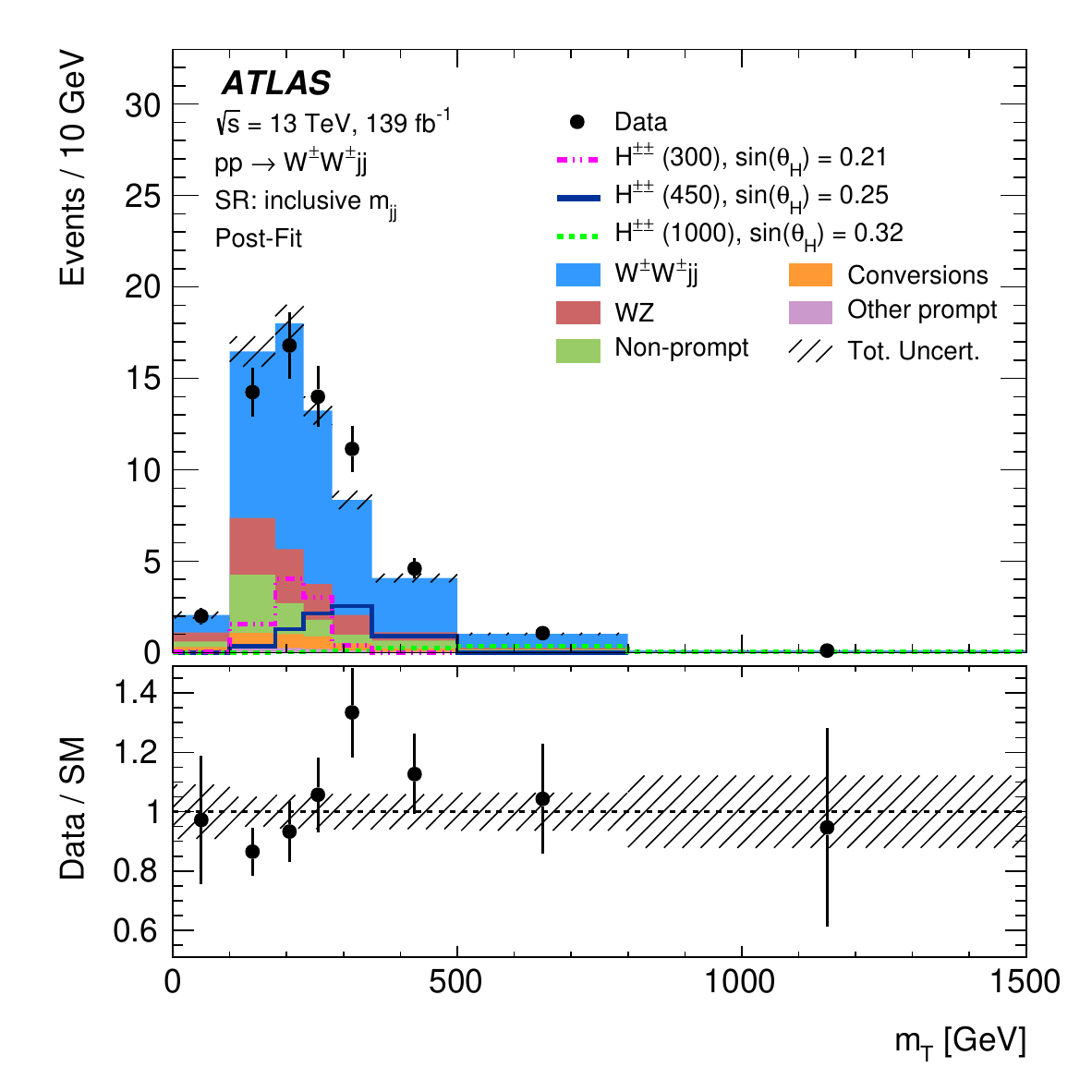}
\label{fig:hpp_mt}
}
\end{center}
\caption{Post-fit \protect\subref{fig:hpp_mjj} $\mjj$ and \protect\subref{fig:hpp_mt} $\mt$ inclusive distributions in the signal region for the SM background-only hypothesis. Data are shown as black markers with vertical error bars representing the statistical uncertainty. Filled histograms show contributions of various SM processes, with the hatched band representing the total uncertainty of the model. The lines show the prediction of the $\mathrm{H}^{\pm\pm}_5$ model for values of $\sin\theta_{\mathrm{H}} = 0.21$ and $m_{\mathrm{H}^{\pm\pm}_{5}} = 300$~GeV, $\sin\theta_{\mathrm{H}} = 0.25$ and $m_{\mathrm{H}^{\pm\pm}_{5}} = 450$~GeV, and $\sin\theta_{\mathrm{H}} = 0.32$ and $m_{\mathrm{H}^{\pm\pm}_{5}} = 1000$~GeV, where the $\sin\theta_{\mathrm{H}}$ values correspond to the expected 95\% CL limits for the respective mass points. The last bin of each distribution includes overflow events. The lower panel shows the ratio of the data to the SM prediction.}
\label{fig:hpp_signal_inclusive}
\end{figure}

The post-fit inclusive distributions in the signal region for the SM background-only hypothesis are shown in Figure~\ref{fig:hpp_signal_inclusive} as a function of $\mjj$ and $\mt$. The contributions from the $\mathrm{H}^{\pm\pm}_5$ signal simulation for the mass points at $300$, $450$, and $1000$~GeV are shown for illustration purposes.

\begin{figure}[htbp]
\centering
\subfloat[]{
\includegraphics[width=.45\textwidth]{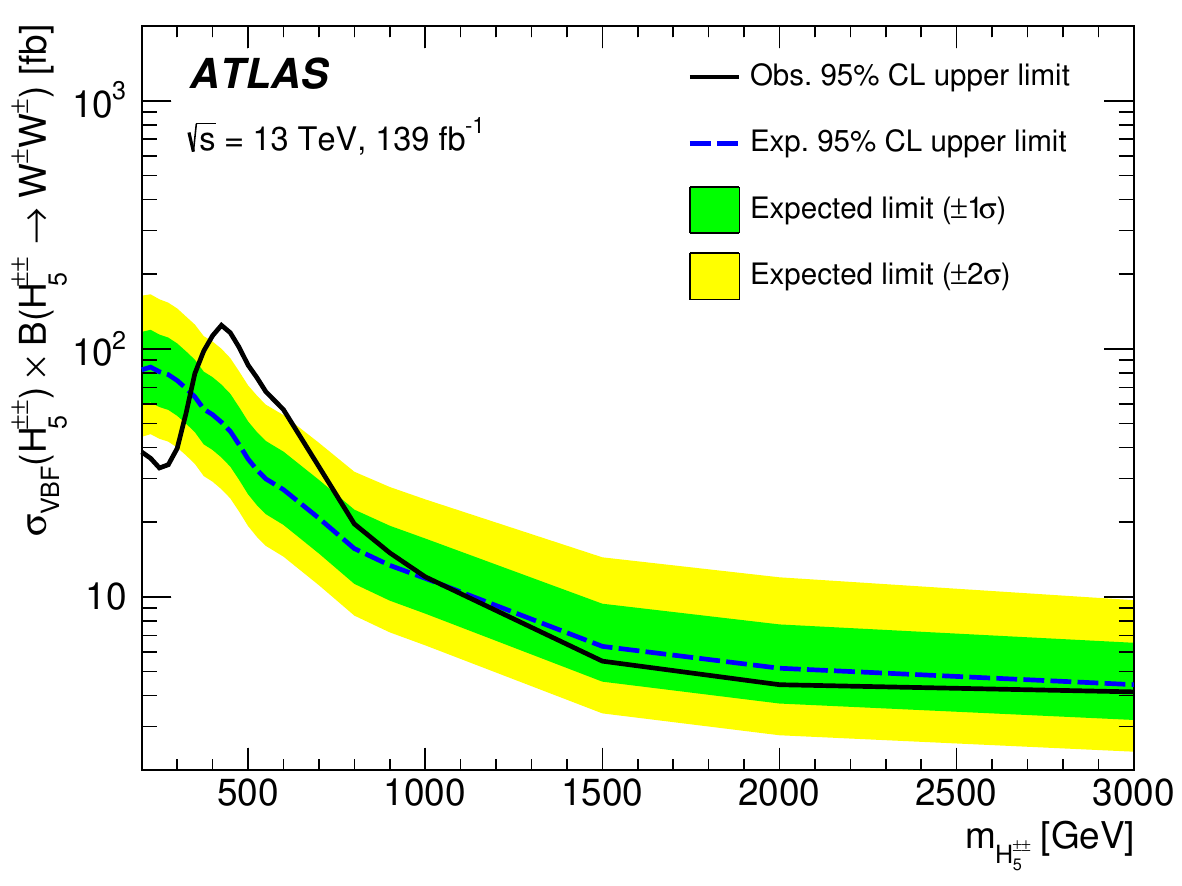}
\label{fig:hpp_limits_cl95}
}
\subfloat[]{
\includegraphics[width=0.45\textwidth]{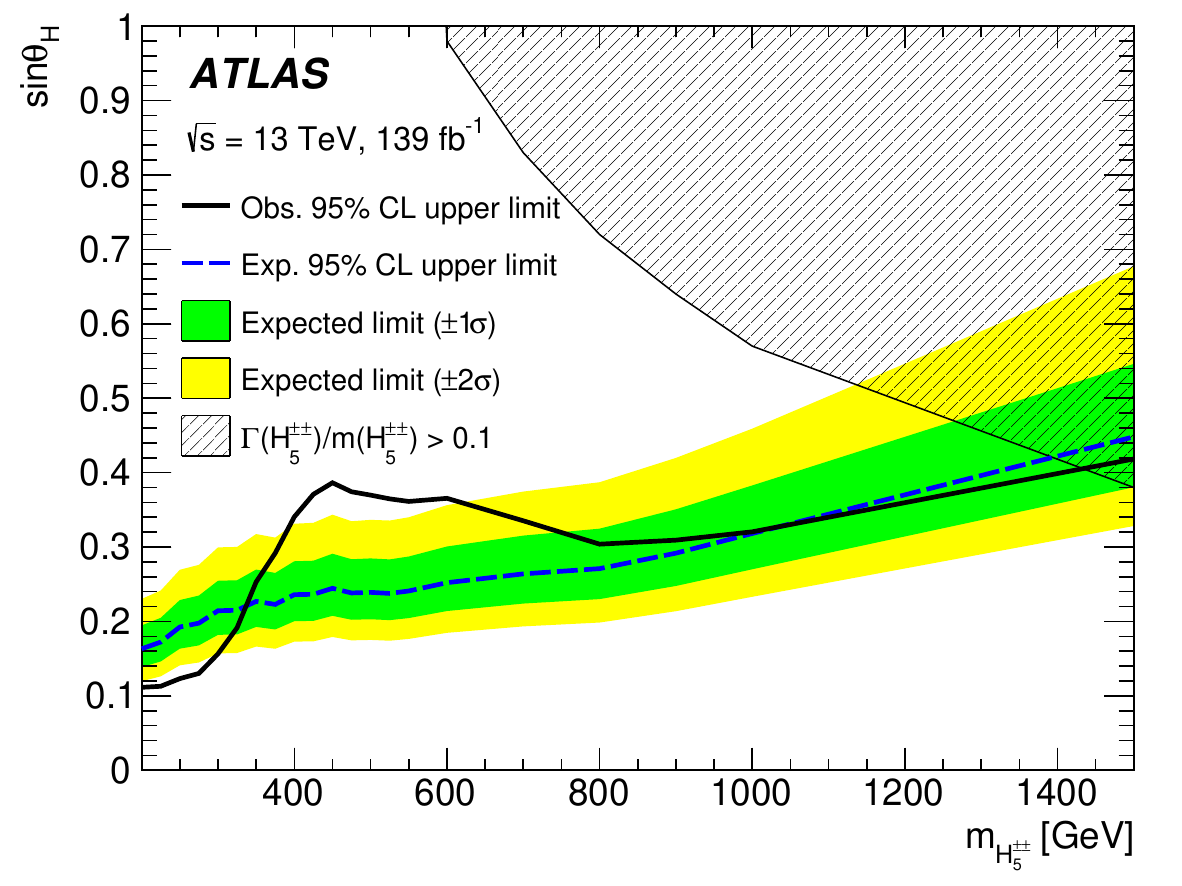}
\label{fig:hpp_limits_sin}
}
\caption{Expected and observed exclusion limits at 95\% CL for \protect\subref{fig:hpp_limits_cl95} $\sigma_{\mathrm{VBF}}(\mathrm{H}^{\pm\pm}_5) \times \mathcal{B}(\mathrm{H}^{\pm\pm}_5 \rightarrow W^{\pm}W^{\pm})$ as a function of $m_{\mathrm{H}^{\pm\pm}_{5}}$, and for \protect\subref{fig:hpp_limits_sin} $\sin\theta_{\mathrm{H}}$ as a function of $m_{\mathrm{H}^{\pm\pm}_{5}}$ in the GM model. The green (yellow) band is the 68\% (95\%) confidence interval
around the median expected limit. The exclusion limits for $\sin\theta_{\mathrm{H}}$ are shown up to $m_{\mathrm{H}^{\pm\pm}_{5}} = 1500$~GeV given the low sensitivity in the GM model above that mass. The hatched region covers the parameter space where the intrinsic width of the H++ boson would be larger than 10\% of the mass and is disfavoured in the GM model~\cite{LHCHXSWG-2015-001}. }
\label{fig:hpp_limits}
\end{figure}
 
Upper limits at 95\% CL on the product of the cross section and branching fraction $\sigma_{\mathrm{VBF}}(\mathrm{H}^{\pm\pm}_5) \times \mathcal{B}(\mathrm{H}^{\pm\pm}_5 \rightarrow W^{\pm}W^{\pm})$ for vector boson fusion production of doubly charged Higgs bosons as a function of mass from 200 to 3000~GeV are presented in Figure~\ref{fig:hpp_limits}. The corresponding limits on the $\sin\theta_{\mathrm{H}}$ parameter of the GM model as a function of $m_{\mathrm{H}^{\pm\pm}_{5}}$ are shown in Figure~\ref{fig:hpp_limits}. The black hatched region represents the parameter space for which the total width of the $\mathrm{H}^{\pm\pm}_5$ exceeds 10\% of $m_{\mathrm{H}^{\pm\pm}_{5}}$, where the model is not applicable due to perturbativity and vacuum stability requirements~\cite{LHCHXSWG-2015-001}. The observed 95\% CL limits exclude $\sin\theta_{\mathrm{H}}$ parameter values greater than 0.11--0.42 for the $m_{\mathrm{H}^{\pm\pm}_{5}}$ between 200 and 1500~GeV. These results show a local excess of events over the SM prediction at a resonance mass of around 450~GeV in Figure~\ref{fig:hpp_limits_cl95}, consistent with the results obtained in the SM part of the analysis. This excess is also observed in the $\mt$ inclusive distribution shown in Figure~\ref{fig:hpp_mt}. The significance of the excess has been evaluated for the different mass points in terms of the local \(p\)-value. The excess is largest for the $450$~GeV mass point, for which a \(p\)-value of $5.5\times10^{-4}$ is obtained, corresponding to a local significance of 3.3 standard deviations. The fit of SM background
plus $\mathrm{H}^{\pm\pm}_5$ signal to the data for this mass point yields the value of the normalisation factor for the EW \sswwjj process $\mu^{\mathrm{EW}\,\sswwjj} = 0.79 \pm 0.15$. This value is much smaller than the one reported in Section~\ref{cross_sec_results}, since a better description of the \mt distribution in data is achieved here by compensating with a $\mathrm{H}^{\pm\pm}_5$ contribution.
The excess for the $450$~GeV mass point would correspond to $\sigma_{\mathrm{VBF}}(\mathrm{H}^{\pm\pm}_5) \times \mathcal{B}(\mathrm{H}^{\pm\pm}_5 \rightarrow W^{\pm}W^{\pm}) = 72 \pm 25$~$\mathrm{fb}$. The global significance of the excess was also evaluated~\cite{Gross:2010qma}; it yields a global \(p\)-value of $5.4\times10^{-3}$, corresponding to a global significance of 2.5 standard deviations.
In a similar, though less sensitive, search performed by the CMS Collaboration~\cite{CMS-HIG-20-017}, no significant excess was observed in this mass region with respect to the SM background.


\FloatBarrier
 
\section{Conclusion}

 
Measurements of the fiducial and differential $W^\pm W^\pm jj$ production cross sections at $\sqrt{s}=13$~TeV using $139$~fb$^{-1}$ of data recorded by the ATLAS detector at the LHC are presented.
The measured fiducial cross section of the electroweak production of \sswwjj  in the leptonic (muon and electron) decay modes is \(\sigma^{\mathrm{EW}}_\mathrm{fid} = \sigmaew~\fb\), and for inclusive production is \(\sigma^{\mathrm{EW+Int+QCD}}_\mathrm{fid} = \sigmaewqcd~\fb\). These are the most precise fiducial cross section measurements of \sswwjj production to date. The corresponding theoretical predictions obtained from \MGNLO+\HERWIG, \(2.53  \pm 0.04\,\text{(PDF)} ~^{+0.22}_{-0.19}\,\text{(scale)}~\fb\) and \(2.92 \pm 0.05\,\text{(PDF)}  ~^{+0.34}_{-0.27}\,\text{(scale)}~\fb\), are consistent with the measurement.
Differential cross sections of the electroweak and inclusive \sswwjj production as a function of \mll, \mt, \mjj, \Ngap, and \Zep are obtained. These distributions are well described by theoretical predictions except for the \mt distribution, where for the electroweak (inclusive) production a \(p\)-value of 0.043 (0.012) is found, characterising the agreement between the measurement and nominal prediction from \MGNLO+\HERWIG.
 
The data are used to set upper and lower limits at 95\% CL on anomalous quartic gauge couplings. The limits are set on individual EFT dimension-8 operator coefficients $f_\mathrm{S02}/\Lambda^4$, $f_\mathrm{S1}/\Lambda^4$, $f_\mathrm{M0}/\Lambda^4$, $f_\mathrm{M1}/\Lambda^4$, $f_\mathrm{M7}/\Lambda^4$, $f_\mathrm{T0}/\Lambda^4$, $f_\mathrm{T1}/\Lambda^4$, and $f_\mathrm{T2}/\Lambda^4$, where $f_\mathrm{S02}$ denotes a simultaneous variation of the operators $O_\mathrm{S0}$ and $O_\mathrm{S2}$ by the same amount. These constraints are competitive with those obtained by the CMS Collaboration using the same final state, based on the 137~fb$^{-1}$ dataset. In addition, two-dimensional limits are set on pairs of operators.
 
The results are also interpreted in the context of the Georgi--Machacek model, yielding upper limits at 95\% CL on the $\sin\theta_{\mathrm{H}}$ parameter as a function of $m_{\mathrm{H}^{\pm\pm}_{5}}$. These limits exclude $\sin\theta_{\mathrm{H}}$ parameter values greater than 0.11--0.42 for $m_{\mathrm{H}^{\pm\pm}_{5}}$ values from 200 to 1500~GeV.
Upper limits at 95\% CL on $\sigma_{\mathrm{VBF}}(\mathrm{H}^{\pm\pm}_5) \times \mathcal{B}(\mathrm{H}^{\pm\pm}_5 \rightarrow W^{\pm}W^{\pm})$ are extracted for $m_{\mathrm{H}^{\pm\pm}_{5}}$ ranging from 200 to 3000~GeV. There is a local excess of events over the SM prediction at a resonance mass of around 450~GeV. The global significance of this excess is 2.5 standard deviations.


\FloatBarrier
 
\section*{Acknowledgements}


We thank CERN for the very successful operation of the LHC, as well as the
support staff from our institutions without whom ATLAS could not be
operated efficiently.
 
We acknowledge the support of
ANPCyT, Argentina;
YerPhI, Armenia;
ARC, Australia;
BMWFW and FWF, Austria;
ANAS, Azerbaijan;
CNPq and FAPESP, Brazil;
NSERC, NRC and CFI, Canada;
CERN;
ANID, Chile;
CAS, MOST and NSFC, China;
Minciencias, Colombia;
MEYS CR, Czech Republic;
DNRF and DNSRC, Denmark;
IN2P3-CNRS and CEA-DRF/IRFU, France;
SRNSFG, Georgia;
BMBF, HGF and MPG, Germany;
GSRI, Greece;
RGC and Hong Kong SAR, China;
ISF and Benoziyo Center, Israel;
INFN, Italy;
MEXT and JSPS, Japan;
CNRST, Morocco;
NWO, Netherlands;
RCN, Norway;
MEiN, Poland;
FCT, Portugal;
MNE/IFA, Romania;
MESTD, Serbia;
MSSR, Slovakia;
ARRS and MIZ\v{S}, Slovenia;
DSI/NRF, South Africa;
MICINN, Spain;
SRC and Wallenberg Foundation, Sweden;
SERI, SNSF and Cantons of Bern and Geneva, Switzerland;
MOST, Taipei;
TENMAK, T\"urkiye;
STFC, United Kingdom;
DOE and NSF, United States of America.
In addition, individual groups and members have received support from
BCKDF, CANARIE, CRC and DRAC, Canada;
PRIMUS 21/SCI/017 and UNCE SCI/013, Czech Republic;
COST, ERC, ERDF, Horizon 2020, ICSC-NextGenerationEU and Marie Sk{\l}odowska-Curie Actions, European Union;
Investissements d'Avenir Labex, Investissements d'Avenir Idex and ANR, France;
DFG and AvH Foundation, Germany;
Herakleitos, Thales and Aristeia programmes co-financed by EU-ESF and the Greek NSRF, Greece;
BSF-NSF and MINERVA, Israel;
Norwegian Financial Mechanism 2014-2021, Norway;
NCN and NAWA, Poland;
La Caixa Banking Foundation, CERCA Programme Generalitat de Catalunya and PROMETEO and GenT Programmes Generalitat Valenciana, Spain;
G\"{o}ran Gustafssons Stiftelse, Sweden;
The Royal Society and Leverhulme Trust, United Kingdom.
 
The crucial computing support from all WLCG partners is acknowledged gratefully, in particular from CERN, the ATLAS Tier-1 facilities at TRIUMF/SFU (Canada), NDGF (Denmark, Norway, Sweden), CC-IN2P3 (France), KIT/GridKA (Germany), INFN-CNAF (Italy), NL-T1 (Netherlands), PIC (Spain), RAL (UK) and BNL (USA), the Tier-2 facilities worldwide and large non-WLCG resource providers. Major contributors of computing resources are listed in Ref.~\cite{ATL-SOFT-PUB-2023-001}.


\clearpage
 
\printbibliography
 
\clearpage
 
\begin{flushleft}
\hypersetup{urlcolor=black}
{\Large The ATLAS Collaboration}

\bigskip

\AtlasOrcid[0000-0002-6665-4934]{G.~Aad}$^\textrm{\scriptsize 102}$,
\AtlasOrcid[0000-0002-5888-2734]{B.~Abbott}$^\textrm{\scriptsize 120}$,
\AtlasOrcid[0000-0002-1002-1652]{K.~Abeling}$^\textrm{\scriptsize 55}$,
\AtlasOrcid[0000-0001-5763-2760]{N.J.~Abicht}$^\textrm{\scriptsize 49}$,
\AtlasOrcid[0000-0002-8496-9294]{S.H.~Abidi}$^\textrm{\scriptsize 29}$,
\AtlasOrcid[0000-0002-9987-2292]{A.~Aboulhorma}$^\textrm{\scriptsize 35e}$,
\AtlasOrcid[0000-0001-5329-6640]{H.~Abramowicz}$^\textrm{\scriptsize 151}$,
\AtlasOrcid[0000-0002-1599-2896]{H.~Abreu}$^\textrm{\scriptsize 150}$,
\AtlasOrcid[0000-0003-0403-3697]{Y.~Abulaiti}$^\textrm{\scriptsize 117}$,
\AtlasOrcid[0000-0003-0762-7204]{A.C.~Abusleme~Hoffman}$^\textrm{\scriptsize 137a}$
% ...
\bigskip
\\

% ...
$^{29}$Physics Department, Brookhaven National Laboratory, Upton NY; United States of America.\\
% ...
$^{35}$$^{(a)}$Facult\'e des Sciences Ain Chock, R\'eseau Universitaire de Physique des Hautes Energies - Universit\'e Hassan II, Casablanca;$^{(b)}$Facult\'{e} des Sciences, Universit\'{e} Ibn-Tofail, K\'{e}nitra;$^{(c)}$Facult\'e des Sciences Semlalia, Universit\'e Cadi Ayyad, LPHEA-Marrakech;$^{(d)}$LPMR, Facult\'e des Sciences, Universit\'e Mohamed Premier, Oujda;$^{(e)}$Facult\'e des sciences, Universit\'e Mohammed V, Rabat;$^{(f)}$Institute of Applied Physics, Mohammed VI Polytechnic University, Ben Guerir; Morocco.\\
% ...
$^{49}$Fakult\"{a}t Physik , Technische Universit{\"a}t Dortmund, Dortmund; Germany.\\
% ...
$^{55}$II. Physikalisches Institut, Georg-August-Universit\"{a}t G\"ottingen, G\"ottingen; Germany.\\
% ...
$^{102}$CPPM, Aix-Marseille Universit\'e, CNRS/IN2P3, Marseille; France.\\
% ...
$^{117}$Department of Physics, New York University, New York NY; United States of America.\\
% ...
$^{120}$Homer L. Dodge Department of Physics and Astronomy, University of Oklahoma, Norman OK; United States of America.\\
% ...
$^{137}$$^{(a)}$Departamento de F\'isica, Pontificia Universidad Cat\'olica de Chile, Santiago;$^{(b)}$Millennium Institute for Subatomic physics at high energy frontier (SAPHIR), Santiago;$^{(c)}$Instituto de Investigaci\'on Multidisciplinario en Ciencia y Tecnolog\'ia, y Departamento de F\'isica, Universidad de La Serena;$^{(d)}$Universidad Andres Bello, Department of Physics, Santiago;$^{(e)}$Instituto de Alta Investigaci\'on, Universidad de Tarapac\'a, Arica;$^{(f)}$Departamento de F\'isica, Universidad T\'ecnica Federico Santa Mar\'ia, Valpara\'iso; Chile.\\
% ...
$^{150}$Department of Physics, Technion, Israel Institute of Technology, Haifa; Israel.\\
$^{151}$Raymond and Beverly Sackler School of Physics and Astronomy, Tel Aviv University, Tel Aviv; Israel.\\
% ...
$^{a}$ Also Affiliated with an institute covered by a cooperation agreement with CERN.\\
$^{b}$ Also at An-Najah National University, Nablus; Palestine.\\
$^{c}$ Also at Borough of Manhattan Community College, City University of New York, New York NY; United States of America.\\
$^{d}$ Also at Center for High Energy Physics, Peking University; China.\\
$^{e}$ Also at Center for Interdisciplinary Research and Innovation (CIRI-AUTH), Thessaloniki; Greece.\\
$^{f}$ Also at Centro Studi e Ricerche Enrico Fermi; Italy.\\
% ...

\end{flushleft}


\FloatBarrier

\end{document}